\pdfoutput = 1

\documentclass[11pt]{article}
\usepackage[utf8]{inputenc}
\usepackage[sort&compress,square,comma,numbers]{natbib}

\usepackage{fancyhdr}

\usepackage{verbatim}
\usepackage{caption}
\usepackage{subcaption}
\usepackage{amssymb}
\usepackage{amsmath}
\usepackage{tabularx}
\usepackage{color}
\usepackage{stmaryrd}
\usepackage{multirow}
\usepackage{mathtools}
\usepackage{hyperref}
\usepackage{epsfig,latexsym}

\usepackage{amsmath,amssymb,amsthm,graphicx,verbatim,bm,fancyhdr,bbm,enumerate,color, tabularx}
\usepackage{hyperref}

\usepackage[mathlines,switch]{lineno}

\DeclareMathAlphabet{\mathpzc}{OT1}{pzc}{m}{it}  

\newtheorem*{conjecture*}{Conjecture}

\theoremstyle{definition}
\theoremstyle{plain}
\newtheorem{theorem}{Theorem}
\newtheorem{proposition}[theorem]{Proposition}

\theoremstyle{remark}

\newtheorem*{remark*}{Remark}

\newtheorem*{terminology*}{Terminology}
\newtheorem*{notation*}{Notation}

\let\oldequation\equation
\let\oldendequation\endequation
\renewenvironment{equation}
  {\linenomathNonumbers\oldequation}
  {\oldendequation\endlinenomath}

\renewcommand{\P}{\mathbb{P}}

\newcommand{\E}{\mathbb{E}}

\let\oldalign\align
\let\oldendalign\endalign

\renewenvironment{align}
  {\linenomathNonumbers\oldalign}
  {\oldendalign\endlinenomath}

\title{
Exact site frequency spectra of neutrally evolving tumors: \\ a transition between power laws reveals \\ a signature of cell viability
}

\author{Einar Bjarki Gunnarsson$^{1}$ \and Kevin Leder$^1$ \and Jasmine Foo$^{2}$}

\date{%
    \footnotesize $^1$Department of Industrial and Systems Engineering, University of Minnesota, Twin Cities, MN 55455, USA. \\[2pt]%
    $^2$School of Mathematics, University of Minnesota, Twin Cities, MN 55455, USA.  \\[-200pt]
}

\renewenvironment{abstract}
 {\small
  \begin{center}
  \bfseries \abstractname\vspace{-.5em}\vspace{0pt}
  \end{center}
  \list{}{%
    \setlength{\leftmargin}{16mm}
    \setlength{\rightmargin}{\leftmargin}%
  }%
  \item\relax}
 {\endlist}

\begin{document}


\begin{center}
{\bf\large Accepted author manuscript (Theoretical Population Biology)}
\end{center}

\vspace*{-12pt}

\begingroup
\let\newpage\relax
\maketitle
\endgroup

\maketitle

\vspace*{-12pt}

\vspace*{-12pt}

\begin{abstract}
The site frequency spectrum (SFS) is a popular summary statistic of genomic data.
While the SFS of a constant-sized population undergoing neutral mutations has been extensively studied in population genetics, the rapidly growing amount of cancer genomic data has attracted interest in the spectrum of an exponentially growing population.
Recent theoretical results have generally dealt with special or limiting cases, such as considering only cells with an infinite line of descent, assuming deterministic tumor growth, or taking large-time or large-population limits. 
In this work, we derive exact expressions for the expected SFS of a cell population that evolves according to a stochastic branching process,
first for cells with an infinite line of descent and then for the total population,
evaluated either at a fixed time (fixed-time spectrum) or at the stochastic time at which the population reaches a certain size (fixed-size spectrum).
We find that while the rate of mutation scales the SFS of the total population linearly, the rates of cell birth and cell death change the shape of the spectrum at the small-frequency end, inducing a transition between a $1/j^2$ power-law spectrum and a $1/j$ spectrum as cell viability decreases.
We show that this insight can in principle be used to estimate the ratio between the rate of cell death and cell birth, as well as the mutation rate, using the site frequency spectrum alone.
Although the discussion is framed in terms of tumor dynamics, our results apply to any exponentially growing population of individuals undergoing neutral mutations.  \vspace*{6pt}
\end{abstract}

 \\[-3pt]
 

\noindent 
{\em Licensing:} 
This is an accepted author manuscript deposited under the terms of a Creative Commons Attribution, NonCommercial, NoDerivatives license (CC BY-NC-ND). \vspace*{-3pt}
 
\section{Introduction}

The study of genetic variation driven by neutral mutations
has a long history in population genetics \citep{kimura1968genetic}.
Usually, the population is assumed to be of a large constant size $N$,
and reproduction follows either the Wright-Fisher model (nonoverlapping generations) or the Moran model (overlapping generations) \citep{durrett2008probability}.
Neutral mutations occur at rate $u$ per individual per time unit, and each new mutation is assumed to be unique (the {\em infinite-sites model} of \citet{kimura1969number}).
This framework gives rise to a sample-based theory of tracing genealogies of 
extant individuals backwards in time via the {\em coalescent}
\citep{kingman1982genealogy,kingman1982coalescent}.
A popular summary statistic of genomic data 
is the {\em site frequency spectrum} (SFS), which records the frequencies of mutations in a population or population sample.
Under the Moran model with neutral mutations, the expected number of mutations found in $j$ cells of a sample of size $n$ is $\E[\xi_j] = (Nu)(1/j)$ \citep{durrett2008probability}, and any linear combination of the form $\sum_{j=1}^{n-1} j c_j \xi_j$ with $\sum_{j=1}^{n-1} c_j = 1$ is an unbiased estimator of $\theta := Nu$, the population-scaled mutation rate \citep{zeng2006statistical,achaz2009frequency}.\footnote{In Theorem 1.33 of \citet{durrett2008probability}, the result is given as $\E[\xi_j]= (2Nu)(1/j)$ for a population of size $2N$.}
Prominent estimators of this form include Watterson's $\theta_W$ \citep{watterson1975number}, Tajima's $\theta_\pi$ \citep{tajima1989statistical}, Fu and Li's $\xi_1$ \citep{fu1993statistical} and Fay and Wu's $\theta_H$ \citep{fay2000hitchhiking},
and these estimators form the basis of several statistical tests of neutral evolution vs.~evolution under selection \citep{zeng2006statistical,achaz2009frequency}.
In this way, the site frequency spectrum has provided a simple means of understanding the evolutionary history of populations from genomic data.

Cancer can be viewed as its own evolutionary process, operating at the somatic level.
Cancer initiation is usually understood to be a series of mutational events that culminates in malignant cells able to proliferate uncontrollably
\citep{armitage1954age,armitage1957two,knudson1971mutation,nowell1976clonal}.
Such ``driver'' mutations are complemented by more frequent neutral or ``passenger'' mutations \citep{tomasetti2013half,bozic2010accumulation}, that have no functional role in the evolution to malignancy, but contribute to the genetic diversity characteristic of cancer \citep{burrell2013causes,vogelstein2013cancer,mcgranahan2017clonal}.
The dominant paradigm of tumor progression has been that of sequential clonal expansion of driver mutations.
However, several recent works suggest that a neutral evolution model, under which all driver mutations are already present in the tumor-initiating cell,
is sufficient to explain the intratumoral heterogeneity in many cancers, see e.g.~\citet{sottoriva2015big}, \citet{ling2015extremely} and \citet{williams2016identification}, and the reviews by \citet{venkatesan2016tumor} and \cite{davis2017tumor}.
A simple test of neutral tumor evolution based on the site frequency spectrum was proposed in \citet{williams2016identification}, which has since generated debate e.g.~surrounding its significance level and statistical power (see e.g.~\citet{mcdonald2018currently,tarabichi2018neutral,bozic2019measuring}, with author responses in \citet{werner2018reply,heide2018reply}).
The authors of \citet{williams2016identification} subsequently suggested a Bayesian framework for detecting tumor subclones evolving under selection \citep{williams2018quantification}, and more recent approaches to that problem include \citet{dinh2020statistical} and \citet{caravagna2020subclonal}.
These works and the surrounding debate
are indicative both of the fact that increased attention is being paid to the role of 
neutral evolution
in cancer,
and that efforts are just underway to develop robust methods of inferring 
the evolutionary history of tumors \citep{turajlic2019resolving}.

While the constant-sized models of population genetics are appropriate for understanding early cancer development in small tissue compartments, exponential growth models are more relevant for understanding long-run tumor progression \citep{durrett2015branching,ohtsuki2017forward}.
In this work, we will employ a stochastic branching process model in which tumor cells divide at rate $r_0$ and die at rate $d_0$, with net birth rate $\lambda_0 := r_0-d_0>0$, and $w$ neutral mutations accumulate on average per cell division.
Let $p_0 := d_0/r_0$ be the {\em extinction probability} of the tumor, and let $q_0 := 1-p_0 = \lambda_0/r_0$ be its {\em survival probability} (Section \ref{sec:branchingprocessdynamics}).
We will show in \eqref{eq:skeletonfixedsizespectrum} of Section \ref{sec:skeletonresults} that if we only consider cells with an infinite line of descent, then at the time the number of cells becomes $N$, the expected number of mutations found in $j$ cells is $\xi_j = (w/q_0)N \cdot 1/(j(j+1))$ for $2 \leq j \leq N-1$.
This SFS differs from the SFS of the constant-sized Moran model
of population genetics in two important ways:
The spectrum now follows a $1/j^2$ power law as opposed to a $1/j$ law, and it now depends on the growth parameters $r_0$ and $d_0$ via the survival probability $q_0$.
Cumulative versions of the $1/j^2$ spectrum have previously been established by  \citet{durrett2013population,durrett2015branching}, \citet{bozic2016quantifying} and \citet{williams2016identification}, as we outline in more detail in Section \ref{sec:discussion}.
In addition, \citet{williams2016identification}, \citet{bozic2016quantifying} and \citet{ling2015extremely} have used the $1/j^2$ spectrum to infer the ratio $w/q_0$ from tumor data,
but extracting information about the mutation parameter $w$ and the growth parameter $q_0$ separately seemingly requires different tools.
In a recent work by \citet{werner2020measuring}, the authors measured pairwise mutational differences between the ancestors of spatially separated tumor bulk samples,
and they developed a coalescent-based approach for estimating $w$ and a function of $p_0$ given by \eqref{eq:kappadef} below.

Tumor evolution is commonly characterized by low cell viability, i.e.~a large extinction probability $p_0$.
For example, in a modeling study of cancer recurrence, \citet{avanzini2019cancer} collected clinical estimates of the tumor volume doubling time
and the time between cell divisions 
in metastatic breast cancer, colorectal cancer, head \& neck cancer, lung cancer and prostate cancer.
Based on the collected data, they computed a typical net growth rate ($\lambda_0$) and division rate ($r_0$) for metastases of each cancer type, which lead them to estimate $p_0=1-\lambda_0/r_0$ 
as $0.90, 0.97, 0.95, 0.97$ and $0.76$, respectively.
Similarly, in an investigation of targeted combination therapy, \citet{bozic2013evolutionary} estimated an average net growth rate of $\lambda_0=0.01$ per day for 21 melanoma lesions, which they combined with a typical division rate of $r_0=0.14$ per day \citep{rew2000cell} to compute a typical death rate of $d_0=0.13$ per day.
These estimates suggest a typical extinction probability of $p_0=0.93$ for the melanoma lesions.
Finally, \citet{bozic2016quantifying} used their cumulative version of the $1/j^2$ spectrum to estimate $w/q_0$ from the SFS of mutations at cell frequency $24\%-50\%$ in 
colorectal cancer.\footnote{\citet{bozic2016quantifying} estimated $w/q_0$ from the {\em variant allele frequency} (VAF) spectrum, which records the proportion of chromosomes carrying the mutations, as opposed to the proportion of cells carrying the mutations. They considered allele frequencies $12\%-25\%$, which translates to cell frequencies $24\%-50\%$ in the simplified setting where all cells are diploid and no cell is mutated at the same site on both chromosomes.}
For microsatellite stable (MSS) tumor samples, they combined their median estimate of $w/q_0$ with an independent estimate of the 
mutation rate \citep{jones2008comparative}
to obtain $p_0$ as 0.997.
It should be emphasized that even for a given cancer type, there is substantial heterogeneity between individual tumors (as the clinical data collected in Table 1 of \citet{avanzini2019cancer} indicate), so these values should only be taken as rough estimates.
However, these simple estimates do suggest that low cell viability is broadly relevant to tumor evolution.
Low cell viability, and the corresponding high cell turnover, induces a large mutational burden and high genetic diversity, which enhances the adaptability of the tumor under treatment.
It is therefore important to understand how low-viability tumors behave, and to explore how they can potentially be identified from genomic data.

Prior theoretical works on the expected SFS of an exponentially growing tumor population offer only a limited understanding of how $p_0$ affects the spectrum, as these works generally consider only cells with an infinite line of descent, or they consider special cases such as deterministic growth of the tumor bulk or no cell death, which is reasonable when $p_0$ is small.
Moreover, many prior results are given in the large-time or large-population limit, and for practical reasons, the focus is often on mutations of frequency $10\%$ and higher.
These results are discussed in more detail in Section \ref{sec:discussion} below.
Our goal in this work is to gain a more complete understanding of the SFS of an exponentially growing tumor with neutral mutations.
We seek to understand how the spectrum behaves both at small and large frequencies, for all values of $p_0$, and for any population size $N$.
We obtain separate results for cells with an infinite line of descent and for the total population,
evaluated either at a fixed time or at the stochastic time at which the population reaches a certain size, each of which is relevant to tumor data analysis depending on the context.
We observe that while the SFS of cells with an infinite line of descent depends on the mutation rate $w$ and the extinction probability $p_0$ only via the ratio $w/q_0$, the two parameters decouple in the SFS of the total population.
In fact, as $p_0$ increases from 0 to 1, the small-frequency end of the spectrum transitions from the $1/j^2$ power law characteristic of pure-birth exponential growth to the $1/j$ power law characteristic of a constant-sized population.
We investigate simple metrics that quantify this transition, and use one of them to propose a simple estimator for $p_0$, which we subsequently 
evaluate using idealized synthetic single-cell sequencing data.

The rest of the paper is organized as follows. 
In Section \ref{sec:priormodels}, we formulate our branching process model with neutral mutations, define the {\em skeleton subpopulation} of cells with an infinite line of descent, and establish relevant notation.
In Section \ref{sec:prelimskeletonresults}, we analyze the SFS of skeleton cells, and in Section \ref{sec:results}, we analyze the SFS of the total cell population.
In Section \ref{sec:signatures}, we use our theoretical results to propose and evaluate a simple estimator for $p_0$.
In Section \ref{sec:discussion}, we summarize our results, and discuss in detail how they relate to the existing literature.
The proofs of all of our theoretical results are found in appendices at the end of the paper.

                 \begin{table*}[t]
    \centering
\hspace*{-1.4cm}
    \begin{tabular}{|l|l|l|}
    \hline 
    Symbol&Description&Definition \\
    \hline
    $r_0$ & Division rate of tumor cells (per unit time) & Section \ref{sec:branchingprocessdynamics} \\
    $d_0$ & Death rate of tumor cells (per unit time) & Section \ref{sec:branchingprocessdynamics} \\
    $p_0$ & Extinction probability of the tumor (and of a single-cell derived clone) &\eqref{eq:extinctionprob} in Section \ref{sec:branchingprocessdynamics} \\ 
    $q_0$ & Survival probability of the tumor (and of a single-cell derived clone) &\eqref{eq:survprob} in Section \ref{sec:branchingprocessdynamics} \\
    $w$ & Mutation rate (expected number of mutations per cell division) & Section \ref{sec:mutationaccumulation} \\
    $\tilde t_N$ & Fixed time at which the skeleton subpopulation has expected size $N$ &\eqref{eq:fixedtimeskeleton} in Section \ref{sec:skeletonresults} \\
    $\tilde \tau_N$ & Stochastic time at which the skeleton subpopulation first reaches size $N$ &\eqref{eq:stoppingtimeskeleton} in Section \ref{sec:skeletonresults} \\
    $t_N$ & Fixed time at which a surviving tumor cell population has expected size $N$ &\eqref{eq:fixedtime} in Section \ref{sec:generalspectra} \\
    $\tau_N$ & Stochastic time at which the tumor cell population first reaches size $N$ &\eqref{eq:stoppingtime} in Section \ref{sec:generalspectra} \\
    \hline
    \end{tabular}
    \caption{
    Notation used in the paper. 
    }
    \label{table:samplingscheme1}
\end{table*}

\section{Model description} \label{sec:priormodels}

\subsection{Branching process dynamics} \label{sec:branchingprocessdynamics}

We assume that the tumor evolution follows a branching process model in continuous time.
Cells divide into two cells at rate $r_0 > 0$ per unit time and  die at rate $d_0 \geq 0$ per unit time, which means that in a small time interval of length $\Delta t$, a cell divides with probability $r_0 \Delta t$ and dies with probability $d_0 \Delta t$.
Assume $r_0>d_0$ and define $\lambda_0 := r_0-d_0>0$ as the net growth rate.
Let $Z_0(t)$ denote the size of the tumor population at time $t$ and assume $Z_0(0) = 1$, i.e.~the tumor expands from a single tumor-initiating cell.
Define
\begin{linenomath*}
\begin{align*}
    \Omega_\infty := \{Z_0(t)>0 \text{ for all $t >0$}\}    
\end{align*}
\end{linenomath*}
as the event that the tumor does not go extinct, and
\begin{align} \label{eq:extinctionprob}
    p_0 := \P(\Omega_\infty^c) = \P(Z_0(t) = 0 \text{ for some $t>0$})
\end{align}
as the {\em extinction probability} of the tumor.
This probability can be computed as $p_0 = d_0/r_0$ with $0 \leq p_0<1$, see e.g.~Section 3 of \citet{durrett2015branching}.
Note that any clone derived from a single tumor cell gives rise to its own branching process with the same growth parameters $r_0$ and $d_0$ and the same extinction probability.
We also define 
\begin{align} \label{eq:survprob}
    q_0 := 1-p_0 = \lambda_0/r_0
\end{align}
as the {\em survival probability} of the tumor or of a single-cell derived clone.

\subsection{Decomposition into skeleton cells and finite-family cells}

On the nonextinction event $\Omega_\infty$, the cells alive at time $t>0$ 
can be split into two categories, one consisting of cells with an infinite line of descent, i.e.~cells that start clones that do not go extinct, and the other consisting of cells whose descendants eventually go extinct.
We refer to the former cells as {\em skeleton} cells and the latter as {\em finite-family} cells.
An arbitrary tumor cell is a skeleton cell with probability $q_0$, so in the long run, the proportion of skeleton cells in the population is $q_0$.
We can think of skeleton cells as forming the trunk and scaffold branches
of the genealogical branching tree, with finite-family clones growing out from the skeleton as lateral branches, see Figure \ref{fig:skeletondiagram}a.

\subsection{Mutation accumulation} \label{sec:mutationaccumulation}

We next add neutral mutations under the infinite-sites model.
Prior to a cell division, each parental DNA molecule is unwound and separated into two complementary strands.
Each parental strand serves as a template for the construction of a new complementary daughter strand.
The end result is two copies of the DNA molecule, each consisting of one parental and one daughter strand.
Errors in nucleotide pairing during this process can result in one or more point mutations per daughter strand.
We assume that these errors amount to $w/2$ mutations on average per daughter strand, for a total of $w$ mutations on average per cell division.
Note that the only assumptions we make 
on the distribution of the number of mutations is that it is nonnegative and integer-valued with a finite mean.
The point mutation rate has been estimated as $5 \cdot 10^{-10}$ per base pair per cell division \citep{jones2008comparative}, and it is commonly higher in cancer due to genomic instability \citep{burrell2013causes}.
Since the number of base pairs is of order $10^7$ in the exome and $10^9$ in the genome, it makes sense to allow $w$ to be any positive number, i.e.~$w \in (0,\infty)$.
In many works, the convention is to allow at most one mutation per cell division, introducing a probability $u \in (0,1)$ of a new mutation.
Since our analysis only depends on the mean number of mutations $w$ per cell division, it includes this case with $w:=u$.
We assume that the mutation rate is constant throughout tumor evolution, which ignores e.g.~the possibility of an elevated mutation rate over time due to genomic instability.

While our focus in this work is on discrete mutation accumulation, we will also present all of our results in terms of continuous mutation accumulation, another common and biologically relevant assumption.
In the continuous model, neutral mutations occur at rate $\nu>0$ per cell per unit time, at any time throughout the lifetime of the cell.
In other words, each cell undergoes a neutral mutation in a small time interval of length $\Delta t$ with probability $\nu \Delta t$.
The continuous model differs from the discrete model in that at most one mutation occurs at a time, and this mutation occurs in between cell divisions with probability 1.
However, as we will show, the mean behavior of the two models is similar when $\nu = wr_0$, since in the discrete model, each cell accumulates $wr_0\Delta t$ mutations on average 
in a small time interval of length $\Delta t$.

\subsection{Clonal and subclonal mutations}

Before proceeding, we need to make a distinction between {\em clonal} and {\em subclonal} mutations.
A mutation is clonal if it is shared by all tumor cells, while it is subclonal if there is at least one tumor cell without it.
As an example, say the tumor-initiating cell divides into two cells, $A$ and $B$, and that cell $A$ acquires a new mutation.
This mutation is initially subclonal.
However, if the clone started by cell $B$ dies out, the mutation in cell $A$ becomes clonal from that point onward.
While this example demonstrates how clonal mutations can arise post-tumor-initiation, all mutations that accumulate prior to initiation, as the cancer precursor cell evolves to malignancy, also become clonal.
For this reason, the clonal mutations usually tell us more about the events preceding cancer than the dynamics post-initiation, and they can in fact outnumber the subclonal mutations \citep{tomasetti2013half}.
Nevertheless, clonal mutations do appear in the SFS of mutations post-initiation, and they play distinct roles in the fixed-time and fixed-size spectrum, which is why we pay them special attention below.

         \begin{figure*}
             \centering
             \includegraphics[scale=1]{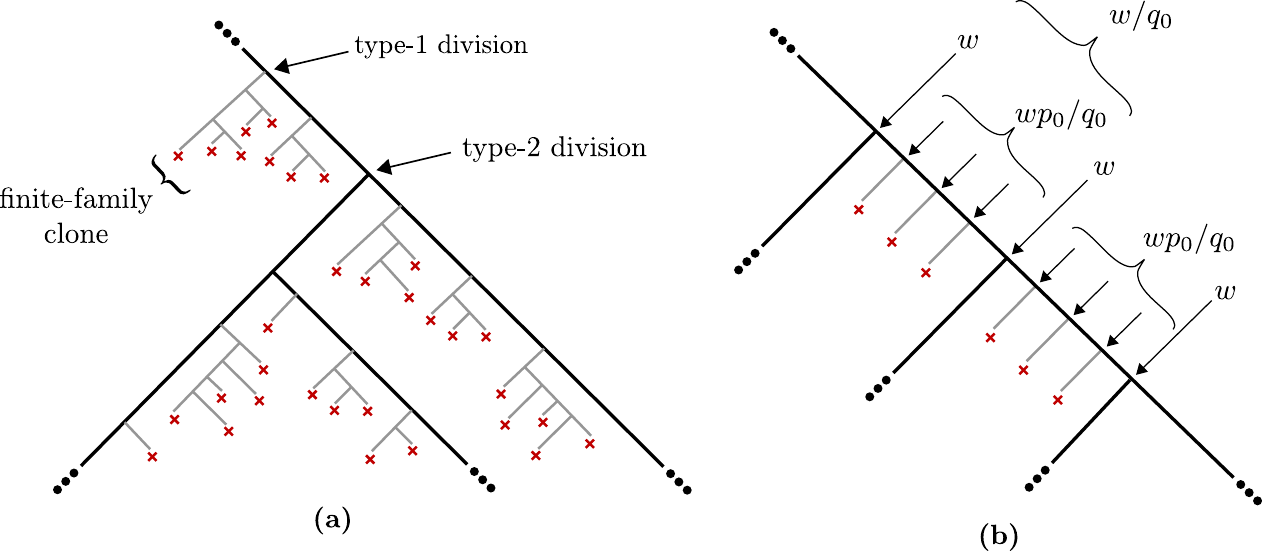}
             \caption{
              Categorization of cells into skeleton cells and finite-family cells, and mutation accumulation on the skeleton.
             {\bf (a)} The skeleton subpopulation, which consists of cells with an infinite line of descent, can be thought of as forming the trunk and scaffold branches of the genealogical branching tree (bold branches).
             Each skeleton cell divides into one skeleton cell and one finite-family cell at rate $2r_0p_0=2d_0$ (type-1 skeleton division, see \eqref{eq:type1skeleton} in Section \ref{sec:skeletonproperties}), in which case a finite-family clone grows out from the skeleton as a lateral branch (light gray branches).
             A skeleton cell divides into two skeleton cells at rate $r_0q_0 = \lambda_0$ (type-2 skeleton division, see \eqref{eq:type2skeleton} in Section \ref{sec:skeletonproperties}), in which case another scaffold branch is added.
             {\bf (b)} In between two type-2 skeleton divisions, the expected number of mutations that accumulate due to type-1 divisions is $wp_0/q_0$ by \eqref{eq:type1mutations}.
             Each type-2 division adds $w$ mutations on average and starts a new skeleton population size level.
             Thus, the expected number of mutations that accumulate on the skeleton per size level is $w+wp_0/q_0=w/q_0$.
             }
             \label{fig:skeletondiagram}
         \end{figure*}

\section{Site frequency spectrum of skeleton cells} \label{sec:prelimskeletonresults}

In this section, we establish the expected fixed-time and fixed-size spectrum of skeleton cells.
The reason we are interested in analyzing skeleton cells separately is twofold:
\begin{itemize}
    \item When $p_0=0$ (no cell death), all cells are skeleton cells, and the SFS of the total population is the SFS of the skeleton.
    More generally, when $p_0$ is small, the total population spectrum is well-approximated by the simpler skeleton spectrum.
    \item When the extinction probability $p_0$ is large, finite-family cells affect the SFS of the total population at the small-frequency end.
    However, the large-frequency end is still characterized by the skeleton spectrum, as we demonstrate in Section \ref{sec:results} below.
\end{itemize}

\subsection{Effective rates of cell division and mutation} \label{sec:skeletonproperties}

When the tumor is conditioned on nonextinction, the probability that the tumor-initiating cell divides during the first $\Delta t$ units of time, and that exactly one of the two daughter cells survives, i.e.~starts a clone that does not go extinct, is 
\begin{align} \label{eq:type1skeleton}
    &\frac{\P(\text{division in $[0, \Delta t]$, one offspring survives})}{\P(\Omega_\infty)} = \frac{r_0 \Delta t \cdot 2p_0(1-p_0) }{1-p_0} = 2r_0p_0 \Delta t,    
\end{align}
and the probability of a division in $[0,\Delta t]$ where both daughter cells survive is
\begin{align} \label{eq:type2skeleton}
    &\frac{\P(\text{division in $[0, \Delta t]$, both offspring survive})}{\P(\Omega_\infty)} = \frac{r_0 \Delta t \cdot (1-p_0)^2}{1-p_0} = r_0q_0 \Delta t.
\end{align}
Since each skeleton cell starts a clone that does not go extinct, we can conclude that a skeleton cell divides into one skeleton cell and one finite-family cell at rate $2r_0p_0=2d_0$ per unit time (type-1 skeleton division), and it divides into two skeleton cells at rate $r_0q_0=\lambda_0$ per unit time (type-2 skeleton division).
The probability that a skeleton division is type-2 is
\begin{align} \label{eq:kappadef}
    \kappa_0 := \frac{r_0q_0}{r_0q_0+2r_0p_0} = \frac{1-p_0}{1+p_0}.
\end{align}
This probability can also be computed directly as follows: $(1-p_0)^2$ is the probability that both daughter cells survive, and $1-p_0^2$ is the probability that at least one of them does, so $(1-p_0)^2/(1-p_0^2) = (1-p_0)/(1+p_0)$ is the probability that a skeleton division is type-2.

Let $\tilde Z_0(t)$ denote the number of skeleton cells at time $t$, conditional on the nonextinction event $\Omega_\infty$.
Since type-1 divisions do not affect the size of the skeleton, we can think of the type-2 divisions as the ``effective'' divisions.
More precisely, $(\tilde Z_0(t))_{t \geq 0}$ is a pure-birth exponential growth process, known as a {\em Yule process}, with birth rate $\lambda_0$ and mean size $\E[\tilde Z_0(t)] = e^{\lambda_0t}$ at time $t$ \citep{durrett2015branching,o1993yule}.
Type-1 divisions do contribute to neutral mutation accumulation however.
Indeed, each type-1 division adds $w/2$ mutations on average to the skeleton, and each type-2 division adds $w$ mutations on average.
The rate at which mutations accumulate on the skeleton is then
\begin{align} \label{eq:mutrateskeleton}
    (w/2) \cdot 2r_0p_0 + w \cdot r_0(1-p_0) = wr_0
\end{align}
per skeleton cell per unit time, which equals the mutation rate for the original, unconditioned process $(Z_0(t))_{t \geq 0}$.
The mutation rate per type-2 division, or the {\em effective} mutation rate, is on the other hand
\begin{align} \label{eq:normalizedmutationrate}
    wr_0/\lambda_0 = w/q_0
\end{align}
per unit time.
We can also think of mutations as accumulating across skeleton population size levels as follows.
A type-2 division increases the size of the skeleton population by one, and it adds $w$ mutations on average.
Upon the type-2 division, 
the number of type-1 divisions before the next type-2 division has the geometric distribution with support $\{0,1,2\ldots\}$ and success probability $\kappa_0$ given by \eqref{eq:kappadef}.
It follows that in between the two type-2 divisions, the expected number of mutations that accumulate on the skeleton is
\begin{align} \label{eq:type1mutations}
    (w/2) \cdot (1/\kappa_0-1) = (w/2) \cdot 2p_0/q_0 = wp_0/q_0.
\end{align}
At each population size level, the skeleton therefore accumulates $w$ mutations on average due to the type-2 division that starts the level, and $wp_0/q_0$ mutations on average due to type-1 divisions that occur before the next type-2 division that changes levels.
We thus obtain
\begin{align} \label{eq:mutperpopulationlevel}
    w+wp_0/q_0 = w/q_0
\end{align}
mutations per level (Figure \ref{fig:skeletondiagram}b).
The effective mutation rate $w/q_0$
plays a key role in the SFS of the skeleton, with the continuous-time viewpoint in \eqref{eq:normalizedmutationrate} applying to the fixed-time spectrum, and the population-size-level viewpoint in \eqref{eq:mutperpopulationlevel} applying to the fixed-size spectrum.

\subsection{Expected fixed-time and fixed-size skeleton spectrum} \label{sec:skeletonresults}

Let $\tilde{S}_{j}(t)$ denote the number of mutations that are found in $j \geq 1$ skeleton cells at time $t$, conditional on the nonextinction event $\Omega_\infty$.
This is the site frequency spectrum of skeleton cells.
For any integer $N \geq 1$, define
\begin{align} \label{eq:fixedtimeskeleton}
    \tilde{t}_N := \log(N)/\lambda_0
\end{align}
as the (fixed) time at which the skeleton has expected size $N$, i.e.~$e^{\lambda_0 \tilde{t}_N} = N$, and define
\begin{align} \label{eq:stoppingtimeskeleton}
\tilde\tau_N := \inf\{t \geq 0: \tilde{Z}_0(t) = N\}    
\end{align}
as the (stochastic) time at which the skeleton reaches size $N$.
In Proposition \ref{prop:skeletonspectrum} below, we provide the expected SFS of the skeleton evaluated both at time $\tilde{t}_N$ (fixed-time spectrum) and at time $\tilde\tau_N$ (fixed-size spectrum).
Both the fixed-time and fixed-size spectrum can be relevant to tumor data analysis depending on the context.
For example, {\em in vitro} cell culture experiments and {\em in vivo} mouse experiments are often conducted over a fixed time period, in which case the fixed-time spectrum would apply.
In the clinic, however, the size of a tumor sample is more readily estimated than its age, in which case the fixed-size spectrum is more relevant \citep{komarova2007fixed}.
It is therefore useful to understand both spectra and to what extent they differ.

\begin{proposition} \label{prop:skeletonspectrum}
\begin{enumerate}[(1)]
    \item Define $\tilde t_N$ as in \eqref{eq:fixedtimeskeleton}. Then, for any $N \geq 1$ and any $j \geq 1$,
    \begin{align} \label{eq:skeletonfixedtimespectrum}
    \begin{array}{lll}
    \begin{split}
                 \E[\tilde{S}_j(\tilde t_N)]  &= \textstyle (w/q_0) N \cdot \int_0^{1-1/N} (1-y)y^{j-1} dy \\[3pt]
         &= \textstyle (w/q_0)N \cdot \big(1-\frac1N)^j \big(\frac1{j(j+1)}+\frac1N \frac1{j+1}\big).
    \end{split}
\end{array}
    \end{align}
For fixed $j \geq 1$, then as $N \to \infty$,
\begin{align} \label{eq:asymptspectrumskeleton}
    \textstyle \E[\tilde{S}_j(\tilde t_N)] &\sim 
    (w/q_0)N \cdot 1/(j(j+1)),
\end{align}
where $f(y) \sim g(y)$ as $y \to \infty$ means $\lim_{y \to \infty} f(y)/g(y) = 1$.
    \item Define $\tilde\tau_N$ as in \eqref{eq:stoppingtimeskeleton}. Then, for any $N \geq 2$,
        \begin{align} \label{eq:skeletonfixedsizespectrum}
                  & \E[\tilde S_j(\tilde\tau_N)] = \begin{cases} (w/q_0)N \cdot 1/(j(j+1))-(wp_0/q_0)\delta_{1,j}, & 1 \leq j \leq N-1, \\
                  wp_0/q_0=w/q_0-w, &  j = N, \end{cases}
    \end{align}
    where 
    $\delta_{\ell,m}=1$ if $\ell = m$ and $\delta_{\ell,m}=0$ otherwise. 
\end{enumerate}
\end{proposition}

\begin{proof}
    Appendix \ref{app:proofprop1}.
\end{proof}

         \begin{figure*}[t]
             \centering
             \includegraphics[scale=1]{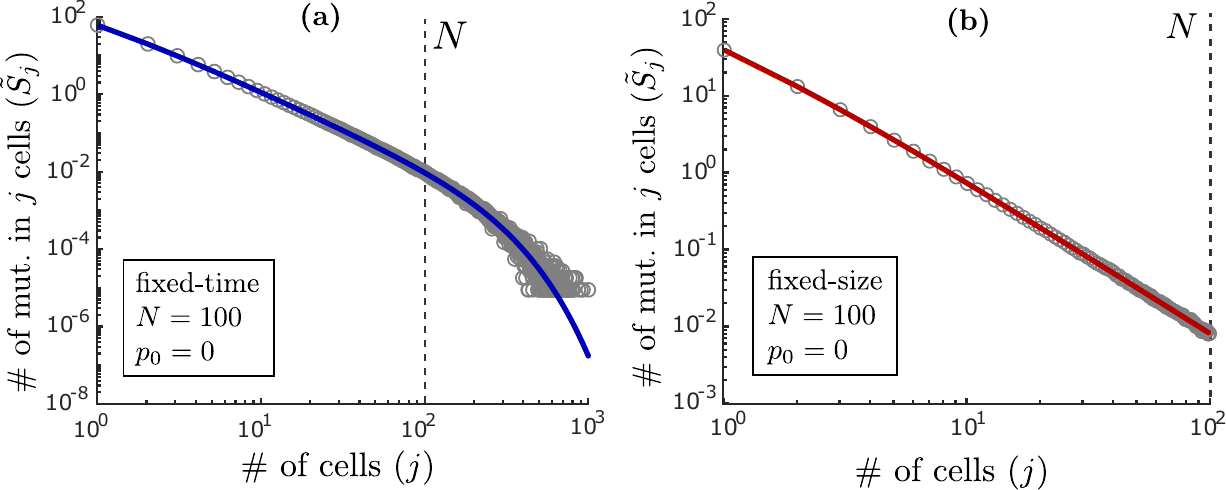}
             \caption{
             Comparison between the expected SFS of the skeleton, as derived in Proposition \ref{prop:skeletonspectrum}, and simulation results.
             {\bf (a)} The expected fixed-time spectrum \eqref{eq:skeletonfixedtimespectrum} of Proposition \ref{prop:skeletonspectrum} (solid blue line) shows good agreement with the average spectrum of simulated tumors (grey dots).
             In this example, the extinction probability is $p_0=0$, the mutation rate is $w=1$, and the expected size of the skeleton is $N=100$, which is also the expected size of the tumor since $p_0=0$.
             We generated $10^5$ tumors with $p_0=0$ and $w=1$ and stopped each simulation at the fixed time $\tilde{t}_N$ with $N=100$ as defined by \eqref{eq:fixedtimeskeleton}.
             At each cell division, the number of mutations acquired by each daughter cell was generated as a Poisson random variable with mean $w/2$.
             {\bf (b)} The expected fixed-size spectrum \eqref{eq:skeletonfixedsizespectrum} of Proposition \ref{prop:skeletonspectrum} (solid red line) shows good agreement with the average spectrum of simulated tumors (grey dots).
             We again generated $10^5$ tumors with $p_0=0$ and $w=1$, but this time, we stopped each simulation when the tumor reached size $N=100$, i.e.~at the stochastic time $\tilde{\tau}_N$ defined by \eqref{eq:stoppingtimeskeleton}.
             The fundamental difference between the fixed-time and fixed-size spectrum is that the skeleton size at time $\tilde t_N$ is variable, while it is always $N$ at time $\tilde \tau_N$.
             As a result, the fixed-size spectrum is restricted to $j=1,\ldots,N$, while the fixed-time spectrum has nonzero mass at values $j>N$.
             }
             \label{fig:skeletonspectrafig}
         \end{figure*}

\begin{figure*}
             \centering
             \includegraphics[scale=1]{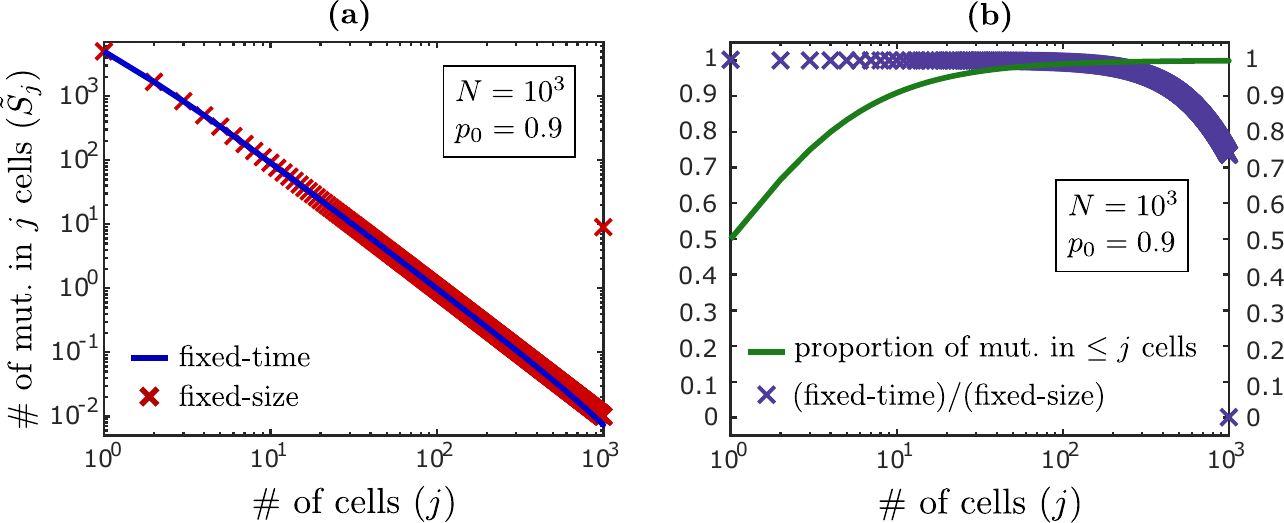}
             \caption{
             Comparison between the expected fixed-time spectrum \eqref{eq:skeletonfixedtimespectrum} and fixed-size spectrum \eqref{eq:skeletonfixedsizespectrum} of the skeleton as derived in Proposition \ref{prop:skeletonspectrum}.
             {\bf (a)} The fixed-time spectrum \eqref{eq:skeletonfixedtimespectrum} of the skeleton (blue curve) is a good approximation of the fixed-size spectrum \eqref{eq:skeletonfixedsizespectrum} (red crosses) for mutations at frequencies $j \ll N$, given parameters $N=10^3$, $p_0=0.9$ and $w=1$.
             The two spectra diverge at the very largest frequencies, and the difference is substantial at $j=N$, since clonal mutations are concentrated at $j=N$ in the fixed-size spectrum, while they are scattered in the fixed-time spectrum.
             {\bf (b)} Here, we show the ratio of the fixed-time spectrum to the fixed-size spectrum (purple crosses), and the proportion of mutations found in $\leq j$ skeleton cells in the fixed-size spectrum (green curve).
             The fixed-time and fixed-size spectrum are virtually the same on $j \ll N$, where almost all mutations are found.
             }
             \label{fig:skeletonspectrafig2}
         \end{figure*}

Analogous results for continuous mutation accumulation are presented in Appendix \ref{app:continuousmutation}.
In Figure \ref{fig:skeletonspectrafig}, we compare our fixed-time \eqref{eq:skeletonfixedtimespectrum}
and fixed-size \eqref{eq:skeletonfixedsizespectrum} results
with simulation results for $w=1$, $p_0=0$ and $N= 100$.
In this example, there are no clonal mutations, since $p_0=0$.
The fundamental difference between the fixed-time and fixed-size spectrum is that the skeleton size at time $\tilde t_N$ is variable, while it is always $N$ at time $\tilde\tau_N$.
The fixed-time spectrum therefore has nonzero mass at $j>N$, due to instances in which the skeleton is larger than $N$ at time $\tilde t_N$.
It is however natural to ask how the fixed-time spectrum restricted to $j=1,\ldots,N$ relates to the fixed-size spectrum.
By \eqref{eq:skeletonfixedsizespectrum}, the fixed-size spectrum follows the power law $(w/q_0)N \cdot 1/(j(j+1))$ exactly on $j=2,\ldots,N-1$, and asymptotically as $N \to \infty$ for $j=1$.
By \eqref{eq:asymptspectrumskeleton}, the fixed-time spectrum converges to the same power law for fixed $j \geq 1$ as $N \to \infty$, which means that it follows this power law when $N$ is large and $j \ll N$.
In Figure \ref{fig:skeletonspectrafig2}, we compare the fixed-time \eqref{eq:skeletonfixedtimespectrum} and fixed-size \eqref{eq:skeletonfixedsizespectrum} spectrum for $N=10^3$ and $p_0=0.9$.
As expected, the two spectra agree on $j \ll N$, while the fixed-time spectrum deviates from the fixed-size spectrum at the very largest frequencies (for $j$ of order $N$).
In Figure \ref{fig:skeletonspectrafig2}b, we show that almost all mutations are found on $j \ll N$. 
In this example, the difference between the two spectra is within 1\% on the range $j=1,\ldots,150$, on which 99.3\% of mutations are found.

We next note the sharp discontinuity at $j=N$ in the fixed-size spectrum of Figure \ref{fig:skeletonspectrafig2}a, which does not appear in the fixed-time spectrum.
This is due to the distinct ways in which clonal mutations manifest in the two spectra.
In the fixed-time spectrum, clonal mutations can appear at any value of $j$, depending on the skeleton size at time $\tilde t_N$.
In the fixed-size spectrum, all these mutations are concentrated at $j=N$, which creates a significant point mass at $j=N$.
Note that by \eqref{eq:skeletonfixedsizespectrum}, the expected number of mutations found in $j=N-1$ skeleton cells is given by $(w/q_0)(1/(N-1))$ in the fixed-size spectrum, which is of order $1/N$ as $N \to \infty$, while clonal mutations are given by the constant $wp_0/q_0$, independently of $N$.

The observed difference between the fixed-time and fixed-size spectrum at the very largest frequencies reflects the unbounded range of the fixed-time spectrum.
When $N$ is large, this difference can be alleviated by computing the SFS of mutations found in a given {\em proportion} of cells (as opposed to a given {\em number} of cells), as this normalizes both spectra to the frequency range $[0,1]$.
To justify this claim, we need to combine our results with results previously obtained by \citet{durrett2013population,durrett2015branching} and \citet{bozic2016quantifying}, as we discuss in detail in Section \ref{sec:discussion} below.

\subsection{Proportion of mutations found in one cell} \label{sec:additionalmetricsskeleton}

We conclude this section by computing two simple and important metrics derived from the site frequency spectrum of the skeleton.
Note first that by 
Proposition \ref{prop:skeletonspectrum}, 
the expected number of mutations found in a single skeleton cell is
\begin{align} \label{eq:nummutonecellskeleton}
\begin{split}
        & \E[\tilde{S}_1(\tilde t_N)] = (1/2)(w/q_0)(N^2-1)/N \sim (1/2)(w/q_0)N, \\ 
        & \E[\tilde{S}_1(\tilde \tau_N)] = (1/2)(w/q_0)N-wp_0/q_0 \sim (1/2)(w/q_0)N, 
\end{split}
\end{align}
as $N \to \infty$.
Next, let $\tilde{M}_{j}(t) := \sum_{k \geq j} \tilde{S}_{k}(t)$ denote the number of mutations found in $\geq j$ skeleton cells at time $t$.
Again, by 
Proposition \ref{prop:skeletonspectrum}, 
the expected total number of mutations on the skeleton is
\begin{align} \label{eq:totalnummutskeleton}
\begin{split}
    \E[\tilde M_1(\tilde t_N)] &= (w/q_0) (N-1) \sim (w/q_0) N,  \\
    \E[\tilde M_1(\tilde \tau_N)] &= (w/q_0)(N-1) \sim (w/q_0)N,
\end{split}
\end{align}
as $N \to \infty$.
Expressions \eqref{eq:nummutonecellskeleton} and \eqref{eq:totalnummutskeleton} suggest that for large $N$, half the mutations discovered at time $\tilde t_N$ or $\tilde \tau_N$
are found in only one cell.
This is a consequence of the pure-birth exponential growth of the skeleton.
Indeed, note that if we only consider the effective type-2 skeleton divisions, the total number of divisions required to reach generation $k$ is $\sum_{j=0}^{k-1} 2^j = 2^k-1$. 
The expected total number of mutations in generation $k$ is then $(w/q_0)(2^k-1)$, which is \eqref{eq:totalnummutskeleton} with $N=2^k$ the number of cells in generation $k$.
An additional $2^k$ divisions are required to reach generation $k+1$, so each generation roughly doubles the total number of mutations.
Of course, our model is stochastic, it operates in continuous time, and generations may overlap, but this simple discrete argument gives intuition as to why half the mutations are found in one cell, and more generally why most mutations are found at the smallest frequencies.

\section{Site frequency spectrum of total population and transition between power laws} \label{sec:results}

When the extinction probability $p_0$ is small, the SFS of the total population $(Z_0(t))_{t \geq 0}$ is well-approximated by the SFS of the skeleton $(\tilde Z_0(t))_{t \geq 0}$.
However, tumor evolution is commonly characterized by a large extinction probability, as was discussed in the introduction.
In this section, we  investigate the expected fixed-time and fixed-size spectrum of the total population $(Z_0(t))_{t \geq 0}$ for all values of $p_0$.
We show that as $p_0$ increases, the small-frequency end of the spectrum starts to deviate from the skeleton spectrum of Section \ref{sec:prelimskeletonresults}, and that as $p_0$ approaches 1, it transitions to the spectrum of a constant-sized population.

\subsection{Expected fixed-time and fixed-size total population spectrum} \label{sec:generalspectra}

Let $S_{j}(t)$ denote the number of mutations found in $j \geq 1$ cells at time $t$, the site frequency spectrum of the total cell population.
We wish to compute the mean of $S_j(t)$ conditioned on the tumor surviving to time $t$.
We can compute the probability of 
this survival event as
\[
\P(Z_0(t)>0) = q_0 e^{\lambda_0t} /(e^{\lambda_0t}-p_0), \quad t \geq 0,
\]
see \eqref{eq:probnonextinbytimet} of  Appendix \ref{app:proofprop2}, 
and the expected size of a tumor that survives to time $t$ as
\begin{align} \label{eq:sizetumorconditional}
    \E[Z_0(t)|Z_0(t)>0] = (e^{\lambda_0t}-p_0)/q_0, \quad t \geq 0.    
\end{align}
Note that $\E[Z_0(t)|Z_0(t)>0] \sim e^{\lambda_0t}/q_0$ as $t \to \infty$, which means that the long-run expected growth of a tumor conditioned on survival is exponential, and the initial value of the exponential growth function is given by $1/q_0$.
An interesting consequence of the conditioning on survival is that if $r_0$ and $d_0$ are increased by the same amount (so that $\lambda_0=r_0-d_0$ stays fixed), the survival probability $q_0=\lambda_0/r_0$ will decrease, while the size of a tumor conditioned on survival will increase.
Now, for any integer $N \geq 1$, define
\begin{align} \label{eq:fixedtime}
    {t}_N := \log(q_0N+p_0)/\lambda_0
\end{align}
as the (fixed) time at which a surviving tumor has expected size $N$, i.e.~$(e^{\lambda_0 {t}_N}-p_0)/q_0 = N$,
and define
\begin{align} \label{eq:stoppingtime}
\tau_N := \inf\{t \geq 0: {Z}_0(t) = N \}    
\end{align}
as the (stochastic) time at which the tumor reaches size $N$, with $\inf \varnothing = \infty$.
In Proposition \ref{prop:spectrum}, we provide the expected SFS of $(Z_0(t))_{t \geq 0}$ at time $t_N$ and $\tau_N$, conditioned on the survival events $\{Z_0(t_N)>0\}$ and $\{\tau_N < \infty\}$, respectively.
In this case, we cannot obtain an explicit expression for the fixed-size spectrum, and provide instead a computational expression.
Note however that for the case $p_0=0$ of no cell death, the fixed-size spectrum follows the power law $wN \cdot 1/(j(j+1))$ exactly on $j=2,\ldots,N-1$, by \eqref{eq:skeletonfixedsizespectrum} of Proposition \ref{prop:skeletonspectrum}.

\begin{proposition} \label{prop:spectrum}
\begin{enumerate}[(1)]
    \item Define $t_N$ as in \eqref{eq:fixedtime}. Then, for any $N \geq 1$ and any $j \geq 1$,
    \begin{align} \label{eq:fixedtimeresultgeneral}
    \begin{split}
        & \E[S_j(t_N)|Z_0(t_N)>0] = \textstyle wN \int_0^{1-1/N} (1-p_0y)^{-1} (1-y) y^{j-1} dy. 
    \end{split}
\end{align}
For fixed $j \geq 1$, then as $N \to \infty$,
\begin{align} \label{eq:asymptspectrumgeneral}
\begin{split}
    \E[S_j(t_N)|Z_0(t_N)>0] &\sim \textstyle wN \cdot \int_0^{1} (1-p_0y)^{-1} (1-y)y^{j-1} dy \\
    &= \textstyle  wN \cdot \sum_{k=0}^\infty \frac{p_0^k}{(j+k)(j+k+1)},
\end{split}
\end{align}
where $f(y) \sim g(y)$ as $y \to \infty$ means $\lim_{y \to \infty} f(y)/g(y) = 1$.
    \item Define $\tau_N$ as in \eqref{eq:stoppingtime}, let ${\cal S} := \{(\ell,m): \ell,m \geq 0 \text{ and } \ell+m \leq N\}$ and $A := \{(0,0)\} \cup \{(r,s): r,s \geq 0 \text{ and } r+s = N\}$. Then, for any $N \geq 2$ and any $1 \leq j \leq N$, 
        \begin{align} \label{eq:fixedsizeresultgeneral}
            \E[S_j(\tau_N)|\tau_N<\infty] 
            &= (w/q_0) \cdot \textstyle \sum_{k=1}^{N-1} (1-p_0^{N-k}) \cdot h^{(j,N-j)}_{(1,k)}, 
    \end{align}
     where for each $(r,s) \in A$, the vector $\big(h_{(\ell,m)}^{(r,s)}\big)_{(\ell,m) \in {\cal S}}$
    solves the system
    \begin{align} \label{eq:linearsystem}
    \begin{split}
        & (\ell+m) (1+p_0) h_{(\ell,m)}^{(r,s)} = \ell  h_{(\ell+1,m)}^{(r,s)} + \ell p_0  h_{(\ell-1,m)}^{(r,s)} + m  h_{(\ell,m+1)}^{(r,s)} + m p_0  h_{(\ell,m-1)}^{(r,s)}
    \end{split}
    \end{align}
    for $\ell,m \geq 1$ and $\ell+m<N$, with boundary conditions given by \eqref{eq:boundaryconditions} in Appendix \ref{app:proofprop2}.
\end{enumerate}
\end{proposition}

\begin{proof}
     Appendix \ref{app:proofprop2}.
\end{proof}

                           \begin{figure*}[t]
             \centering
             \includegraphics[scale=1]{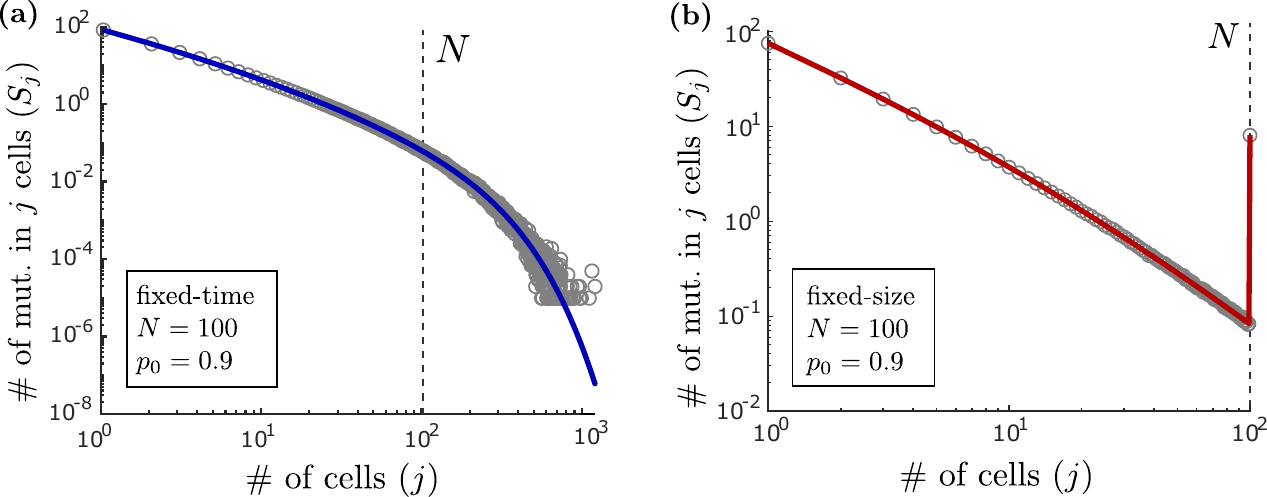}
             \caption{
             Comparison between the expected SFS of the total population, as derived in Proposition \ref{prop:spectrum}, and simulation results.
             {\bf (a)} The expected fixed-time spectrum \eqref{eq:fixedtimeresultgeneral} of Proposition \ref{prop:spectrum} (solid blue line) shows good agreement with the average spectrum of simulated tumors (grey dots).
             In this example, the extinction probability is $p_0=0.9$, the mutation rate is $w=1$, and the expected tumor size is $N=100$.
             We generated $10^5$ tumors with $p_0=0.9$ and $w=1$ and stopped each simulation at the fixed time $t_N$ with $N=100$ as defined by \eqref{eq:fixedtime}.
             At each cell division, the number of mutations acquired by each daughter cell was generated as a Poisson random variable with mean $w/2$.
             {\bf (b)} The expected fixed-size spectrum \eqref{eq:fixedsizeresultgeneral} of Proposition \ref{prop:spectrum} (solid red line) shows good agreement with the average spectrum of simulated tumors (grey dots).
             We again generated $10^5$ tumors with $p_0=0.9$ and $w=1$, but this time, we stopped each simulation when the tumor reached size $N=100$, i.e.~at the stochastic time $\tau_N$ defined by \eqref{eq:stoppingtime}.
             Note the discontinuity in the fixed-size spectrum at $j=N$, which is due to clonal mutations.
             }
             \label{fig:generalspectrafig}
         \end{figure*}

                   \begin{figure*}
             \centering
             \includegraphics[scale=1]{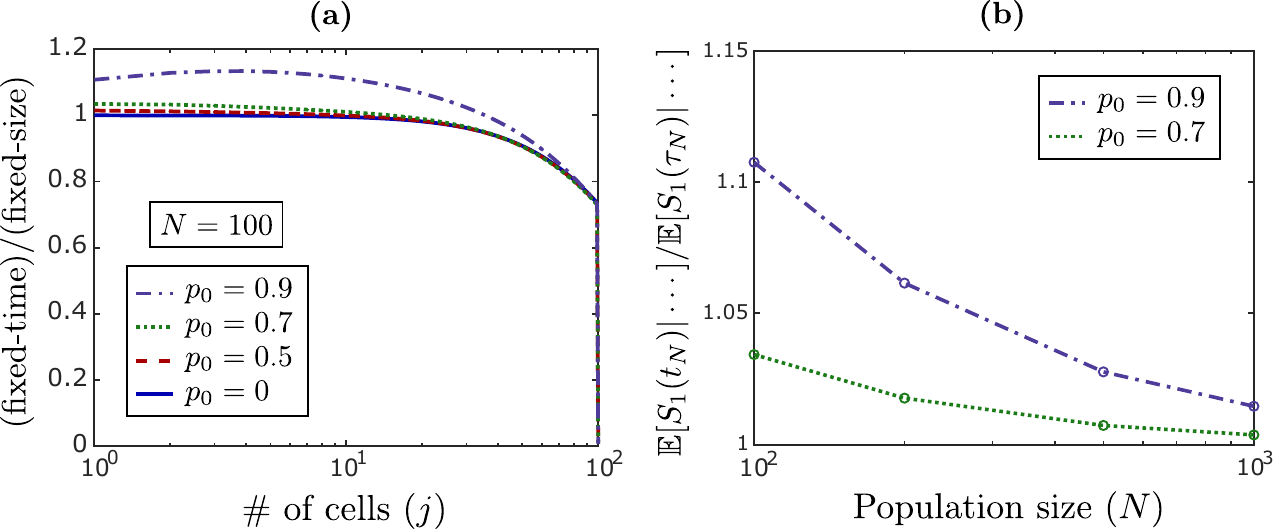}
             \caption{
                          Comparison between the expected fixed-time spectrum \eqref{eq:fixedtimeresultgeneral} and fixed-size spectrum \eqref{eq:fixedsizeresultgeneral} of the total population as derived in Proposition \ref{prop:spectrum}.
             {\bf (a)} Here, we show the ratio between the fixed-time spectrum \eqref{eq:fixedtimeresultgeneral} and the fixed-size spectrum \eqref{eq:fixedsizeresultgeneral} for $N=100$, $p_0 \in \{0,0.5,0.7,0.9\}$ and $w=1$.
             The fixed-time spectrum is a good approximation of the fixed-size spectrum on $j \ll N$ for smaller values of $p_0$, but the two spectra start to diverge as $p_0$ increases.
             {\bf (b)} Here, we show the ratio between the expected number of mutations found in one cell for the fixed-time spectrum and the fixed-size spectrum, as a function of the population size $N$, for $p_0 \in \{0.7,0.9\}$ and $w=1$.
             As $N$ increases, the difference between the two spectra reduces both for $p_0=0.7$ and $p_0=0.9$.
             }
             \label{fig:comparegeneralspectra}
         \end{figure*}

Analogous results for continuous mutation accumulation are presented in Appendix \ref{app:continuousmutation}.
In Figure \ref{fig:generalspectrafig}, we compare our fixed-time \eqref{eq:fixedtimeresultgeneral} and fixed-size \eqref{eq:fixedsizeresultgeneral} results with simulation results for $w=1$, $p_0=0.9$ and $N= 100$.
The fundamental difference between the fixed-time and fixed-size spectrum is the same as we observed in Section \ref{sec:skeletonresults}.
In Figure \ref{fig:comparegeneralspectra}a, we compare the two spectra in more detail for $N=100$ and $p_0 \in \{0,0.5,0.7,0.9\}$.
The fixed-time spectrum is a good approximation of the fixed-size spectrum on $j \ll N$ for all but $p_0=0.9$, in which case there is a significant difference even on $j \ll N$.
In Figure \ref{fig:comparegeneralspectra}b, we show that this difference reduces as $N$ increases, with the expected number of mutations found in one cell being 1.46\% off for $p=0.9$ and $N=1000$.
The number of cells in 1 ${\rm cm}^3$ of tumor tissue is around $10^7-10^9$ \citep{del2009does}, in which case the fixed-time spectrum can generally be expected to be a good approximation of the fixed-size spectrum on $j \ll N$.
The two spectra may diverge, however, when applied to smaller tumor samples (e.g.~1 ${\rm mm}^3$ or smaller) or when $p_0$ is very close to 1.

In Appendix \ref{app:comparetwospectra}, we present heuristic calculations that indicate how the fixed-size spectrum can be approximated by the fixed-time spectrum on $j \ll N$ as $N \to \infty$.
The key insight is that once the tumor has reached a large size, its growth becomes essentially deterministic with exponential rate $\lambda_0$, which allows us to approximate the probabilities $h_{(1,k)}^{(j,N-j)}$ in \eqref{eq:fixedsizeresultgeneral} by continuous-time probabilities following \citet{iwasa2006evolution}.
For added intuition, we refer to our discussion in Section \ref{sec:lawoflargenumbers} below and accompanying calculations in Appendix \ref{app:lawoflargenumbers}, where we conjecture a law of large numbers for the fixed-time and fixed-size spectrum, whose limits agree in the mean with the asymptotic expected fixed-time spectrum \eqref{eq:asymptspectrumgeneral}.

For each $j=1,\ldots,N$, computing $\E[S_j(\tau_N)|\tau_N<\infty]$ according to \eqref{eq:fixedsizeresultgeneral} requires solving a linear system of the form \eqref{eq:linearsystem}, which has order $N^2/2$ equations.
A more general version of this system arises in the study of the number of wild-type and mutant cells under a two-type (Luria-Delbr\"{u}ck)
population 
model stopped at a certain size, see e.g.~\citet{komarova2007fixed}.
The coefficient matrix of the system is sparse and banded, and it has a certain structure which allows one to solve it in
$O(N^3)$ arithmetic operations \citep{george1973nested}, compared to $O\big((N^2)^3\big) = O(N^6)$ for Gaussian elimination.
Solving the system directly is still not computationally feasible for the largest values of $N$, in which case one must develop an approximate solution.
See \citet{komarova2007fixed} for a partial differential equations approach.
Our observations suggest another simple approach:
For $N$ sufficiently large, one can approximate the fixed-size spectrum by the fixed-time spectrum on $j \ll N$.
For $j$ of order $N$, one can apply the $1/j^2$ skeleton law of \eqref{eq:skeletonfixedsizespectrum}, since as we discuss next, the skeleton will be responsible for the large-frequency mutations.

\subsection{Transition between power laws} \label{sec:transitionbetweenpowerlaws}

By Proposition \ref{prop:skeletonspectrum} of Section \ref{sec:skeletonresults}, the SFS of the skeleton depends on the mutation rate $w$ and the extinction probability $p_0$ only through the effective mutation rate $w/q_0$.
By Proposition \ref{prop:spectrum}, however, the two parameters decouple in the SFS of the total population.
The mutation rate $w$ scales the total population spectrum linearly, and the same is true of a large population size $N$, but the dependence on $p_0$ is more complex.
To better understand how $p_0$ affects the spectrum, recall first that by \eqref{eq:asymptspectrumgeneral}, the asymptotic expected fixed-time spectrum is given by
\[
        \E[S_j(t_N)|Z_0(t_N)>0] \sim \textstyle  wN \cdot \sum_{k=0}^\infty \frac{p_0^k}{(j+k)(j+k+1)}, \quad N \to \infty.
\]
In Appendix \ref{app:asymptotics}, we show that for fixed $0 < p_0 < 1$, sending $j \to \infty$ in this expression yields
\begin{align} \label{eq:transitionlargefreq}
    \textstyle wN \cdot \sum_{k=0}^\infty \frac{p_0^k}{(j+k)(j+k+1)} \sim (w/q_0)N \cdot 1/(j(j+1)), \quad j \to \infty,
\end{align}
which is the $1/j^2$ power law given in \eqref{eq:asymptspectrumskeleton}-\eqref{eq:skeletonfixedsizespectrum} of
Proposition \ref{prop:skeletonspectrum}.
We also show that for fixed $j \geq 1$, sending $p_0 \to 1$ in the same expression yields
\begin{align} \label{eq:largep0limit}
    \textstyle  wN \cdot \sum_{k=0}^\infty \frac{p_0^k}{(j+k)(j+k+1)} \textstyle  \sim wN \cdot 1/j, \quad p_0 \to 1,
\end{align}
which is the $1/j$ power law of the constant-sized Moran model of population genetics, see Theorem 1.33 of \citet{durrett2008probability}.
When referring to results from \citet{durrett2008probability}, note that there, the population is assumed to have size $2N$, which is a common convention in population genetics.

  \begin{figure*}
             \centering
\hspace*{-0.9cm}
             \includegraphics[scale=1]{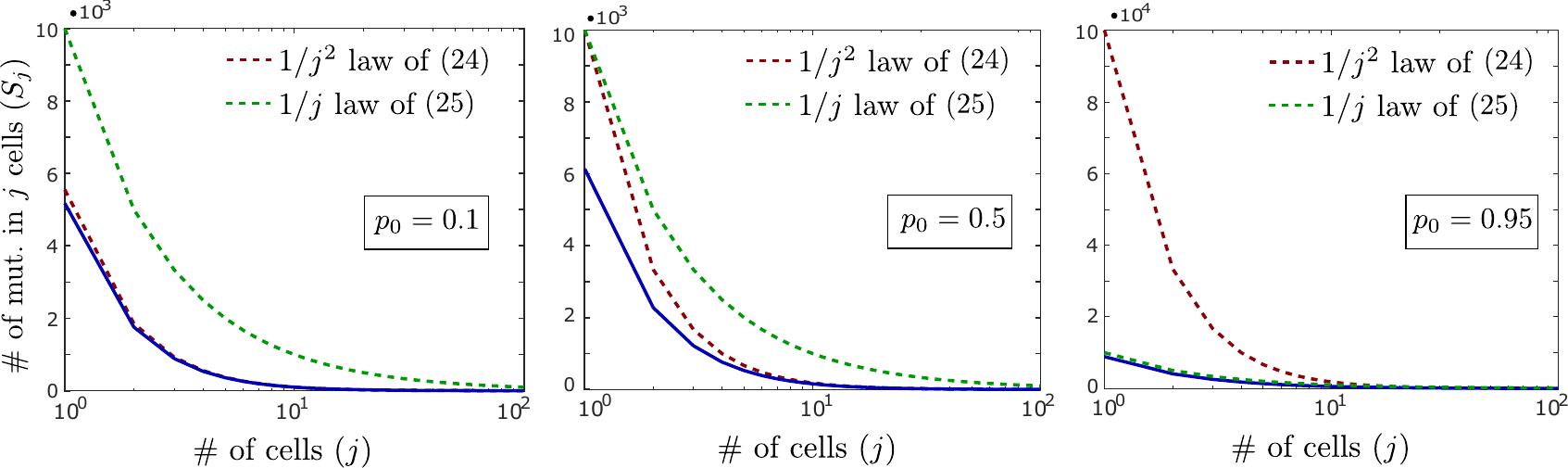}
             \caption{
             The small-frequency end of the total population spectrum transitions between two power laws as $p_0$ increases from 0 to 1.
             When $p_0$ is small, the expected fixed-time spectrum \eqref{eq:fixedtimeresultgeneral} of Proposition \ref{prop:spectrum} (solid blue line) approximately follows the the $1/j^2$ law of \eqref{eq:transitionlargefreq} (dotted red line).
             As $p_0$ increases to 1, the small-frequency end of the spectrum transitions to the $1/j$ law of \eqref{eq:largep0limit} (dotted green line), while the $1/j^2$ law increases according to $(w/q_0)N \cdot 1/(j(j+1))$.
             Note that the $y$-axis is on a linear scale, and that the scale of the rightmost panel is ten times larger than the scale of the other two.
             Also note that the $1/j$ law of \eqref{eq:largep0limit} is fixed as a function of $p_0$ and is therefore the same curve in all panels.
             Parameters are $N=10^4$ and $w=1$, and the spectrum is shown
             only at the smallest frequencies $j=1,\ldots,100$.
             }
             \label{fig:deviation}
         \end{figure*}

                        \begin{figure*}[t]
             \centering
             \includegraphics[scale=1]{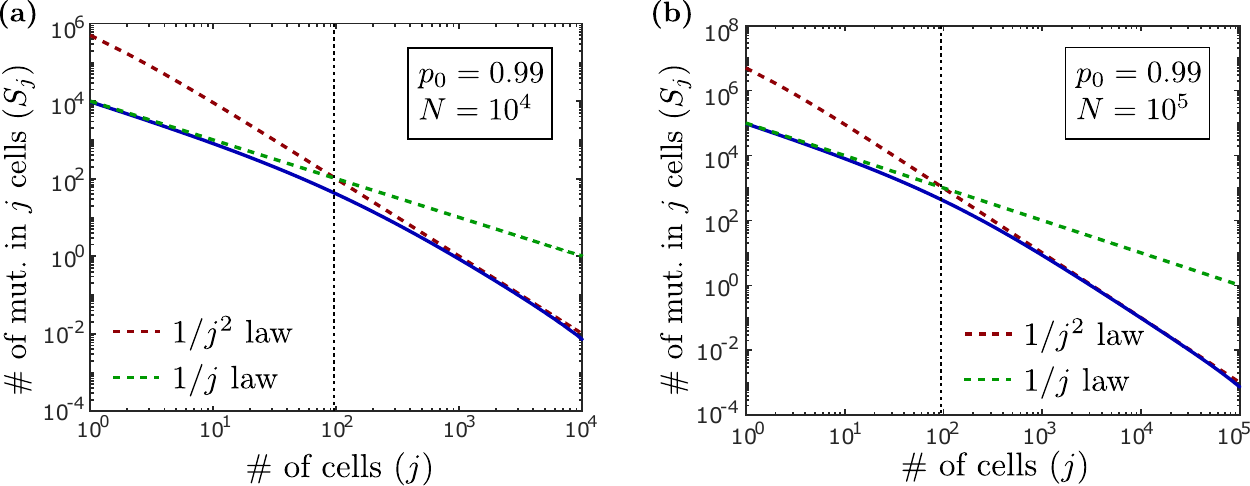}
             \caption{
             For a fixed, large value of the extinction probability $p_0$, the expected fixed-time spectrum \eqref{eq:fixedtimeresultgeneral} of Proposition \ref{prop:spectrum} (solid blue line) transitions from the $1/j^2$ power law of \eqref{eq:transitionlargefreq} (dotted red line) at the large-frequency end to the $1/j$ power law of \eqref{eq:largep0limit} (dotted green line) at the small-frequency end.
             The dotted vertical line shows the intersecting point $j=1/q_0-1$ of the two power laws in \eqref{eq:transitionlargefreq} and \eqref{eq:largep0limit}.
             In {\bf (a)}, the parameters are $N=10^4$, $p_0=0.99$ and $w=1$, and in {\bf (b)}, we increase the expected tumor size to $N=10^5$.
             Note that the transition between power laws occurs around the same value of $j$ in both cases, $j=1/q_0-1$.
             }
             \label{fig:transition}
         \end{figure*}

Recall that when $p_0=0$, the SFS of the total population $(Z_0(t))_{t \geq 0}$ is the SFS of the skeleton $(\tilde Z_0(t))_{t \geq  0}$.
Expressions \eqref{eq:transitionlargefreq} and \eqref{eq:largep0limit} suggest that when $p_0>0$, the SFS of $(Z_0(t))_{t \geq 0}$ continues to follow the $1/j^2$ skeleton law at the large-frequency end, while a deviation starts to occur
at the small-frequency end.
In fact, as $p_0$ approaches 1, the small-frequency end transitions to the $1/j$ law of the constant-sized Moran model.
This is illustrated in Figure \ref{fig:deviation} for $p_0 \in \{0.1,0.5,0.95\}$ and $N=10^4$.
Note that the $y$-axis is on a linear scale, and that the scale of the rightmost panel is ten times larger than the scale of the other two.
Also note that the $1/j$ law in \eqref{eq:largep0limit} is independent of $p_0$, so it is the same curve in all panels.
For small $p_0$, the fixed-time spectrum \eqref{eq:fixedtimeresultgeneral} is well-approximated by the $1/j^2$ law \eqref{eq:transitionlargefreq}, and each lies below the $1/j$ law \eqref{eq:largep0limit}.
To see why, note that the $1/j$ law gives $wN$ as the number of mutations found in one cell, while the $1/j^2$ law gives $(1/2)(w/q_0)N$, which is $(1/2)wN$ for $p_0=0$.
As $p_0$ increases to 1, the fixed-time spectrum \eqref{eq:fixedtimeresultgeneral} transitions to the $1/j$ law, while the $1/j^2$ law increases according to $(w/q_0)N \cdot 1/(j(j+1))$, diverging further and further from the fixed-time spectrum.
In Figure \ref{fig:transition}a, we show how for a fixed, large value of $p_0$ ($p_0=0.99$), the fixed-time spectrum \eqref{eq:fixedtimeresultgeneral} transitions from the $1/j^2$ law at the large-frequency end to the $1/j$ law at the small-frequency end.
Note that the power laws of \eqref{eq:transitionlargefreq} and \eqref{eq:largep0limit} intersect at
\[
    j = 1/q_0-1,
\]
which gives an indication of the frequency at which the transition occurs.
This intersecting point is independent of $N$, as is illustrated in Figure \ref{fig:transition}b (dotted vertical line).

To understand this transition between power laws, note that mutations that occur early in the evolution of a large tumor are only detected if they occur on the skeleton, which is why mutations at large frequencies in a large tumor follow the $1/j^2$ skeleton spectrum.
For mutations that occur late, we need to consider both skeleton cells and finite-family cells.
When $p_0$ is large, most cells are finite-family cells (their long-run proportion is $p_0$), and as $p_0$ approaches 1, finite-family clones start to behave like a critical branching process with net growth rate 0.
This is why late mutations
follow the $1/j$ law of a constant-sized population.
Thus, even though the branching process dynamics of cell division and cell death are the same throughout the evolution of the tumor, from the perspective of mutation accumulation, the tumor effectively behaves like a pure-birth exponential growth process initially, and more like a constant-sized process at the end, assuming a large extinction probability $p_0$.
The transition between power laws is the transition between these two growth regimes.

Note finally that as we observed for the skeleton spectrum in Section \ref{sec:skeletonresults}, the fixed-time spectrum \eqref{eq:fixedtimeresultgeneral} deviates from the $1/j^2$ law at the very largest frequencies (for $j$ of order $N$) in Figure \ref{fig:transition}, due to the variability in tumor size at time $t_N$.
It is therefore more correct to say that for fixed large $p_0$, the fixed-time spectrum transitions from the skeleton spectrum at the large-frequency end to the constant-sized spectrum at the small-frequency end, without reference to the power laws.
It remains true, however, that as $p_0$ increases from 0 to 1, the small-frequency end of the spectrum transitions between the two power laws.
         
\subsection{Total mutational burden of the tumor} \label{sec:nummutgeneral}

We next wish to quantify how $p_0$ affects overall mutation accumulation.
To this end, we derive in Proposition \ref{prop:nummut} the expected total mutational burden of the tumor, both under the fixed-time and fixed-size spectrum.
This quantity indicates the genetic diversity of the tumor, which has implications e.g.~for its adaptability under treatment.
It also enables us to compute a normalized version of the SFS, which can be useful for parameter estimation, as is discussed further in Section \ref{sec:signatures} below.
Before stating the proposition, we define $M_j(t) := \sum_{k \geq j} S_k(t)$ as the number of mutations found in $\geq j$ cells at time $t$.

\begin{proposition} \label{prop:nummut}
\begin{enumerate}[(1)]
    \item For $0 < p_0 < 1$, the expected total number of mutations in the fixed-time spectrum is given by
\begin{align} \label{eq:totalnummutgeneral}
\begin{split}
\E[M_1(t_N)|Z_0(t_N)>0] &= -wN \cdot (1/p_0)\log(q_0+p_0/N) \\
    &\sim -wN \cdot \log(q_0)/p_0, \quad N \to \infty.
\end{split}
\end{align}
For $p_0=0$, $\E[M_1(t_N)] = w(N-1) \sim wN$ as $N \to \infty$ by \eqref{eq:totalnummutskeleton}.
\item Define ${\cal S}$ and $A$ as in Proposition \ref{prop:spectrum}. The expected total number of mutations in the fixed-size spectrum is given by
\begin{align} \label{eq:totalnummutgeneralfixedsize}
    & \textstyle \E[M_1(\tau_N)|\tau_N<\infty] = \textstyle (w/q_0) \cdot \sum_{k=1}^{N-1} (1-p_0^{N-k}) \big(1-h_{(1,k)}^{(0,N)}-h_{(1,k)}^{(0,0)}\big),
\end{align}
where for each $(r,s) \in A$, the vector $\big(h_{(\ell,m)}^{(r,s)}\big)_{(\ell,m) \in {\cal S}}$
    solves the linear system \eqref{eq:linearsystem} of Proposition \ref{prop:spectrum}.
    For $p_0=0$, $\E[M_1(\tau_N)]=w(N-1) \sim wN$ as $N \to \infty$ by \eqref{eq:totalnummutskeleton}.
\end{enumerate}
\end{proposition}

\begin{proof}
   Appendix  \ref{app:proofprop3}.
\end{proof}

\begin{figure*}
             \centering
             \includegraphics[scale=1]{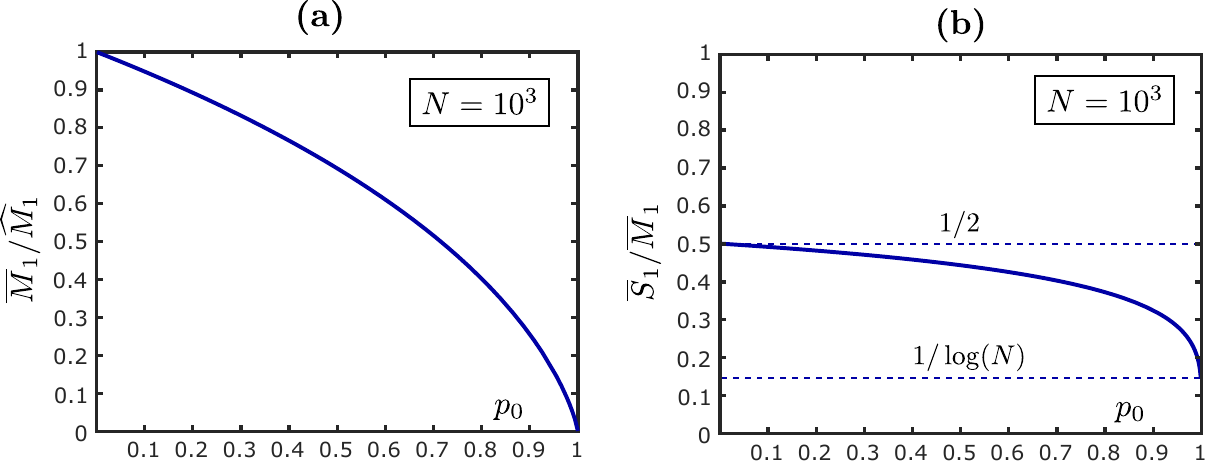}
             \caption{
            Graphs of the metrics $\overline{M}_1/\widehat{M}_1$ and $\overline{S}_1/\overline{M}_1$ given by \eqref{eq:totalnummutratio} and \eqref{eq:relativeproportiononecellgeneral}, as a function of $p_0$.
             {\bf (a)} Ratio between the expected total number of mutations in the fixed-time spectrum, $\overline M_1 := \E[M_1(t_N)|Z_0(t_N)>0]$ given by \eqref{eq:totalnummutgeneral}, and the simple estimate $\widehat{M}_1$ of \eqref{eq:totalnummutoverestimate}, derived from the $1/j^2 $ skeleton law of \eqref{eq:transitionlargefreq}, as a function of $p_0$ for $N=10^3$.
             The two estimates $\overline{M}_1$ and $\widehat{M}_1$ agree for $p_0=0$, but $\widehat{M}_1$ becomes a significant overestimate of $\overline{M}_1$ as $p_0$ increases.
             {\bf (b)} The proportion of mutations found in one cell, $\overline{S}_1/\overline{M}_1 = \varphi_N(p_0)$ as given by \eqref{eq:relativeproportiononecellgeneral}, quantifies the transition between the $1/j^2$ and $1/j$ power laws at the small-frequency end of the spectrum.
             Note that the transition between power laws accelerates as $p_0$ increases.
             }
             \label{fig:metrics1}
         \end{figure*}

Now, for ease of notation, write $\overline M_1 := \E[M_1(t_N)|Z_0(t_N)>0]$ for the expected total mutational burden under the fixed-time spectrum.
We are interested in comparing $\overline M_1$ with the expected number of mutations under the $1/j^2$ skeleton law of 
\eqref{eq:transitionlargefreq}. To this end, define
\begin{align} \label{eq:totalnummutoverestimate}
    \widehat{M}_1 := (w/q_0)N,
\end{align}
following \eqref{eq:totalnummutskeleton}.
This simple estimate has been used e.g.~in \citet{ling2015extremely}, where the authors estimate
the total number of mutations in a hepatocellular carcinoma (HCC) tumor under a few different assumptions on tumor evolution.
The ratio between $\overline M_1$ and $\widehat{M}_1$ is, for $0<p_0<1$,
\begin{align} \label{eq:totalnummutratio}
\begin{split}
        \overline M_1/\widehat{M}_1 &= - (q_0/p_0) \cdot \log(q_0+p_0/N) \\
        &\sim -(q_0/p_0) \cdot \log(q_0), \quad N \to \infty.
\end{split}
\end{align}
In Figure \ref{fig:metrics1}a, we show this ratio as a function of $p_0$ for $N=1000$.
The ratio is decreasing in $p_0$, it converges to 1 as $p_0 \to 0$, and it converges to 0 as $p_0 \to 1$.
To give some examples, in the $N \to \infty$ limit, $ \overline M_1/\widehat{M}_1 = 0.46$ for $p_0=0.75$ and $ \overline M_1/\widehat{M}_1 = 0.047$ for $p_0=0.99$.
Thus, if one is interested in estimating the total number of mutations in an exponentially growing tumor, using the simple expression \eqref{eq:totalnummutoverestimate} implied by the skeleton spectrum will result in a significant overestimate when $p_0$ is large.
Indeed, as $p_0$ increases, finite-family cells start to dominate the population, and they accumulate mutations less efficiently than skeleton cells.

\subsection{Proportion of mutations found in one cell} \label{sec:nummutonecellgeneral}

At the extremes $p_0=0$ and $p_0=1$, the small-frequency end of the SFS is characterized by the $1/j^2$ skeleton law and the $1/j$ constant-sized law, respectively.
To better understand how the small-frequency end behaves for intermediate values of $p_0$, we next determine the relative proportion of mutations found at the very smallest frequency, i.e.~in one cell.
This metric quantifies the transition between the $1/j^2$ and $1/j$ power laws at the small-frequency end, and it enables us to propose a simple estimator for $p_0$ in Section \ref{sec:signatures} below.

In Appendix \ref{app:s1derivation}, we show that in the fixed-time spectrum, the expected number of mutations found in one cell is given by, for $0 < p_0 < 1$,
\begin{align} \label{eq:nummutonecell}
\begin{split}
    \E[S_1(t_N)|Z_0(t_N)>0]    &= wN \cdot \textstyle (1/p_0)\big(1-1/N+(q_0/p_0)\log(q_0+p_0/N)\big) \\
    &\sim \textstyle wN \cdot (1/p_0)(1+(q_0/p_0)\log(q_0)), \quad N \to \infty.
\end{split}
\end{align}
As in Section \ref{sec:nummutgeneral}, we write $\overline{S}_1 := \E[S_1(t_N)|Z_0(t_N)>0]$ for ease of notation.
We then define $\varphi_N(p_0) := \overline{S}_1/\overline{M}_1$ as the proportion of mutations found in one cell.
By \eqref{eq:totalnummutgeneral} and \eqref{eq:nummutonecell}, 
\begin{align} \label{eq:relativeproportiononecellgeneral}
\begin{split}
\varphi_N(p_0)         &= \textstyle -\big({1-1/N+(q_0/p_0)\log(q_0+p_0/N)}\big)\,\big/\,{\log(q_0+p_0/N)} \\
    &\sim -(1+(q_0/p_0)\log(q_0))/\log(q_0), \quad N \to \infty. 
\end{split}
\end{align}
In Figure \ref{fig:metrics1}b, we show $\varphi_N = \overline{S}_1/\overline{M}_1$ as a function of $p_0$ for $N=1000$.
The function is strictly decreasing in $p_0$, it converges to $0.50$ as $p_0 \to 0$, and it converges to $(1-1/N)/\log(N)$ as $p_0 \to 1$, which is of order $1/\log(N)$ for $N$ large.
In the $p_0 \to 0$ regime, the SFS of the total population $(Z_0(t))_{t \geq 0}$ is the SFS of the skeleton $(\tilde Z_0(t))_{t\geq0}$, in which case half the mutations are found in one cell by Section \ref{sec:additionalmetricsskeleton}.
In the $p_0 \to 1$ regime, the SFS of $(Z_0(t))_{t \geq 0}$ is the SFS of the constant-sized Moran model, in which case the expected number of mutations is of order $wN \sum_{j=1}^N 1/j \sim wN \log(N)$ for $N$ large, and the proportion of mutations found in one cell is of order $1/\log(N)$.
Note that the rate of change of $\varphi_N(p_0)$ increases as $p_0$ increases (Figure \ref{fig:metrics1}b).
This implies that the deviation from the $1/j^2$ skeleton law at the small-frequency end is initially slow for small values of $p_0$, but it accelerates as $p_0$ increases and transitions quickly to the $1/j$ law for large values of $p_0$.
It also implies that $\overline{S}_1/\overline{M}_1$ is more useful for distinguishing larger values of $p_0$ than smaller values, as will become more apparent in Section \ref{sec:signatures} below.

 \begin{figure*}
             \centering
             \includegraphics[scale=1]{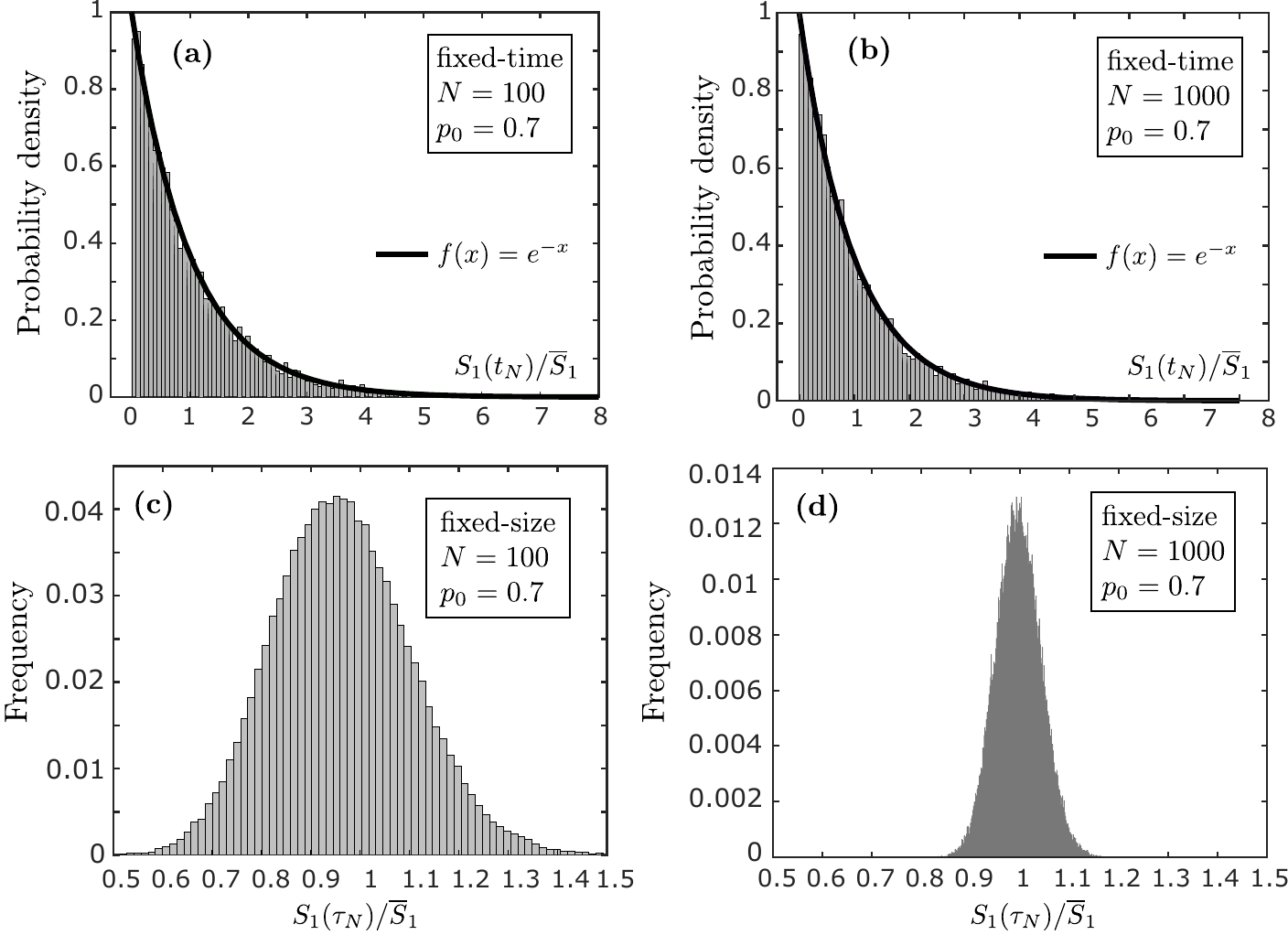}
             \caption{
             Simulation results support the conjectured laws of large numbers \eqref{eq:lawoflargenumbersgeneralfixedtime} and \eqref{eq:lawoflargenumbersgeneralfixedsize} for the fixed-time and fixed-size spectrum respectively.
             {\bf (a)} Histogram of $S_1(t_N)\big/\overline{S}_1$ over $10^4$ simulation runs with $N=100$, $p=0.7$ and $w=1$, where $t_N$ is defined by \eqref{eq:fixedtime}, and $\overline{S}_1 := \E[S_1(t_N)|Z_0(t_N)>0]$ as given by \eqref{eq:nummutonecell}.
             The $y$-axis is normalized so as to approximate the density of the underlying probability distribution.
             By comparison with $x \mapsto e^{-x}$, we see that $S_1(t_N)\big/\overline{S}_1$ appears to be a mean-1 exponential random variable, which is consistent with the conjectured law of large numbers \eqref{eq:lawoflargenumbersgeneralfixedtime}.
             {\bf (b)} When the population size is increased to $N=1000$, $S_1(t_N)\big/\overline{S}_1$ retains the mean-1 exponential distribution, consistent with \eqref{eq:lawoflargenumbersgeneralfixedtime}.
             {\bf (c)} Histogram of $S_1(\tau_N)/\overline{S}_1$ over $10^4$ simulation runs with $N=100$, $p=0.7$ and $w=1$, where $\tau_N$ is defined by \eqref{eq:stoppingtime}.
             {\bf (d)} Same as in (c), except now $N=1000$. Together, (c) and (d) indicate that the ratio $S_1(\tau_N)\big/\overline{S}_1$ concentrates around 1 as $N$ increases, which is consistent with the conjectured law of large numbers \eqref{eq:lawoflargenumbersgeneralfixedsize}.
             }
             \label{fig:lawoflargenumbers}
         \end{figure*}

\subsection{Spectra of individual large tumors (laws of large numbers)} \label{sec:lawoflargenumbers}

The results of Proposition \ref{prop:spectrum} hold in expectation, meaning that they apply to an average SFS computed over a large number of tumors.
If we want to use these results to understand the evolutionary history of individual tumors,
we need to know more about how well they apply on a tumor-by-tumor basis.
It is well-known that conditional on the nonextinction event $\Omega_\infty$, $Z_0(t) \sim Ye^{\lambda_0t}$ as $t \to \infty$ almost surely, where $Y$ follows the exponential distribution with mean $1/q_0$ (Theorem  1 of \citet{durrett2015branching}).
In other words, the tumor population $Z_0(t)$ eventually grows at exponential rate $\lambda_0$, but the initial value of the exponential growth function is random and depends on the individual tumor.
We can use this fact to formulate laws of large numbers for the fixed-time and fixed-size spectrum, which we state formally as conjectures.
In Appendix \ref{app:lawoflargenumbers}, we present simple calculations in support of these conjectures, and we also prove an analogous result \eqref{eq:llnsimplified} for a simplified, semideterministic version of our model.

 \begin{figure*}[t]
             \centering
             \includegraphics[scale=1]{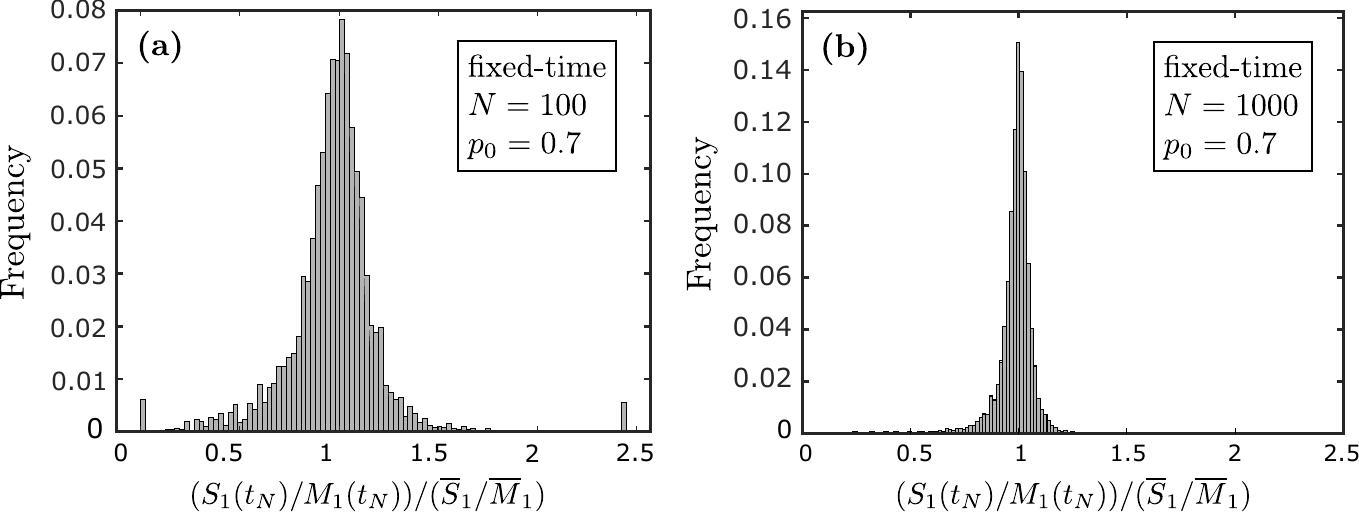}
             \caption{
             Simulation results indicate that if the SFS is normalized by the total number of mutations, the fixed-time and fixed-size spectrum obey the same law of large numbers.
             {\bf (a)} Histogram of $(S_1(t_N)/M_1(t_N))\big/(\overline{S}_1/\overline{M}_1)$
             over $10^4$ simulation runs with $N=100$, $p=0.7$ and $w=1$, where $t_N$ is defined by \eqref{eq:fixedtime}, and the expected ratio $\overline{S}_1/\overline{M}_1$ is given by \eqref{eq:relativeproportiononecellgeneral}.
             Note the point masses at 0 and 2.44, which represent simulation runs where $S_1(t_N)/M_1(t_N)=0$ and $S_1(t_N)/M_1(t_N)=1$, respectively.
            {\bf (b)} Same as in (a), except now, $N=1000$.
            Together, (a) and (b) indicate that as $N$ increases, $S_1(t_N)/M_1(t_N)$ concentrates around $\overline{S}_1/\overline{M}_1$.
            This in turn suggests that if the fixed-time spectrum is normalized by the total number of mutations $M_1(t_N)$, it obeys the same law of large numbers as the normalized version of the fixed-size spectrum.
             }
             \label{fig:lawoflargenumbers2}
         \end{figure*}

\begin{conjecture*}
\begin{enumerate}[(1)]
\item Define $t_N$ as in \eqref{eq:fixedtime}.
    Then, there exists an exponential random variable $X$ with mean 1 so that for fixed $j \geq 1$, conditional on $\Omega_\infty$, 
        \begin{align} \label{eq:lawoflargenumbersgeneralfixedtime}
    \textstyle S_j(t_N) \sim X \cdot wN \cdot \int_0^1 (1-p_0y)^{-1} (1-y) y^{j-1} dy
    \end{align}
    as $N \to \infty$ almost surely.
    \item Define $\tau_N$ as in \eqref{eq:stoppingtime}.
    Then, for fixed $j \geq 1$, conditional on $\Omega_\infty$,
\begin{align} \label{eq:lawoflargenumbersgeneralfixedsize}
    \textstyle S_j(\tau_N) \sim wN \cdot  \int_0^1 (1-p_0y)^{-1} (1-y) y^{j-1} dy
\end{align}
    as $N \to \infty$ almost surely.
\end{enumerate}
\end{conjecture*}
Both the fixed-time conjecture \eqref{eq:lawoflargenumbersgeneralfixedtime} and the fixed-size conjecture \eqref{eq:lawoflargenumbersgeneralfixedsize} agree with simulation results, see Figures \ref{fig:lawoflargenumbers} and \ref{fig:lawoflargenumbers2}.
The main difference between \eqref{eq:lawoflargenumbersgeneralfixedtime} and \eqref{eq:lawoflargenumbersgeneralfixedsize} is that the right-hand-side of \eqref{eq:lawoflargenumbersgeneralfixedtime} is stochastic, while the right-hand side of \eqref{eq:lawoflargenumbersgeneralfixedsize} is a constant.
The former expression has a random scaling factor $X$, which captures the variability in tumor size at time $t_N$, whereas the fact that the tumor size is always $N$ at time $\tau_N$ eliminates this variability in the latter expression.
Note that since $\E[X]=1$, the right-hand sides of \eqref{eq:lawoflargenumbersgeneralfixedtime} and \eqref{eq:lawoflargenumbersgeneralfixedsize} agree in the mean, and this mean agrees with the right-hand side of the asymptotic spectrum \eqref{eq:asymptspectrumgeneral} of Proposition \ref{prop:spectrum}.
Importantly, the scaling factor $X$ in  \eqref{eq:lawoflargenumbersgeneralfixedtime} is independent of $j$, which indicates that the proportion of mutations found in $j$ cells is the same in \eqref{eq:asymptspectrumgeneral}, \eqref{eq:lawoflargenumbersgeneralfixedtime} and \eqref{eq:lawoflargenumbersgeneralfixedsize}.
In other words, according to these conjectures, if we normalize the SFS with the total number of mutations, $S_j(t)/M_1(t)$, the fixed-time and fixed-size spectrum of an individual large tumor will be completely characterized by the asymptotic expected spectrum \eqref{eq:asymptspectrumgeneral} of Proposition \ref{prop:spectrum} (Figure \ref{fig:lawoflargenumbers2}).
This can be useful for parameter estimation, as we discuss next.

\section{Signatures of cell viability} \label{sec:signatures}

In this section, we use our theoretical results to propose a simple estimator for the extinction probability $p_0$, based on extracting one or more spatially separated subclones from a tumor.
By a subclone, we mean all currently living descendants of a given common ancestor, i.e.~all leaves of the branching tree started by a given tumor cell.
Since every clone or subclone derived from a single tumor cell obeys the same branching process dynamics as the overall tumor, all of our previous results can be applied to individual subclones.

We make the strong assumption that each cell in each sampled subclone can be single-cell sequenced so that all its mutations are captured, even at the smallest frequencies, which is beyond current sequencing technology.
Our main purpose with this section is to show how the information contained in the small-frequency end of the SFS 
can in principle be used to 
decouple the mutation rate $w$ and the extinction probability $p_0$.
We discuss practical considerations and potential alternative approaches in Section \ref{sec:discussion} below.

Say that we sample a subclone of size $n$.
For $1 \leq j \leq n-1$, let $s_j$ be the number of mutations found in $j$ cells of the subclone, and let $m_1 := \sum_{j=1}^{n-1} s_j$ be the total number of mutations.
Here, we ignore mutations found in all cells of the subclone, since they include (i) mutations that accumulate prior to tumor initiation, (ii) mutations that occur post-tumor-initiation but prior to initiation of the subclone, and (iii) mutations that occur post-subclone-initiation but still end up in all subclone cells.
Let $\overline{s}_1$ be the expected number of mutations found in one subclone cell under the fixed-time spectrum, and let $\overline{m}_1$ be the expected total number of mutations under the fixed-time spectrum.
By \eqref{eq:relativeproportiononecellgeneral},
we can write
\[
    \overline{s}_1/\overline{m}_1 = \varphi_n(p_0),
\]
where $\varphi_n(p_0)$ is continuous and strictly decreasing in $p_0$.
In particular, $\varphi_n(p_0)$ is invertible.
This implies that given $\overline{s}_1$ and $\overline{m}_1$, $p_0$ can be recovered from this expression via
\[
    p_0 = \varphi_n^{-1}(\overline{s}_1/\overline{m}_1).
\]
For the sampled values $s_1$ and $m_1$, this suggests the following estimator for $p_0$:
\begin{align} \label{eq:estimatordef}
    \widehat{p}_0(n) := \varphi_n^{-1}(s_1/m_1).
\end{align}
Of course, if the subclone is sampled at a certain size, it makes more sense to use the fixed-size spectrum than the fixed-time spectrum.
In addition, $m_1$ excludes mutations found in all subclone cells, while $\overline{m}_1$ includes some of these mutations.
These potential sources of error are minor and can easily be resolved if necessary, as we discuss in more detail below.

                  \begin{figure*}[t]
             \centering
\hspace*{-0.5cm}
             \includegraphics[scale=1]{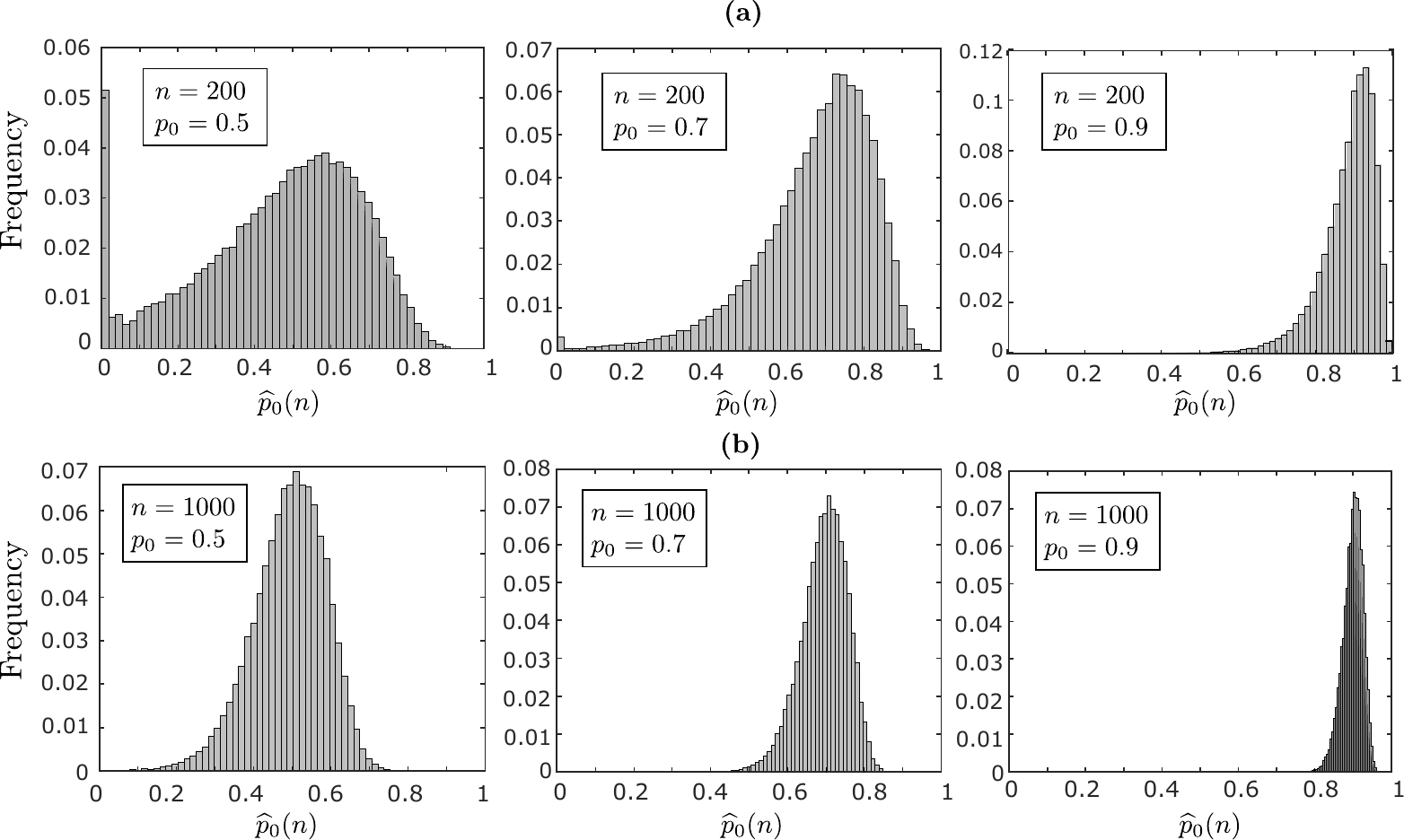}
             \caption{
             Histogram of the estimator $\widehat{p}_0(n)$ defined in \eqref{eq:estimatordef} computed across $10^5$ synthetic subclone samples of size $n$, given true values $p_0 \in \{0.5,0.7,0.9\}$ and $w=1$.
             In {\bf (a)}, the subclone size is $n=200$, and in {\bf (b)}, the size is $n=1000$.
             Together, (a) and (b) indicate that as the size of the subclone increases, the estimator $\widehat{p}_0(n)$ concentrates around the true value of $p_0$, which in turn indicates statistical consistency of the estimator.
             }
             \label{fig:samplingscheme1}
         \end{figure*}

Note that the ratio of expected values $\overline{s}_1/\overline{m}_1$ takes values in $[(1-1/n)/\log(n),1/2]$ by Section \ref{sec:nummutonecellgeneral},
whereas due to stochasticity, the sampled ratio $s_1/m_1$ can take any value in $[0,1]$.
To complete the estimator in \eqref{eq:estimatordef}, we therefore extend the definition of $\varphi_n^{-1}$ by setting
\begin{align} \label{eq:phidefextension}
    \varphi_n^{-1}(x)  := \begin{cases} 0, & 1/2 \leq x \leq 1, \\ 1, & 0 \leq x \leq (1-1/n)/\log(n). \end{cases}
\end{align}
For example, if we observe a ratio $s_1/m_1$ larger than 1/2, we default to the estimate $\widehat{p}_0(n) = 0$, since the expected ratio $\overline{s}_1/\overline{m}_1$ is largest (and equal to $1/2$) 
for $p_0=0$.
Once $p_0$ has been estimated, an estimate for the mutation rate $w$ can be obtained from \eqref{eq:nummutonecell} or \eqref{eq:totalnummutgeneral}.

The estimator $\widehat{p}_0(n)$ has the benefit of being simple to define and to compute.
However, as was mentioned above, it may make more sense to use the fixed-size spectrum than the fixed-time spectrum, and $\overline{m}_1$ includes clonal mutations that arise post-subclone-initiation, whereas $m_1$ excludes these mutations.
The second potential source of error is minor,
and it can easily be removed simply by subtracting from $\overline{m}_1$ the contribution from mutations shared by all subclone cells.
The first potential source of error is also likely to be insignificant when $n$ is large, since by the conjectured laws of large numbers in Section \ref{sec:lawoflargenumbers}, the normalized spectrum of an individual large subclone is robust to whether it is observed at a fixed time or a fixed size.
For smaller values of $n$, one can replace $\overline{s}_1$ and $\overline{m}_1$
by the corresponding quantities \eqref{eq:fixedsizeresultgeneral} and \eqref{eq:totalnummutgeneralfixedsize} for the fixed-size spectrum, which we denote here by $\overline{\overline{s}}_1$ and $\overline{\overline{m}}_1$.
It remains true that we can write $\overline{\overline{s}}_1/\overline{\overline{m}}_1 = \psi_n(p_0)$
for some function $\psi_n$ of $p_0$, which allows us to define an estimator for $p_0$ as before.
However, the fixed-size estimator has to be obtained numerically, e.g.~by precomputing $\psi_n(p_0)$ over a grid of values for $p_0$, and minimizing the error between the observed ratio $s_1/m_1$ and the expected ratio $\overline{\overline{s}}_1/\overline{\overline{m}}_1$ over the grid.

                 \begin{table*}[t]
    \centering
    \begin{tabular}{|c|c|c|c|c|c|c|}
    \hline 
    $n$&$p_0$& mean & median &st.~dev. & defaults to 0 (\%) & defaults to 1 (\%) \\
    \hline
    200 &0.5 & 0.4694 & 0.5041 & 0.2099 & 0.0459 & 0 \\
    & 0.7 & 0.6755 & 0.7045 & 0.1505 & 0.0026 & 0 \\
    & 0.9 & 0.8839 & 0.9001 & 0.0725 & 0 & 0.00002 \\
    \hline
    1000& 0.5 & 0.4921 & 0.5000 & 0.0948 & 0.00002 & 0 \\
    &0.7 & 0.6957 & 0.7010 & 0.0603 & 0 & 0 \\
    &0.9 & 0.8979 & 0.9009 & 0.0266 & 0 & 0 \\
    \hline
    \end{tabular}
    \caption{Performance metrics for the estimator $\widehat{p}_0(n)$ computed from the data that underlies Figure \ref{fig:samplingscheme1}. 
    Both for $n=200$ and $n=1000$, the median of $\widehat{p}_0(n)$ accurately recovers the true value of $p_0$.
    In addition, the estimator improves in terms of standard deviation both as $n$ increases and as $p_0$ increases.
    }
    \label{table:samplingscheme1}
\end{table*}

To evaluate $\widehat{p}_0(n)$, we use computer simulations to generate multiple independent subclones of size $n$ with true extinction probability $p_0$, and for each generated subclone, we compute the estimate $\widehat{p}_0(n)$.
In Figure \ref{fig:samplingscheme1}a, we show a histogram for $\widehat{p}_0(n)$ across $10^5$ synthetic subclone samples of size $n=200$ with true extinction probabilities $p_0 \in \{0.5,0.7,0.9\}$.
In Table \ref{table:samplingscheme1}, we show performance metrics for $\widehat{p}_0(n)$ computed across the $10^5$ samples.
For $p_0=0.5$, the estimator defaults to $\widehat{p}_0(n) = 0$ for 4.6\% of the subclone samples, for $p_0 = 0.7$, it defaults to 0 in 0.3\% of cases, and for $p_0=0.9$, it never defaults to 0.
For all values of $p_0$, the median estimate of $\widehat{p}_0(n)$ accurately recovers the true value, and as $p_0$ increases, the quality of the estimate improves in terms of standard deviation.
In Figure \ref{fig:samplingscheme1}b, we increase the subclone size to $n=1000$ and observe a marked improvement in the quality of $\widehat{p}_0(n)$.
This indicates statistical consistency of the estimator, meaning that $\widehat{p}_0(n) \to p_0$ in probability as $n \to \infty$, which would also be a direct consequence of the conjectured laws of large numbers \eqref{eq:lawoflargenumbersgeneralfixedtime}-\eqref{eq:lawoflargenumbersgeneralfixedsize} and the continuous mapping theorem.
In other words, the estimator appears to recover the true value of $p_0$ with arbitrarily high precision given a sufficiently large subclone.
Note that the size of $n$ required to return a high-precision estimate becomes smaller as $p_0$ increases, making $\widehat{p}_0(n)$ especially useful when $p_0$ is large.
Indeed, as we remarked in Section \ref{sec:nummutonecellgeneral}, the rate of change of the expected ratio $\overline{s}_1/\overline{m}_1$ increases as $p_0$ increases, making it more useful for distinguishing between larger values of $p_0$ than smaller values.

\begin{figure*}[t]
             \centering
\hspace*{-0.5cm}
             \includegraphics[scale=1]{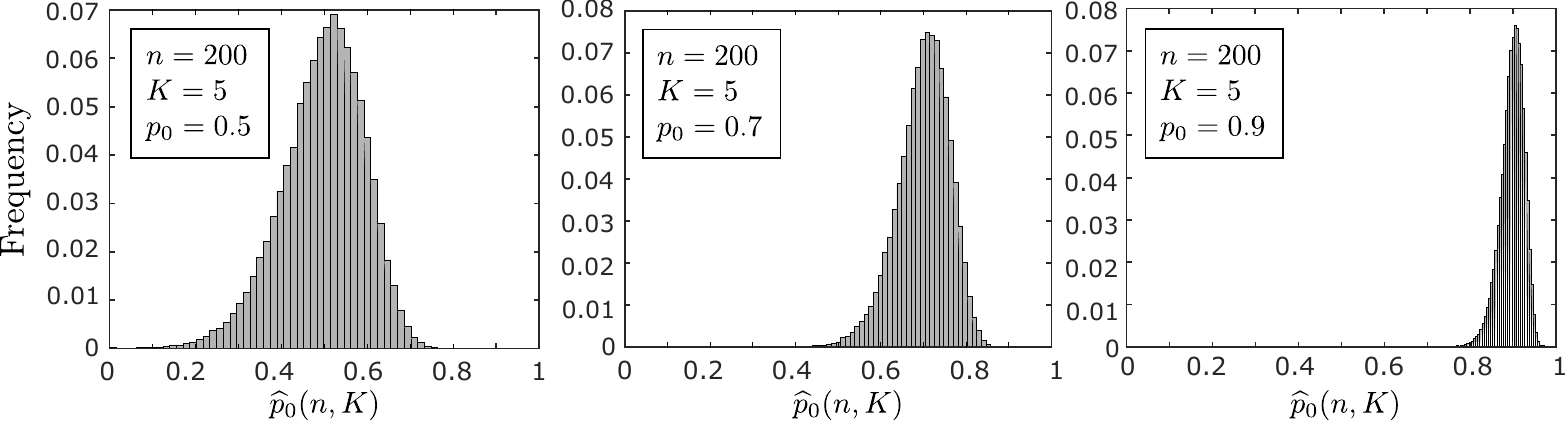}
\caption{
            Histogram of the estimator $\widehat{p}_0(n,K)$ of \eqref{eq:estimator2def} computed across $10^5$ synthetic samples of $K=5$ subclones of size $n=200$, given true values $p_0 \in \{0.5,0.7,0.9\}$ and $w=1$.
            Note the similarity between these histograms and the histograms in (b) of Figure \ref{fig:samplingscheme1}, which indicates that sampling five subclones of size 200 gives a comparable estimate of $p_0$ to sampling a single subclone of size 1000 in these examples.
             }
             \label{fig:samplingscheme2}
         \end{figure*}

Whenever it is possible to do multi-region sampling, there may be benefits to extracting multiple small, spatially separated subclones over a single large one.
In this more general setting, we sample $K \geq 1$ subclones of size $n$.
For $1 \leq j \leq n-1$, let $s_j^{k}$ be the number of mutations found in $j$ cells of subclone number $k$, and let $m_1^{k} := \sum_{j=1}^{n-1} s_j^{k}$ be the total number of mutations in subclone $k$.
We replace $s_1$ and $m_1$ in the definition of $\widehat{p}_0(n)$ in \eqref{eq:estimatordef} by the sums $\sum_{k=1}^K s_1^{k}$ and $\sum_{k=1}^K m_1^{k}$ to obtain the estimator
\begin{align} \label{eq:estimator2def}
    \textstyle \widehat{p}_0(n,K) := \varphi_n^{-1}\big(\sum_{k=1}^K s_1^{k}\big/\sum_{k=1}^K m_1^{k}\big).
\end{align}
Of course, $\widehat{p}_0(n,1) = \widehat{p}_0(n)$.
In Figure \ref{fig:samplingscheme2}, we show a histogram for $\widehat{p}_0(n,K)$ evaluated across $10^5$ synthetic samples, each sample consisting of $K=5$ independent subclones of size $n=200$.
In Table \ref{table:samplingscheme2}, we show performance metrics computed across the $10^5$ samples.
Qualitatively, the histograms in Figure \ref{fig:samplingscheme2} are very similar to the histograms of Figure \ref{fig:samplingscheme1}b, and quantitatively, the performance metrics in Table \ref{table:samplingscheme2} mimic those for the $n=1000$ case in Table \ref{table:samplingscheme1}.
In other words, the quality of the estimate of $p_0$ obtained from sampling one subclone of size 1000 is comparable to the one obtained from sampling five subclones of size 200.
In this scenario, it may make more sense to extract multiple small subclones than one large one,
since it is impossible to tell from a single subclone sample alone whether the tumor as a whole can be considered as evolving neutrally.
Should there be differences in the subclone dynamics, a multiregion sample may tease this out, and should the dynamics be the same, the estimate one obtains for $p_0$ will be of comparable quality to the single large subclone case.

                 \begin{table*}[b]
    \centering
    \begin{tabular}{|c|c|c|c|c|c|c|c|}
    \hline 
    $K$&$n$&$p_0$& mean & median &st.~dev. & defaults to 0 (\%) & defaults to 1 (\%) \\
    \hline
    5&200 &0.5 & 0.4965 & 0.5050 & 0.0957 & 0.00013 & 0 \\
    && 0.7 & 0.7000 & 0.7055 & 0.0612 & 0 & 0 \\
    && 0.9 & 0.8983 & 0.9014 & 0.0279 & 0 & 0 \\
    \hline
    \end{tabular}
    \caption{
    Performance metrics for the the estimator $\widehat{p}_0(n,K)$ computed from the data that underlies Figure \ref{fig:samplingscheme2}. 
    Note the similarity between these metrics and the lower half of Table \ref{table:samplingscheme1}.
    }
    \label{table:samplingscheme2}
\end{table*}

\section{Discussion} \label{sec:discussion}

In this work, we have established exact expressions for the expected site frequency spectrum of a tumor, or more generally any population, that evolves according to a branching process with neutral mutations under the infinite-sites assumption of population genetics.
We first considered the skeleton subpopulation, consisting of cells with an infinite line of descent, and obtained explicit expressions for the SFS of the skeleton evaluated both at a fixed time and a fixed size.
We then examined the total population, deriving an explicit expression for the fixed-time spectrum and a computational expression for the fixed-size spectrum.
Our results apply to mutations at small and large frequencies, to tumor tissue samples and tumor subclones of any size, and to all values of the extinction probability $p_0$, even values as large as $p_0=0.90$ and above, which are broadly relevant for cancer.
We now discuss in detail how our results relate to results previously obtained in the literature.

We begin by stating skeleton results established by  \citet{durrett2013population,durrett2015branching}, \citet{bozic2016quantifying} and \citet{williams2016identification}.
The result in \citet{bozic2016quantifying} says that for fixed $0 < f < 1$, the expected number of {\em subclonal} mutations found in a {\em proportion} $\geq f$ of cells at time $t$ is, as $t \to \infty$,
\begin{align} \label{eq:1fspectrumour}
(w/q_0)(1/f-1).
\end{align}
Note that this is a {\em cumulative} spectrum of mutations with frequency at least $f$.
A similar result appears in \citet{williams2016identification} under a deterministic growth model.
Durrett's result \citep{durrett2013population,durrett2015branching}, which preceded the other two, is given under continuous mutation accumulation, and it includes clonal mutations, which by \eqref{eq:skeletonspectrumfixedsizecontinuous} of Appendix \ref{app:continuousmutation} requires adding $\nu/\lambda_0$ ($ = w/q_0$) mutations to \eqref{eq:1fspectrumour}.
This yields $(\nu/\lambda_0)(1/f)$ as the cumulative spectrum, see Theorem 1 of \citet{durrett2013population} and Theorem 2 of \citet{durrett2015branching}.\footnote{There is an apparent typo in Theorem 1 of \citet{durrett2013population}. The result is written as $(\nu/q_0)(1/f)$ in our notation, but should be $(\nu/\lambda_0)(1/f)$. This is corrected in Theorem 2 of the later notes \citet{durrett2015branching} by the same author.}
Under discrete mutation accumulation, the number of clonal mutations is $wp_0/q_0$ by \eqref{eq:skeletonfixedsizespectrum} of Proposition \ref{prop:skeletonspectrum}, which yields the result $(w/q_0)(1/f)-w$ including clonal mutations.
The difference of $w$ reflects the difference in the number of clonal mutations between the discrete and continuous model, see Appendix \ref{app:continuousmutation}.

To compare the $1/f$ law in \eqref{eq:1fspectrumour} with  our results of Proposition \ref{prop:skeletonspectrum}, recall that by \eqref{eq:skeletonfixedsizespectrum}, the fixed-size skeleton spectrum can be written as $\E[\tilde S_j(\tilde\tau_N)] = (w/q_0)N\cdot1/(j(j+1))-(wp_0/q_0)\delta_{1,j}$ for $j=1,\ldots,N-1$.
Under this spectrum, it is easy to compute the expected number of subclonal mutations found in a proportion $\geq f$ of skeleton cells as
\begin{align} \label{eq:1fcalculation}
\textstyle \E\big[\sum_{j=\lceil Nf\rceil}^{N-1} \tilde S_j(\tilde \tau_N)\big] &= \textstyle (w/q_0)N \cdot \big(\sum_{j=\lceil Nf\rceil}^{N-1} 1/(j(j+1))\big) - (wp_0/q_0) \delta_{1,\lceil Nf\rceil} \nonumber \\
    &= (w/q_0)N \cdot \big(1/\lceil Nf\rceil - 1/N\big)  - (wp_0/q_0) \delta_{1,\lceil Nf\rceil} \nonumber \\
    &\sim (w/q_0)(1/f-1), \quad N \to \infty.
\end{align}
The $1/f$ law in \eqref{eq:1fspectrumour} can therefore be viewed as a cumulative version of our $1/j^2$ law.
Note that \eqref{eq:1fspectrumour} is established in the fixed-time regime in the above cited works, while 
the calculations in \eqref{eq:1fcalculation} show that the $1/f$ law also holds in the fixed-size regime.

To summarize, in the fixed-size regime, we have established exact adherence to the $1/j^2$ law on $j=2,\ldots,N-1$ for any skeleton size $N \geq 2$, and our result implies the cumulative $1/f$ law as $N \to \infty$.
In the fixed-time regime, we have established the exact expression \eqref{eq:skeletonfixedtimespectrum}, which converges to the $1/j^2$ law for $j \ll N$ as $N \to \infty$.
The $1/f$ result in \eqref{eq:1fspectrumour} complements our fixed-time result, 
since it confirms that if we compute the spectrum of mutations found in a certain {\em proportion} of cells, rather than in a certain {\em number} of cells, the fixed-time spectrum converges to the same $1/f$ law as the fixed-size spectrum as $N \to \infty$, including at the very largest frequencies.
It should be emphasized that the $1/f$ law, which has been extensively cited in the literature, is an asymptotic result established for the skeleton.
For the total population, including finite-family cells, the $1/f$ law is applicable only to mutations at large frequencies in a large tumor.
This is particularly the case for low-viability tumors.

The expressions \eqref{eq:fixedtimeresultgeneral} and \eqref{eq:asymptspectrumgeneral} for the expected fixed-time spectrum of Proposition \ref{prop:spectrum} have been previously obtained by \citet{ohtsuki2017forward} under the assumption of deterministic growth of the tumor bulk and stochastic growth of mutant subclones, which is a reasonable approximation when $p_0$ is small.
We have established these expressions for the fully stochastic model and for all values of $p_0$, by evaluating the spectrum at the fixed time $t_N$ at which a surviving tumor has expected size $N$, defined in \eqref{eq:fixedtime}, and conditioning on the survival event $\{Z_0(t_N)>0\}$.
We have also conjectured the laws of large numbers \eqref{eq:lawoflargenumbersgeneralfixedtime} and \eqref{eq:lawoflargenumbersgeneralfixedsize} for the fixed-time and fixed-size spectrum, supported by heuristic calculations and simulations.
We have proved a law of large numbers \eqref{eq:llnsimplified} for the semideterministic model of \citet{ohtsuki2017forward}, see Appendix \ref{app:lawoflargenumbers}, and we plan to prove the fully stochastic results \eqref{eq:lawoflargenumbersgeneralfixedtime}  and \eqref{eq:lawoflargenumbersgeneralfixedsize} in a future work.
\citet{lambert2009allelic} has proved a similar result in the context of a coalescent point process (CPP), a framework under which an extant population is endowed with a coalescent structure that specifies how lineages coalesce when traced backwards in time.
Lambert's result, see his Theorem 2.3, deals with a ranked sample of individuals from a CPP in the large-sample limit, and it has the same form as our conjectured fixed-size law of large numbers \eqref{eq:lawoflargenumbersgeneralfixedsize}, the latter applying to our forwards-in-time branching process stopped at a certain size.
The expected fixed-size spectrum \eqref{eq:fixedsizeresultgeneral} of Proposition \ref{prop:spectrum} is new as far as we know, as well as expressions \eqref{eq:totalnummutgeneral} and \eqref{eq:totalnummutgeneralfixedsize} of Proposition \ref{prop:nummut} for the expected total mutational burden of the tumor.

All of the above results hold under the infinite-sites model of population genetics.
\citet{cheek2018mutation} have recently examined the SFS of an exponentially growing population without this assumption, citing single-cell sequencing results of \citet{kuipers2017single} as motivation. 
They observe that if recurrent mutations are allowed (but no back mutations), and there are $S$ sites in the genome, the expected SFS at time $t$ can be computed as $S \cdot \P(Y(t)=j)$, where $Y(t)$ is the number of mutants at time $t$ in a two-type model of wild-type and mutant cells, each growing at the same rate.
Then, to compute the SFS, one needs the distribution of $Y(t)$, which has been obtained under various assumptions e.g.~by \citet{antal2011exact}, \citet{kessler2013large,kessler2015scaling}, and \citet{keller2015mutant}.
Cheek and Antal obtain SFS results under limits of large time/size and small mutation rate, and their results obey $1/j^2$ power laws at large frequencies.
However, when the mutation rate is sufficiently large compared to the population size, their small-frequency behavior diverges from ours, see e.g.~their Figure 1.
Yet other authors have substituted the infinite-sites model with the {\em infinite-alleles} model, under which each new mutation creates a new type of individual,
see e.g.~\citet{griffiths1988infinite, pakes1989infinite, champagnat2012birth, wu2013modeling}. 
Under this model, the site frequency spectrum is usually replaced by an {\em allele frequency spectrum}, which tracks frequencies of genetically distinct individuals, known as {\em haplotypes} \citep{lambert2009allelic}.

Our complete theoretical results give rise to several important insights.
First of all, whereas the fixed-time and fixed-size skeleton spectrum depends on the mutation rate $w$ and the extinction probability $p_0$ only through the effective mutation rate $w/q_0$, the two parameters decouple in the total population spectrum.
The mutation rate $w$ scales the spectrum linearly, whereas the extinction probability $p_0$ changes its shape at the small-frequency end.
In fact, as $p_0$ increases from 0 to 1, the small-frequency end of the spectrum transitions from the $1/j^2$ power law characteristic of pure-birth exponential growth to the $1/j$ law characteristic of constant-sized populations.
We examined the simple metrics $\overline{M}_1/\widehat{M}_1$ and $\overline{S}_1/\overline{M}_1$ that quantify this transition, where $\widehat{M}_1$ is the expected total mutational burden under the $1/j^2$ power law spectrum.
We saw that $\overline{M}_1/\widehat{M}_1 \to 0$ as $p_0 \to 1$, which suggests that the simple estimate $\widehat{M}_1$ of the total number of mutations, applied e.g.~in \citet{ling2015extremely}, is a significant overestimate of the actual expected number of mutations $\overline{M}_1$ when $p_0$ is large.
We finally used the metric $\overline{S}_1/\overline{M}_1$ to propose a simple estimator for $p_0$, based on sampling one or more spatially separated subclones from a tumor.
This estimator 
accurately recovers the true value of $p_0$ from synthetic single-cell sequencing data,
and it is most accurate when $p_0$ is large.

Our proposed estimator is currently of more theoretical than practical significance.
It assumes that complete subclones of a given size can be reliably extracted from a tumor sample, and that each cell in each subclone can be single-cell sequenced so that all of its mutations are captured.
We have proposed the estimator mainly to emphasize the information contained in the small-frequency end of the spectrum, and to show how it can in principle be used to decouple $w$ and $p_0$ using the SFS alone.
Whether and how this decoupling can be achieved under current and foreseeable limitations of genomic data warrants further investigation.
For example, one can derive an estimator based on $s_j/m_j$ for $j>1$, where $m_j := \sum_{k=j}^{n-1} s_k$, which excludes mutations found at the smallest frequencies.
One can also design a more elaborate estimating procedure, which retains some of the smallest frequencies, but explicitly models the sequencing error.
As it becomes easier to distinguish small-frequency mutations from sequencing errors, our simple estimator may at the very least provide quick and easy identification of low-viability tumors, and complement other more involved techniques.

We finally note that the estimation of $w$ and $p_0$ relies on several assumptions on tumor evolution.
First of all, our model assumes that all mutations are selectively neutral.
While this assumption is likely reasonable for smaller tumor subclones, it may be hard to verify that a subclone estimate is representative for the tumor as a whole.
Our model also assumes exponential growth throughout tumor evolution, whereas e.g.~due to spatial constraints and nutrient availability, the growth may be subexponential both during the early and late stages.
Finally, our model assumes that the mutation rate is constant over time,  and that the infinite-sites assumption holds for the tumor sample being analyzed.
We note that the SFS-based estimates of $w/q_0$ obtained by \citet{williams2016identification} and \citet{bozic2016quantifying} 
are derived from the large-frequency end of the SFS, which reflects early tumor dynamics.
The same is true of the \citet{werner2018reply} method of decoupling $w$ and $p_0$ by tracing genealogies of spatially separated bulk samples,
since the common ancestors of these samples are likely to have existed early in tumor evolution.
Conversely, our suggested approach of utilizing the small-frequency end of the SFS will more reflect late tumor evolution.
On the one hand, in the evolving discussion of tumor evolutionary history inference, it is important to acknowledge that any given estimation procedure may only give a temporally or spatially constrained picture of the dynamics.
On the other hand, utilizing different parts of the SFS, or combining SFS estimates with other estimates, may allow one to glean insights into the dynamics at different stages of tumor evolution, and to possibly assess the validity of any modeling assumptions.

\section*{Acknowledgments}

The authors would like to thank the three anonymous reviewers and the editor for their valuable comments and suggestions.
EBG and KL were supported in part by NSF grant CMMI-1552764. JF was supported in part by NSF grant DMS-1349724. KL and JF were supported in part by Research Council of Norway R\&D Grant 309273. EBG was supported in part with funds from the Norwegian Centennial Chair Program.

\section*{Competing Interests}

The authors declare no competing interests.

\appendix 

\section{Proof of Proposition \ref{prop:skeletonspectrum}} \label{app:proofprop1}

\setcounter{theorem}{0}

In this section, we prove Proposition \ref{prop:skeletonspectrum} on the expected SFS of the skeleton.
We first present a brief outline of the proof.

In part (1), the SFS is observed at the fixed time $\tilde t_N$.
To compute the expected number of mutations that end up in $j \geq 1$ skeleton cells at time $\tilde t_N$, we decompose the time interval $[0, \tilde t_N]$ into infinitesimal intervals $[t,t+\Delta t]$.
We can compute how many mutations occur during each small interval using the expected mutation rate \eqref{eq:mutrateskeleton} in the main text.
Then, to only count the mutations that end up in $j$ skeleton cells at time $\tilde t_N$, we multiply this number by the probability that a single-cell derived skeleton clone has size $j$ at time $\tilde t_N$, which has a known expression.
We finally integrate over time to add up the contributions of the infinitesimal intervals.

In part (2), the SFS is observed at the stochastic time $\tilde \tau_N$, at which the skeleton reaches size $N$.
In this case, we decompose into population size levels instead of into small time intervals.
We know how many mutations accumulate on population size level $k$ by \eqref{eq:mutperpopulationlevel} in the main text.
Then, to get the expected number of mutations that accumulate on level $k$ and end up in $j \geq 1$ skeleton cells at time $\tilde \tau_N$, we need to compute the probability that starting from one skeleton cell carrying a particular mutation and $k-1$ cells without it, $j$ cells carry the mutation when the skeleton reaches size $N$.
To that end, we define a Markov chain that keeps track of how many skeleton cells carry the mutation as the skeleton increases in size, and we compute its hitting probabilities.
We finally sum over $k$ to add up the contributions of each population size level.

\begin{proof}[Proof of Proposition \ref{prop:skeletonspectrum}]
     \begin{enumerate}[(1)]
        \item By \eqref{eq:mutrateskeleton} in the main text, the expected mutation rate per skeleton cell per unit time is $wr_0$.
To stratify mutations based on their frequencies at time $\tilde t_N$, we define
\[
    \tilde p_j(s) := \P(\tilde Z_0(s) = j | \tilde Z_0(0)=1), \quad j \geq 1, s \geq 0,
\]
as the size-distribution at time $s$ of a single-cell derived skeleton clone.
Since $(\tilde Z_0(t))_{t \geq 0}$ is a Yule process with birth rate $\lambda_0$, this distribution has an explicit expression,
\[
    \tilde p_j(s) = (1/e^{\lambda_0s}) (1-1/e^{\lambda_0s})^{j-1}, \quad j \geq 1,
\]
which is the geometric distribution with support $\{1,2,\ldots\}$ and success probability $1/e^{\lambda_0s}$, see e.g.~Section 3 of \citet{durrett2015branching} (the support does not include 0 since skeleton clones do not go extinct).
For $0 \leq t \leq \tilde t_N$, let $\tilde{S}_{j,\tilde t_N}(t)$ denote the number of mutations that accumulate in the time interval $[0,t]$ and are found in $j \geq 1$ skeleton cells at time $\tilde t_N$.
We write $\tilde{S}_{j}(\tilde t_N) := \tilde{S}_{j,\tilde t_N}(\tilde t_N)$ for the site frequency spectrum of the skeleton at time $\tilde t_N$.
If a mutation occurs during an infinitesimal time interval $[t,t+\Delta t]$, the clone started by the cell carrying the mutation has size $j$ at time $\tilde t_N$ with probability $\tilde p_j(\tilde t_N-t)+O(\Delta t)$, where $f(x) = O(x)$ means that there exists $C>0$ so that $|f(x)| \leq Cx$ for sufficiently small $x>0$.
The expected number of mutations that accumulate in $[t,t+\Delta t]$ and are present in $j \geq 1$ skeleton cells at time $\tilde t_N$ is therefore
\[
            \E[\tilde S_{j,\tilde t_N}( t+\Delta t)]-\E[\tilde S_{j,\tilde t_N}( t)] = wr_0 \Delta t \cdot e^{\lambda_0t} \cdot \tilde p_j(\tilde t_N-t) + o(\Delta t),
\]
where we use that $\E[\tilde Z_0(t)] = e^{\lambda_0t}$ is the mean skeleton size at time $t$, and $f(x) = o(x)$ means that $f(x)/x \to 0$ as $x \to 0$.
This calculation is somewhat heuristic in that we have simply multiplied an expected mutation rate by an expected population size, which is in turn multiplied by the probability that a particular mutation ends up in $j$ cells at time $\tilde t_N$.
In our proof of part (1) of Proposition \ref{prop:spectrum} for the total population, we present a more detailed argument which can be used to obtain this expression more rigorously.

Integrating over time, and using that $q_0 = \lambda_0/r_0$ by expression \eqref{eq:survprob} and $N= e^{\lambda_0 \tilde t_N}$ by expression \eqref{eq:fixedtimeskeleton}, we obtain
\begin{linenomath*}
\begin{align*}
            \E[\tilde S_j(\tilde  t_N)] &= \textstyle \int_0^{\tilde t_N} wr_0 \tilde p_j(\tilde t_N-t) e^{\lambda_0t} dt \\
            &= \textstyle (w/q_0)N \cdot \int_0^{\tilde t_N} (e^{\lambda_0t}/N) (1-e^{\lambda_0t}/N)^{j-1} \cdot \lambda_0 (e^{\lambda_0t}/N) dt.
\end{align*}
\end{linenomath*}
Substituting $y := 1-e^{\lambda_0t}/N$, $dy = -\lambda_0 (e^{\lambda_0t}/N) dt$, 
this implies
\begin{linenomath*}
\begin{align*} 
\begin{split}
            \E[\tilde S_j(\tilde  t_N)] &= (w/q_0)N \cdot \textstyle \int_0^{1-1/N} (1-y) y^{j-1} dy \\
            &= \textstyle (w/q_0)N \cdot \big(1-\frac1N)^j \big(\frac1{j(j+1)}+\frac1N \frac1{j+1}\big),    
\end{split}
\end{align*}
\end{linenomath*}
the desired result. Clearly, for fixed $j \geq 1$, then as $N \to \infty$,
\begin{align}  \label{eq:1j2lawYule}
    \E[\tilde S_j(\tilde t_N)] \sim (w/q_0)N \cdot 1/(j(j+1)).
\end{align}
The asymptotic expression \eqref{eq:1j2lawYule} can also be derived more heuristically as follows, which gives another way of interpreting the expression.
If the skeleton is observed at a large time $t$, the age $s$ of an arbitrary mutation has approximate density $\lambda_0 e^{-\lambda_0s}$.
A mutation with age $s$ at time $t$ is found in $j \geq 1$ skeleton cells at time $t$ with probability $\tilde p_j(s)$.
The probability that an arbitrary mutation is found in $j \geq 1$ skeleton cells at time $t$ is therefore, as $t \to \infty$,
\begin{linenomath*}
\begin{align*}
    \textstyle \int_0^\infty \tilde p_j(s) \cdot \lambda_0 e^{-\lambda_0s}ds = \int_0^1 (1-y)y^{j-1} dy = 1/(j(j+1)),
\end{align*}
\end{linenomath*}
using the substitution $y := 1-e^{-\lambda_0s}$, $dy = \lambda_0e^{-\lambda_0s}ds$.
Next, we can compute the expected total number of mutations up until time $\tilde t_N$ via
\begin{linenomath*}
\begin{align*}
    \textstyle \int_0^{\tilde t_N} wr_0 e^{\lambda_0s}ds = (w/q_0)(N-1) \sim (w/q_0)N,
\end{align*}
\end{linenomath*}
which is given as \eqref{eq:totalnummutskeleton} in the main text.
Finally, we can obtain \eqref{eq:1j2lawYule} as the expected total number of mutations up until time $\tilde t_N$ multiplied by the probability $1/(j(j+1))$ of finding an arbitrary mutation in $j$ skeleton cells at time $\tilde t_N$ as $N \to \infty$.

The distribution $j \mapsto 1/(j(j+1))$ is a special case of the {\em Yule-Simon distribution},
which was originally computed by \citet{yule1925ii} as the distribution of the number of species within a genus, where a species mutates to a new species within the same genus at some rate $s$, and a genus mutates to a new genus at some rate $g$.
In the previous paragraph, we have adapted Yule's basic argument to our setting with $g=s=\lambda_0$.
We refer to \citet{simkin2011re} for a comprehensive discussion of how the Yule-Simon distribution and variants thereof have appeared in a wide variety of scientific contexts since its original conception.
         \item In  \eqref{eq:type1mutations} of Section \ref{sec:skeletonproperties}, we showed that on average, $wp_0/q_0$ mutations accumulate on type-1 divisions in between two type-2 divisions, while type-2 divisions add $w$ mutations and change the skeleton population size level.
         Since the type-1 mutations on level $k=1$ are the clonal mutations, the expected number of clonal mutations is $wp_0/q_0$, which is the $j=N$ case of the desired result.
         For $k = 2,\ldots,N-1$, the expected number of mutations on level $k$ is $w+wp_0/q_0 = w/q_0$, which includes the type-2 division that starts the level.
         
         For $1 \leq j \leq N-1$, let $\tilde h_{(1,k-1)}^j$ be the probability that starting with one skeleton cell carrying a particular mutation and $k-1$ cells without it, $j$ cells carry the mutation when the skeleton reaches size $N$.
         Since for levels $k=2,\ldots,N-1$, there are $w/q_0$ mutations on average per level, and each mutation on level $k$ contributes $\tilde h_{(1,k-1)}^j$ to the expected number of mutations found in $j$ skeleton cells at level $N$, we obtain for $1 \leq j \leq N-1$,
         \begin{linenomath*}
         \begin{align*}
                          \textstyle \E[\tilde S_j(\tilde\tau_N)] &= \textstyle  (w/q_0) \sum_{k=2}^{N-1} \tilde h_{(1,k-1)}^j + w\delta_{1,j} \\
                          &= \textstyle  (w/q_0) \sum_{k=1}^{N-2} \tilde h_{(1,k)}^j + w\delta_{1,j}, 
         \end{align*}
         \end{linenomath*}
         where the extra $w\delta_{1,j}$ term is due to mutations that occur on the final type-2 division that changes levels from $N-1$ to $N$, each of which is found in one skeleton cell.
         As was the case for the proof of part (1), the fact that for each level $k$, we can simply multiply the expected number of mutations with the probability that each particular mutation ends up in $j$ cells when the skeleton reaches size $N$ can be justified more rigorously using an argument similar to the one we present in part (2) of Proposition \ref{prop:spectrum} in Appendix \ref{app:proofprop2}.

         It remains to compute the probabilities $\tilde h_{(1,k)}^j$.
         To this end, define a two-dimensional discrete-time Markov chain on the state space $\{(\ell,m): \ell,m \geq 1, \ell+m \leq N\}$, where $\ell$ is the number of skeleton cells carrying a particular mutation and $m$ is the number of cells without it.
         Since each skeleton cell, with or without the mutation, divides into two cells at rate $\lambda_0$, the transition probabilities for this chain are given by
         \begin{linenomath*}
         \begin{align*}
             &(\ell,m) \to (\ell+1,m) \quad\text{w.p.}\quad \ell/(\ell+m), \\
             &(\ell,m) \to (\ell,m+1) \quad\text{w.p.}\quad m/(\ell+m),
         \end{align*}
         \end{linenomath*}
         for $\ell,m \geq 1$ and $\ell+m<N$. The states $(\ell,N-\ell)$ for $1 \leq \ell \leq N-1$ are absorbing. A diagram for this Markov chain is shown in Figure \ref{fig:mchain_skeleton}.
         
         Let $\tilde h_{(\ell,m)}^r$ denote the probability that the above chain is absorbed in state $(r,N-r)$ when started from state $(\ell,m)$.
            It is immediate that $\tilde h_{(\ell,m)}^r = 0$ if $\ell>r$ or $m>N-r$.
         For $(\ell,m)$ with $\ell \leq r$, $m \leq N-r$ and $\ell+m < N$, by conditioning on whether the first transition out of state $(\ell,m)$ is to $(\ell+1,m)$ or $(\ell,m+1)$, we obtain the following recursion for $\tilde h_{(\ell,m)}^r$:
         \begin{align} \label{eq:recursionskeleton}
             (\ell+m) \tilde h_{(\ell,m)}^r &= \ell \tilde h_{(\ell+1,m)}^r + m \tilde h_{(\ell,m+1)}^r.
         \end{align}
         The boundary conditions are $\tilde h_{(\ell,N-\ell)}^r = \delta_{\ell,r}$ for $1 \leq \ell \leq N-1$.
         It is actually possible to compute $\tilde h_{(\ell,m)}^r$ directly as the sum of probabilities of all possible paths from $(\ell,m)$ to $(r,N-r)$ without using the above recursion.
         By noting that there are $\binom{N-(\ell+m)}{r-\ell}$ possible paths, and that each path has the same probability, we can obtain
         \begin{align} \label{eq:recursionsolution}
             \tilde h_{(\ell,m)}^r &= \textstyle \binom{N-(\ell+m)}{r-\ell} \cdot \prod_{n=0}^{r-\ell-1} \frac{\ell+n}{\ell+m+n} \cdot \prod_{n=0}^{N-m-r-1} \frac{m+n}{r+m+n},
         \end{align}
         with $\prod_\varnothing := 1$. As verification, it is straightforward to check that \eqref{eq:recursionsolution} solves \eqref{eq:recursionskeleton}.

         \begin{figure*}
             \centering
             \includegraphics[scale=1]{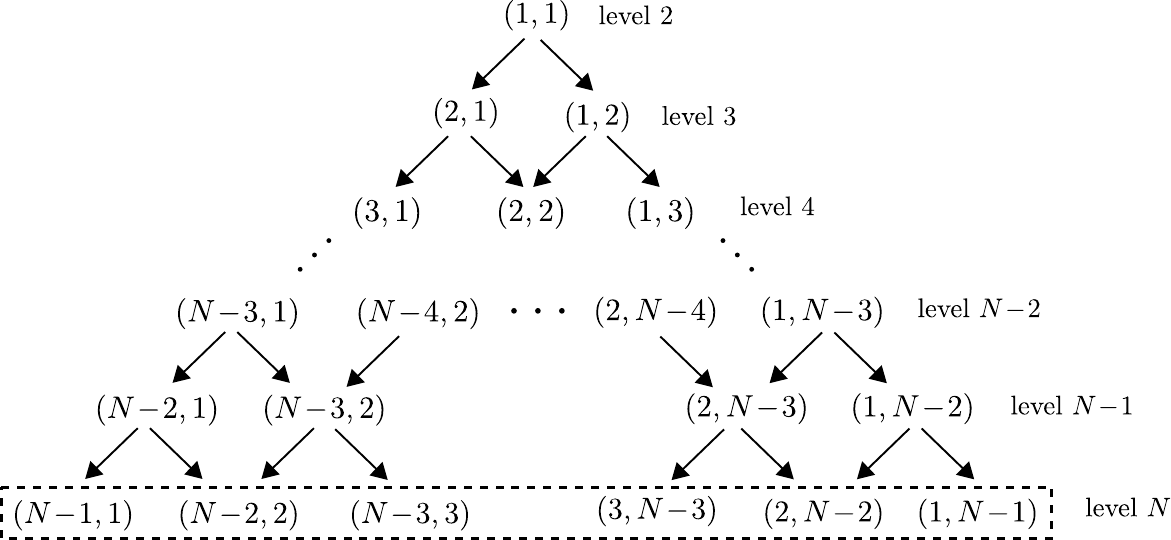}
             \caption{A diagram of the discrete-time Markov chain on $\{(\ell,m): \ell,m \geq 1, \ell+m \leq N\}$, where $\ell$ is the number of skeleton cells carrying a particular mutation and $m$ is the number of cells without it. Each type-2 division increases the skeleton population size level by one, and since skeleton cells do not die, the chain never returns to the lower levels. The states $(\ell,N-\ell)$ with $1 \leq \ell \leq N-1$ are absorbing (dashed box), since we are only interested in the evolution up until level $N$.
             }
             \label{fig:mchain_skeleton}
         \end{figure*}
         
         To obtain $\tilde h_{(1,k)}^j$ for $1 \leq k \leq N-2$ and $1 \leq j \leq N-1$, note first that $\tilde h_{(1,k)}^j=0$ for $k>N-j$.
         For $1 \leq k \leq \min(N-j,N-2)$, we can simplify \eqref{eq:recursionsolution} as follows:
         \begin{linenomath*}
         \begin{align*}
                \tilde h_{(1,k)}^j &= \textstyle \binom{N-(k+1)}{j-1} \cdot \prod_{n=0}^{j-2} \frac{1+n}{k+1+n} \cdot \prod_{n=0}^{N-k-j-1} \frac{k+n}{k+j+n} \\
                &= \textstyle \frac{(N-k-1)!}{(j-1)!(N-k-j)!} \cdot \frac{(j-1)!}{(k+1)\cdots(k+j-1)} \cdot \frac{k \cdots (N-j-1)}{(k+j)\cdots(N-1)} \\
                &= \textstyle k \cdot \frac{(N-k-1)!}{(N-k-j)!} \cdot \frac{ (N-j-1)!}{(N-1)!}  \\
                &= \textstyle \binom{N-j-1}{k-1} \binom{N-1}{k}^{-1}.
         \end{align*}
         \end{linenomath*}
         Thus, for $1 \leq j \leq N-1$,
                  \begin{align} \label{eq:fixedsizesubst}
             \textstyle \E[\tilde S_j(\tilde\tau_N)] &= \textstyle (w/q_0) \sum_{k=1}^{N-2} \tilde h_{(1,k)}^j + w\delta_{1,j} \nonumber \\
             &=(w/q_0) \textstyle \sum_{k=1}^{\min(N-j,N-2)} \binom{N-j-1}{k-1} \binom{N-1}{k}^{-1} + w\delta_{1,j}.
         \end{align}
         For $j=1$, it is easy to compute
                  \begin{linenomath*}
                  \begin{align*}
             \textstyle \E[\tilde S_1(\tilde\tau_N)] &= \textstyle (w/q_0)(1/(N-1)) \big(\sum_{k=1}^{N-2} k\big) + w \\
             &= (1/2)(w/q_0)(N-2)+w \\
             &= (1/2)(w/q_0)N - wp_0/q_0.
         \end{align*}
         \end{linenomath*}
        To simplify \eqref{eq:fixedsizesubst} for $2 \leq j \leq N-1$, we first note that for any positive integers $a$ and $b$,
        \begin{linenomath*}
              \begin{align*}
                  \textstyle \int_0^1 t^{a-1} (1-t)^{b-1} dt = \frac{(a-1)!(b-1)!}{(a+b-1)!},
              \end{align*}
              \end{linenomath*}
             see e.g.~Theorem 1.1.4 of \citet{andrews1999special}. By observing that
             \begin{linenomath*}
             \begin{align*}
                 \textstyle N \int_0^1 t^k (1-t)^{N-1-k} dt = \binom{N-1}{k}^{-1},
             \end{align*}
             \end{linenomath*}
             we can rewrite \eqref{eq:fixedsizesubst} as follows:
             \begin{linenomath*}
             \begin{align*}
                 \E[\tilde S_{j}(\tilde\tau_N)] &= \textstyle (w/q_0)N \cdot \int_0^1 \big(\sum_{k=1}^{N-j} \binom{N-j-1}{k-1} t^k (1-t)^{N-1-k}\big)dt \\
                 &= \textstyle (w/q_0)N \cdot \textstyle\int_0^1 t(1-t)^{j-1} \big(\sum_{k=1}^{N-j} \binom{N-j-1}{k-1} t^{k-1} (1-t)^{(N-j-1)-(k-1)}\big)dt.
             \end{align*}
             \end{linenomath*}
             The sum inside the integral is the total probability mass ($=1$) of the binomial distribution with number of trials $N-j-1$ and success probability $t$. Therefore, for $2 \leq j \leq N-1$,
             \begin{linenomath*}
                          \begin{align*}
                 \E[\tilde S_{j}(\tilde\tau_N)] &= \textstyle (w/q_0)N \cdot \int_0^1 t(1-t)^{j-1} dt \\
                 &= (w/q_0)N \cdot 1/(j(j+1)).
             \end{align*}
             \end{linenomath*}
             This concludes the proof. \qedhere
             \end{enumerate}
\end{proof}

\section{Proof of Proposition \ref{prop:spectrum}} \label{app:proofprop2}

In this section, we prove Proposition \ref{prop:spectrum} on the expected SFS of the total population.
The proof strategy is the same as in the proof of Proposition \ref{prop:skeletonspectrum} for the skeleton.
There are some added complications, however, since the tumor as a whole follows a birth-death process, whereas the skeleton subpopulation follows a pure-birth process.

In part (1), we wish to compute the expected number of mutations that accumulate in an infinitesimal time interval and end up in $j \geq 1$ tumor cells at the fixed time $t_N$.
In this case, we need to condition on survival up until time $t_N$, which was not necessary in the same computation for the skeleton.
In the proof below, we first give an informal argument for how to handle the conditioning on survival, and then present detailed calculations at the end of the proof.
Another complication is that the size distribution of a single-cell derived clone now has a more complex form than for the skeleton, which translates into more work simplifying the integral that results from adding up the infinitesimal interval contributions.

In part (2), we decompose into population size levels.
To compute how many mutations accumulate on population size level $k$, we first compute how many times the population hits this level.
This computation was not necessary for the skeleton, since the skeleton only increases in size.
We then need to compute the probability that starting from one cell carrying a particular mutation and $k-1$ cells without it, $j$ cells carry the mutation when the population reaches size $N$.
This probability can be computed as a hitting probability of a Markov chain, as in the proof of Proposition \ref{prop:skeletonspectrum}, but the Markov chain now has a more complicated structure.
This time, we are not able to compute the hitting probabilities explicitly,
and we instead provide a linear system which determines them.

\begin{proof}[Proof of Proposition \ref{prop:spectrum}]
     \begin{enumerate}[(1)]
         \item We begin by defining
         \[
    p_j(s) := \P(Z_0(s) = j | Z_0(0)=1), \quad j \geq 0, s \geq 0,
\]
as the size-distribution at time $s$ of a single-cell derived clone.
This distribution has an explicit expression: Setting
\[
\textstyle g(t) := \frac{p_0(e^{\lambda_0t}-1)}{e^{\lambda_0t}-p_0} \quad\text{and}\quad h(t) := \frac{e^{\lambda_0t}-1}{e^{\lambda_0t}-p_0},
\]
we can write
\begin{linenomath*}
\begin{align*} 
    & p_0(t) = \P(Z_0(t)=0|Z_0(0)=1)= g(t), 
   \end{align*}
\end{linenomath*} 
and for $j \geq 1$,
    \begin{linenomath*}
\begin{align*} 
    p_j(t) &= \P(Z_0(t)=j|Z_0(0)=1) \\
    &= (1-g(t))(1-h(t))(h(t))^{j-1},
\end{align*}
\end{linenomath*}
see e.g.~(8) of \citet{durrett2015branching}.
Simplifying, we obtain
\begin{align} \label{eq:sizedisgeneral}
\begin{split}
    p_0(t) &= \textstyle \frac{p_0(e^{\lambda_0t}-1)}{e^{\lambda_0t}-p_0},  \\
    p_j(t) &= \textstyle \frac{q_0^2 e^{\lambda_0t}}{(e^{\lambda_0t}-p_0)^2} \cdot \left(\frac{e^{\lambda_0t}-1}{e^{\lambda_0t}-p_0}\right)^{j-1}, \quad j \geq 1.
\end{split}
\end{align}
The probability that a single-cell derived clone is still alive at time $t$ is then given by
\begin{align} \label{eq:probnonextinbytimet}
    \textstyle \P(Z_0(t)>0|Z_0(0)=1) = 1-p_0(t) = {q_0 e^{\lambda_0t}}/({e^{\lambda_0t}-p_0}).
\end{align}
For $0 \leq t \leq t_N$, let $S_{j,t_N}(t)$ denote the number of mutations that accumulate in $[0,t]$ and are found in $j \geq 1$ cells at time $t_N$.
We write $S_{j}(t_N) := S_{j,t_N}(t_N)$ for the site frequency spectrum at time $t_N$.
Say a cell division occurs in an infinitesimal time interval $[t,t+\Delta t]$.
The division results in $w$ mutations on average, each assigned to one of the two daughter cells, and the clone started by this cell has size $j \geq 1$ cells at time $t_N$ with probability $p_j(t_N-t)+O(\Delta t)$.
We wish to show that on the event $\{Z_0(t_N)>0\}$ of survival of the population up until time $t_N$, the expected number of mutations that accumulate in $[t,t+\Delta t]$ and are found in $j \geq 1$ cells at time $t_N$ is
\begin{align} \label{eq:accepttrue}
               \E\big[\big(S_{j,t_N}( t+\Delta t)-S_{j,t_N}( t)\big)1_{\{Z_0(t_N)>0\}}\big] = wr_0\Delta t \cdot e^{\lambda_0t} \cdot p_j(t_N-t) + o(\Delta t),
\end{align}
where we use that $\E[Z_0(t)] = e^{\lambda_0t}$.
It will then follow from \eqref{eq:accepttrue} that 
\begin{align}  \label{eq:accepttrue1}
                &\E\big[S_{j,t_N}( t+\Delta t)-S_{j,t_N}( t)\,\big|\,Z_0(t_N)>0\big] \nonumber \\
            &= wr_0e^{\lambda_0t}\Delta t \cdot p_j(t_N-t)/(1-p_0(t_N)) + o(\Delta t).
\end{align}
Note that \eqref{eq:accepttrue} clearly holds for the semideterministic model in which the tumor bulk grows deterministically at rate $\lambda_0$, mutant clones arise at stochastic rate $wr_0$, and mutant clones grow stochastically.
It is not obvious that \eqref{eq:accepttrue} also holds for our fully stochastic model, since including the event $\{Z_0(t_N)>0\}$ of survival up until time $t_N$ should presumably affect the expected population size at time $t \leq t_N$.
The key is to observe that if a mutation occurs on a cell division at time $t \leq t_N$ and ends up in $j \geq 1$ cells at time $t_N$, the population is automatically alive at time $t_N$.
The relevant survival event in \eqref{eq:accepttrue} is therefore $\{Z_0(t)>0\}$, and the relevant population size factor is $\E[Z_0(t)1_{\{Z_0(t)>0\}}] = \E[Z_0(t)]=e^{\lambda_0t}$.
Another potential concern in establishing \eqref{eq:accepttrue} for our model is that we allow multiple mutations to occur per cell division.
To not distract further from the main calculations, we assume that the reader is willing to accept \eqref{eq:accepttrue} as true for the moment, and we provide a detailed mathematical argument for this expression at the end of the proof.

Using \eqref{eq:accepttrue1}, 
we can integrate over time to obtain
\begin{align} \label{eq:intermediatestep}
    \textstyle & \E\big[S_j(t_N)|Z_0(t_N)>0\big] = (1-p_0(t_N))^{-1} \cdot \textstyle \int_0^{t_N} wr_0 e^{\lambda_0t} p_j(t_N-t) dt.
\end{align}
Focusing on the integral, we write
\begin{linenomath*}
\begin{align*}
& \textstyle \int_0^{t_N} wr_0 e^{\lambda_0t} p_j(t_N-t) dt \\
    &= \textstyle (w/q_0) \cdot \textstyle \int_0^{t_N} \frac{q_0^2 e^{\lambda_0t_N}e^{\lambda_0t}}{(e^{\lambda_0t_N}-p_0e^{\lambda_0t})^2}\cdot \big(\frac{e^{\lambda_0t_N}-e^{\lambda_0t}}{e^{\lambda_0{t_N}}-p_0e^{\lambda_0t}}\big)^{j-1}  \cdot \lambda_0 e^{\lambda_0 t} dt \\
    &= \textstyle w q_0 e^{\lambda_0 t_N} \textstyle \cdot \int_0^{t_N} \frac{e^{\lambda_0t}}{(e^{\lambda_0t_N}-p_0e^{\lambda_0t})^2}\cdot \big(\frac{e^{\lambda_0t_N}-e^{\lambda_0t}}{e^{\lambda_0t_N}-p_0e^{\lambda_0t}}\big)^{j-1}  \cdot \lambda_0 e^{\lambda_0 t} dt.
\end{align*}
\end{linenomath*}
Set $L := e^{\lambda_0 t_N}$. Using the substitution $x := e^{\lambda_0t}$, $dx = \lambda_0e^{\lambda_0t}$, we obtain
\[
    \textstyle \int_0^{t_N} wr_0 e^{\lambda_0t} p_j(t_N-t) dt = \textstyle wq_0L \cdot \textstyle \int_1^{L} \frac{x}{(L-p_0x)^2}\cdot \big(\frac{L-x}{L-p_0x}\big)^{j-1}  dx.
\]
We again change variables, this time $y := (L-x)/(L-p_0x)$, in which case
\begin{linenomath*}
\begin{align*}
\begin{array}{ll}
     &x = L (1-y)/(1-p_0y), \\
     & dx= -\big(q_0 L/(1-p_0y)^2\big)dy, \\
    &L-p_0x = q_0 L/(1-p_0y),
\end{array}
\end{align*}
\end{linenomath*}
and $y = (L-1)/(L-p_0) = 1-q_0/(L-p_0)$ for $x=1$ and $y=0$ for $x=L$, which implies
\begin{align} \label{eq:generalresultgeneralform}
    & \textstyle \int_0^{t_N} wr_0 e^{\lambda_0t} p_j(t_N-t) dt 
=\textstyle w L \cdot \textstyle \int_0^{1-q_0/(L-p_0)} (1-p_0y)^{-1} (1-y) y^{j-1} dy.
\end{align}
We now apply \eqref{eq:intermediatestep} and \eqref{eq:probnonextinbytimet} to see that
\begin{align*}
        &\textstyle \E\big[S_j(t_N)|Z_0(t_N)>0\big] = \textstyle w \cdot \frac{e^{\lambda_0t_N}-p_0}{q_0} \cdot \textstyle \int_0^{1-q_0/(e^{\lambda_0t_N}-p_0)} (1-p_0y)^{-1} (1-y) y^{j-1} dy,    
\end{align*}
and the desired result \eqref{eq:fixedtimeresultgeneral} follows from the fact that $(e^{\lambda_0t_N}-p_0)/q_0=N$ by the definition of $t_N$ in \eqref{eq:fixedtime}. It also follows that for fixed $j \geq 1$,
\begin{align*}
    \E\big[S_j(t_N)|Z_0(t_N)>0\big] \sim wN \cdot \textstyle \int_0^1 (1-p_0y)^{-1} (1-y) y^{j-1} dy
\end{align*}
as $N \to \infty$.
To write the last expression as a sum, note that
\[
    \textstyle (1-p_0y)^{-1} = \sum_{k=0}^\infty p_0^ky^k,
\]
which is valid for all $0 \leq p_0<1$ and $0 \leq y \leq 1$. It follows that
\begin{linenomath*}
\begin{align*}
   \textstyle \int_0^{1} (1-p_0y)^{-1} (1-y) y^{j-1} dy &= \textstyle \sum_{k=0}^\infty p_0^k \big( \int_0^{1} (1-y) y^{j+k-1} dy\big) \\
   &= \textstyle \sum_{k=0}^\infty \frac{p_0^k}{(j+k)(j+k+1)}.
\end{align*}
\end{linenomath*}

We conclude by establishing \eqref{eq:accepttrue} above, which was
\begin{align*}
                & \E\big[(S_{j,t_N}( t+\Delta t)-S_{j,t_N}( t))1_{\{Z_0(t_N)>0\}}\big] = wr_0\Delta t \cdot e^{\lambda_0t} \cdot p_j(t_N-t) + o(\Delta t).
\end{align*}
We decompose according to population size at time $t$. 
Assume that $Z_0(t)=k$ with $k \geq 1$, i.e.~there are $k$ cells at time $t$.
Let $D_{t,\Delta t}$ denote the event that exactly one of the $k$ cells divides in the infinitesimal time interval $[t,t+\Delta t]$, and enumerate the $k+1$ cells after the cell division as $Y_t^1,\ldots,Y_t^{k+1}$, where  $Y_t^1$ and $Y_t^2$ are the two new cells.
Let $W$ denote the number of mutations that occur on the cell division, where $W$ is a nonnegative integer-valued random variable with $\E[W]=w$, independent of $(Z_0(t))_{t \geq0}$.
For $\ell \geq 1$, let $B_{\ell}$ be i.i.d.~with $\P(B_{\ell}=1)=\P(B_{\ell}=2)=1/2$, independent of $(Z_0(t))_{t \geq 0}$ and $W$, and assign mutation number $\ell$ to cell number $B_{\ell}$ for $1 \leq \ell \leq W$.
Finally, let $Y_t^m(s)$ be the number of descendants of cell $Y_t^m$ at time $t+s$, with $Y_t^m(0)=1$.
With this notation, define
 \begin{align*}
                 A_{j,k,\ell}(t) :=\;  &\{Z_0(t)=k\} \cap D_{t,\Delta t} \cap \{\ell \leq W\} \cap \{Y_t^{B_{\ell}}(t_N-t)=j\} \cap \{Z_0(t_N)>0\}.
 \end{align*}
        This is the event that the tumor survives to time $t_N$, that it consists of $k$ cells at time $t$, that exactly one of the $k$ cells divides in $[t,t+\Delta t]$, that at least $\ell$ mutations occur on this division, and that mutation number $\ell$ is found in $j$ cells at time $t_N$.
        The reason we are interested in this event is that we can write
        \begin{align*}
                         &\E\big[(S_{j,t_N}\big( t+\Delta t)-S_{j,t_N}( t)\big)1_{\{Z_0(t_N)>0\}}\big] = \textstyle \sum_{k=1}^\infty \sum_{\ell=1}^\infty \P\big(A_{j,k,\ell}(t)\big) + o(\Delta t),
        \end{align*}
        where the $o(\Delta t)$ term captures the possibility of more than one cell division in $[t,t+\Delta t]$.
        
        To compute $\P\big(A_{j,k,\ell}(t)\big)$, note first that
        \begin{align*}
                        & \P\big(A_{j,k,\ell}(t)\big) = \P\big(\{Z_0(t)=k\} \cap D_{t,\Delta t} \cap \{\ell \leq W\} \cap \{Y_t^{B_{\ell}}(t_N-t)=j\}\big),
        \end{align*}
        since the survival event $\{Z_0(t_N) >0\}$ is implied by the other events. By independence,
        \begin{align*}
                        & \P\big(A_{j,k,\ell}(t)\big) = \P(\ell \leq W) \cdot \P\big(\{Z_0(t)=k\} \cap D_{t,\Delta t} \cap \{Y_t^{B_{\ell}}(t_N-t)=j\}\big).
        \end{align*}
         To analyze the latter probability, note that since $B_{\ell}$ is independent of $(Z_0(t))_{t \geq 0}$, and $\big((Z_0(s))_{s \leq t}, (Y_t^1(s))_{s \geq 0}\big) \stackrel{d}{=} \big((Z_0(s))_{s \leq t}, (Y_t^2(s))_{s \geq 0}\big)$, we can write
         \begin{linenomath*}
        \begin{align*}
            & \P\big(\{Z_0(t)=k\} \cap D_{t,\Delta t} \cap \{Y_t^{B_{\ell}}(t_N-t)=j\} \big)\\
            &= \P\big(\{Z_0(t)=k\} \cap D_{t,\Delta t} \cap \{Y_t^1(t_N-t)=j\}\big).
        \end{align*}
        \end{linenomath*}
        Using the Markov property, we can calculate the latter probability as
        \begin{linenomath*}
        \begin{align*}
            & \P\big(\{Z_0(t)=k\} \cap D_{t,\Delta t} \cap \{Y_t^1(t_N-t)=j\}\big) \\
            &= \P(Z_0(t)=k) \cdot \P(D_{t,\Delta t} | Z_0(t)=k) \cdot \P\big( Y_t^1(t_N-t)=j \big| Z_0(t)=k, D_{t,\Delta t}\big) \\
            &= \P(Z_0(t)=k) \cdot e^{-kr_0\Delta t}kr_0\Delta t \cdot (p_j(t_N-t) + O(\Delta t)).
        \end{align*}
        \end{linenomath*}
        Combining the above, we obtain
        \begin{linenomath*}
        \begin{align*}
             & \E\big[\big(S_{j,t_N}( t+\Delta t)-S_{j,t_N}( t)\big)1_{\{Z_0(t_N)>0\}}\big] \\
             &=\textstyle \sum_{k=1}^\infty \sum_{\ell=1}^\infty \P\big(A_{j,k,\ell}(t)\big) + o(\Delta t) \\
              &= \textstyle r_0p_j(t_N-t)\Delta t \cdot\big(\sum_{\ell=1}^\infty \P(W \geq \ell)\big) \cdot \big(\sum_{k=1}^\infty k\P(Z_0(t)=k)\big) +o(\Delta t) \\
             &= \textstyle wr_0\Delta t \cdot e^{\lambda_0t} \cdot p_j(t_N-t) + o(\Delta t),
        \end{align*}
        \end{linenomath*}
        where we use $\sum_{\ell=1}^\infty \P(W \geq \ell) = \E[W] = w$ and $\sum_{k=1}^\infty k\P(Z_0(t)=k) = \E[Z_0(t)] = e^{\lambda_0t}$.
        This concludes the proof.
\item Let $(X_n)_{n \geq 0}$ denote the discrete-time jump process embedded in $(Z_0(t))_{t \geq 0}$ that only keeps track of changes in population size. More precisely, if $\sigma_n$ is the time of the $n$-th jump of $(Z_0(t))_{t \geq 0}$ for $n \geq 1$, then $X_0 =1$ and $X_n = Z_0(\sigma_n)$ for $n \geq 1$.
Since cells divide at rate $r_0$ and die at rate $d_0$, $(X_n)_{n \geq 0}$ is a simple random walk, absorbed at 0, which moves up with probability~$a := r_0/(r_0+d_0) = 1/(1+p_0)$ and down with probability~$b =1-a = p_0/(1+p_0)$.
         Since we are only interested in what happens until the population either goes extinct or reaches level $N$, we treat $N$ as an absorbing state.
         Define
         \begin{align} \label{eq:stoppingtimediscrmc}
                      T_k := \inf \{n \geq 0: X_n = k\}, \quad 0 \leq k \leq N,
         \end{align}
         as the (discrete) time at which the random walk first hits level $k$, with $\inf \varnothing = \infty$.
         Let $\P_j$ denote the probability measure of $(X_n)_{n \geq 0}$ when started at $X_0=j$.
         By the gambler's ruin formula,
         \begin{align} \label{eq:gamblersruin}
             \P_j(T_k<T_0) = (1-p_0^j)/(1-p_0^k), \quad 0 \leq j \leq k.
         \end{align}
         For $1 \leq k \leq N-1$, let $\Lambda_{k,k+1}$ denote the number of transitions from $k$ to $k+1$,
         \[
             \textstyle \Lambda_{k,k+1} := \sum_{j=0}^\infty 1_{\{Z_j=k,Z_{j+1}=k+1\}},
         \]
         and let $\Lambda_k$ denote the number of visits to $k$,
                          \[
             \textstyle \Lambda_{k} := \sum_{j=0}^\infty 1_{\{Z_j=k\}}.
         \]
         By the strong Markov property and \eqref{eq:gamblersruin}, we can write
         \[
             \E_1[\Lambda_k] = \textstyle \P_1(T_k<T_0) \cdot \E_k[\Lambda_k] = \textstyle \frac{q_0}{1-p_0^k} \cdot \E_k[\Lambda_k].
         \]
         When the chain leaves state $k$, it moves up with probability $1/(1+p_0)$ and down with probability $p_0/(1+p_0)$. Starting from $k+1$, the probability that the chain does not return to $k$ (probability it is absorbed at $N$) is $q_0/(1-p_0^{N-k})$ by \eqref{eq:gamblersruin}, and starting from $k-1$, the probability it does not return to $k$ (probability it is absorbed at 0) is $1-(1-p_0^{k-1})/(1-p_0^k)$ again by \eqref{eq:gamblersruin}. Thus, starting from $k$, $\Lambda_k$ has the geometric distribution with support $\{1,2,\ldots\}$ and success probability
         \[
             \textstyle \frac1{1+p_0} \cdot \frac{q_0}{1-p_0^{N-k}} + \frac{p_0}{1+p_0} \cdot \big(1-\frac{1-p_0^{k-1}}{1-p_0^k}\big) = \frac{q_0(1-p_0^N)}{(1+p_0)(1-p_0^k)(1-p_0^{N-k})}.
         \]
         It follows that
         \begin{align}  \label{eq:numtimesstatek}
         \begin{split}
             &\E_1[\Lambda_k] = \textstyle \frac{(1+p_0)(1-p_0^{N-k})}{1-p_0^N}, \\
             &\textstyle  \E_1[\Lambda_{k,k+1}] = \frac{1}{1+p_0} \cdot \E_1[\Lambda_k] = \frac{1-p_0^{N-k}}{1-p_0^N}.
         \end{split}
         \end{align}
         For $1 \leq k \leq N-1$, define $T_{k,k+1}^i$ as the (discrete) time of the $i$-th transition from $k$ to $k+1$ inductively by
        \[
             T_{k,k+1}^{i} := \inf\{n>T_{k,k+1}^{i-1}: X_{n-1}=k, X_{n}=k+1\}, \quad i \geq 1,
         \]
         with $T_{k,k+1}^0 := 0$ and $\inf \varnothing = \infty$. A transition from $k$ to $k+1$ in $(X_n)_{n \geq 0}$ occurs due to one of the $k$ cells in the original process $(Z_0(t))_{t \geq 0}$ dividing.
         Assume $W_{i,k}$ mutations occur on the $i$-th such transition, where $W_{i,k}$ are i.i.d.~nonnegative integer-valued random variables with $\E[W_{i,k}] = w$, independent of $(X_n)_{n \geq 0}$.
         Enumerate the cells at time $T_{k,k+1}^i$ as  $Y_{i,k}^1,\ldots,Y_{i,k}^{k+1}$.
         By the same argument as laid out in part (1) above,
         we can assume that each mutation is assigned to the first cell.
         Let $Y_{i,k}^m(n)$ be the number of descendants of cell $Y_{i,k}^m$ at time step $T_{k,k+1}^i+n$, with $Y_{i,k}^m(0)=1$.
        Then define the event
        \begin{align*}
                        & A_{j,k,i,\ell} := \big\{ T_N<T_0, \; T_{k,k+1}^i<\infty, \;\ell \leq W_{i,k}, \; Y_{i,k}^{1}(T_N-T_{k,k+1}^i)=j\big\}.
        \end{align*}
        This is the event that the random walk eventually hits level $N$,
        that it transitions at least $i$ times from $k$ to $k+1$ before doing so,
        that at least $\ell$ mutations occur on the $i$-th such transition,
        and that the $\ell$-th mutation is found in $j$ cells at level $N$.
        We can then write
        \begin{linenomath*}
         \begin{align*}
             \E[S_j(\tau_N)|\tau_N<\infty] 
             &= \textstyle\big(\P_1(T_N<T_0)\big)^{-1} \cdot \sum_{k=1}^{N-1} \sum_{i=1}^\infty \sum_{\ell=1}^\infty  \P_1\big( A_{j,k,i,\ell}\big) \\
             &= \textstyle \frac{1-p_0^N}{q_0} \cdot \sum_{k=1}^{N-1} \sum_{i=1}^\infty \sum_{\ell=1}^\infty \P_1\big( A_{j,k,i,\ell}\big).
         \end{align*}
         \end{linenomath*}
         To compute $\P_1\big(A_{j,k,i,\ell}\big)$, note first that by independence,
         \begin{linenomath*}
                 \begin{align*}
            & \P_1\big(T_N<T_0,  T_{k,k+1}^i<\infty, \ell \leq W_{i,k},  Y_{i,k}^{1}(T_N-T_{k,k+1}^i)=j\big) \\
            &= \P\big(\ell \leq W_{i,k}) \cdot \P_1(T_N<T_0, T_{k,k+1}^i<\infty,Y_{i,k}^{1}(T_N-T_{k,k+1}^i)=j\big).
        \end{align*}
        \end{linenomath*}
        By the strong Markov property, 
        \begin{linenomath*}
                \begin{align*}
            & \P_1(T_N<T_0, T_{k,k+1}^i<\infty,Y_{i,k}^1(T_N-T_{k,k+1}^i)=j\big) \\
            &= \P_1(T_N<T_0,Y_{i,k}^1(T_N-T_{k,k+1}^i)=j| T_{k,k+1}^i<\infty\big) \cdot \P_1( T_{k,k+1}^i<\infty)\\
            &= \P_{k+1}\big(T_N<T_0,Y^{1}(T_N)=j\big) \cdot \P_1(T_{k,k+1}^i<\infty),
        \end{align*}
        \end{linenomath*}
        where we restart the chain at the stopping time $T_{k,k+1}^i$ with $k+1$ cells enumerated as $Y^1,\ldots,Y^{k+1}$.
        Define $h_{(1,k)}^{(j,N-j)} := \P_{k+1}(T_N<T_0,Y^{1}(T_N)=j)$ for the moment, i.e.~the probability that starting with one cell carrying a particular mutation and $k$ cells without it, $j$ cells carry the mutation when the population reaches size $N$.
        We can then write
        \begin{linenomath*}
                        \begin{align*}
            & \P_1\big(T_N<T_0, T_{k,k+1}^i<\infty,Y_{i,k}^1(T_N-T_{k,k+1}^i)=j\big) = h_{(1,k)}^{(j,N-j)} \cdot \P_1(T_{k,k+1}^i<\infty).
        \end{align*}
        \end{linenomath*}
        Combining the above, and using $\E_1[\Lambda_{k,k+1}] = ({1-p_0^{N-k}})/({1-p_0^N})$  by \eqref{eq:numtimesstatek}, we obtain
        \begin{linenomath*}
        \begin{align*}
&             \E[S_j(\tau_N)|\tau_N<\infty] \\
             &= \textstyle \frac{1-p_0^N}{q_0} \cdot \textstyle \sum_{k=1}^{N-1} \big(\sum_{i=1}^\infty  \big(\sum_{\ell=1}^\infty \P(W_{i,k} \geq \ell)\big)  \P_1(T_{k,k+1}^i<\infty) \big)  h_{(1,k)}^{(j,N-j)} \\
             &= \textstyle \frac{1-p_0^N}{q_0} \cdot \sum_{k=1}^{N-1}  w \cdot \E_1[\Lambda_{k,k+1}] \cdot  h_{(1,k)}^{(j,N-j)} \\
             &= \textstyle (w/q_0) \cdot \sum_{k=1}^{N-1} (1-p_0^{N-k}) \cdot h_{(1,k)}^{(j,N-j)},
         \end{align*}
         \end{linenomath*}
         which is the desired result.

         It remains to determine how the probabilities $h_{(1,k)}^{(j,N-j)}$ for $1 \leq j \leq N$ can be computed.
         As in the proof of (1) of Proposition \ref{prop:skeletonspectrum}, one can view $h_{(\ell,m)}^{(r,N-r)}$ as the probability of absorption in state $(r,N-r)$, starting from state $(\ell,m)$, for a Markov chain on the state space ${\cal S} := \{(\ell,m): \ell,m \geq 0 \text{ and } \ell+m \leq N\}$, where $\ell$ is the number of cells carrying a particular mutation and $m$ is the number of cells without it. The difference is that now, cells can die, so population level changes can be both up and down.
         The transition probabilities are therefore more complex in this case, and given by
         \begin{linenomath*}
        \begin{align*}
            (\ell,m) \to (\ell+1,m) \quad \text{w.p.}\quad  &{\ell}r_0/({(\ell+m)(r_0+d_0)}) \\
            &= {\ell}/({(\ell+m)(1+p_0)}), \\
            (\ell,m) \to (\ell-1,m) \quad \text{w.p.}\quad  &{\ell}d_0/({(\ell+m)(r_0+d_0)}) \\
            &= {\ell p_0}/({(\ell+m)(1+p_0)}), \\
            (\ell,m) \to (\ell,m+1) \quad \text{w.p.}\quad &{m}r_0/({(\ell+m)(r_0+d_0)}) \\
            &=  {m}/({(\ell+m)(1+p_0)}), \\
            (\ell,m) \to (\ell,m-1) \quad \text{w.p.}\quad  &{m}d_0/({(\ell+m)(r_0+d_0)}) \\
            &=  {m  p_0}/({(\ell+m)(1+p_0)}),
        \end{align*}
        \end{linenomath*}
        for $(\ell,m)$ with $\ell,m \geq 0$ and $0<\ell+m<N$. The states $(0,0)$ and $(\ell,N-\ell)$ with $0 \leq \ell \leq N$ are absorbing. A diagram of the Markov chain is shown in Figure \ref{fig:mchain_general}.
        
        For given $(r,s) \in A := \{(0,0)\} \cup \{(r,s): r,s \geq 0 \text{ and } r+s = N\}$ and $(\ell,m)$ with $\ell,m \geq 0$ and $0<\ell+m < N$, by conditioning on the first transition out of state $(\ell,m$), we can derive the following recursion for $h_{(\ell,m)}^{(r,s)}$:
        \begin{linenomath*}
            \begin{align*}
& (\ell+m) (1+p_0) h_{(\ell,m)}^{(r,s)} 
             = \ell  h_{(\ell+1,m)}^{(r,s)} + \ell p_0  h_{(\ell-1,m)}^{(r,s)} + m  h_{(\ell,m+1)}^{(r,s)} + m p_0 h_{(\ell,m-1)}^{(r,s)}.
    \end{align*}
    \end{linenomath*}
    By the gambler's ruin formula \eqref{eq:gamblersruin}, the boundary conditions are
    \begin{align} \label{eq:boundaryconditions}
    \begin{split}
                & h_{(\ell,0)}^{(N,0)} = 1-h_{(\ell,0)}^{(0,0)} = (1-p_0^\ell)/(1-p_0^N), \quad 0 \leq \ell \leq N, \\
        & h_{(\ell,0)}^{(r,s)} = 0, \quad  0 \leq \ell \leq N,\;(r,s) \notin \{(0,0),(N,0)\},  \\
        & h_{(0,m)}^{(0,N)} = 1-h_{(0,m)}^{(0,0)} = (1-p_0^m)/(1-p_0^N), \quad 0 \leq m \leq N, \\
        & h_{(0,m)}^{(r,s)} = 0, \quad 0 \leq m \leq N, \; (r,s) \notin \{(0,0),(0,N)\}, \\
        & h_{(\ell,N-\ell)}^{(r,N-r)} = \delta_{r,\ell}, \quad 1 \leq \ell \leq N-1, \\
        & h_{(\ell,N-\ell)}^{(r,s)}=0, \quad  1 \leq \ell \leq N-1, \; (r,s) \in \{(0,0),(N,0),(0,N)\}.
    \end{split}
    \end{align}
    This is the desired linear system. \qedhere
 \end{enumerate}
\end{proof}

                  \begin{figure*}[t]
             \centering
             \includegraphics[scale=1]{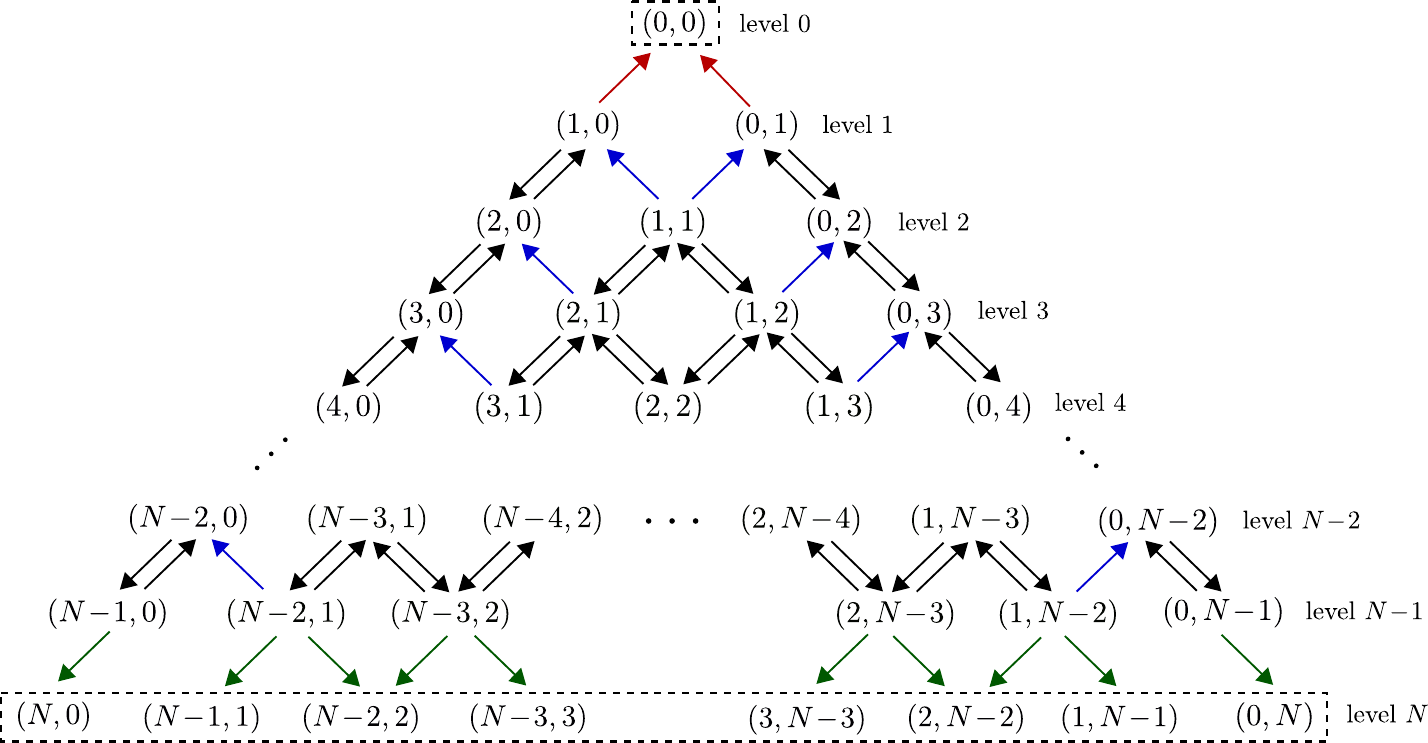}
             \caption{A diagram of the discrete-time Markov chain on ${\cal S} := \{(\ell,m): \ell,m \geq 0, \ell+m \leq N\}$, where $\ell$ is the number of cells carrying a particular mutation and $m$ is the number of cells without it.
             Contrary to the chain for the skeleton process of Proposition \ref{prop:skeletonspectrum} (Figure \ref{fig:mchain_skeleton}), we now incorporate cell death, which means both that population level changes can be up and down, and that we add states of the form $(\ell,m)$ with $\ell=0$ or $m=0$.
             The states $(0,0)$ and $\{(\ell,N-\ell)\}_{0 \leq \ell \leq N}$ are absorbing (dashed boxes), and the states $\{(\ell,m): \ell,m \geq 1, \ell+m<N\}$, $\{(\ell,0)\}_{1 \leq \ell \leq N-1}$ and $\{(0,m)\}_{1 \leq m \leq N-1}$ form their respective communicating classes. Colored arrows indicate transitions out of communicating classes.
             }
             \label{fig:mchain_general}
         \end{figure*}

\section{Results for continuous mutation accumulation} \label{app:continuousmutation}

In this section, we discuss how to derive the expected SFS of the skeleton and the total population under the continuous model of mutation accumulation (see Section \ref{sec:mutationaccumulation}).
This requires minor modifications to the proofs of Propositions \ref{prop:skeletonspectrum} and \ref{prop:spectrum}.

\subsection{Skeleton spectrum}

\begin{figure*}
             \centering
             \includegraphics[scale=1]{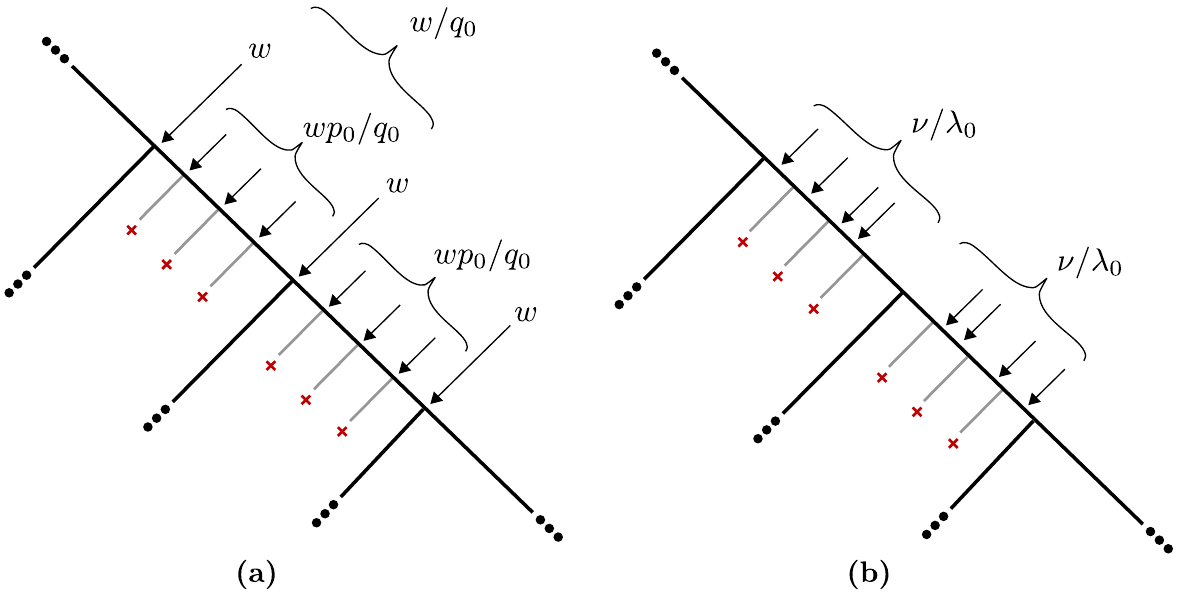}
             \caption{
             Mutation accumulation on the skeleton in the discrete vs.~continuous model of mutation.
             {\bf (a)} In the discrete model, mutations coincide with cell divisions.
             On average, $wp_0/q_0$ mutations accumulate on type-1 divisions in between two type-2 divisions, and $w$ mutations are added on each type-2 division.
             This results in $wp_0/q_0+w=w/q_0$ mutations on average per skeleton population size level, for all but the first level.
             {\bf (b)} In the continuous model, all $\nu/\lambda_0$ $(= w/q_0)$ mutations per level accumulate in between cell divisions.
             The differences between the discrete and continuous model result in slightly different behavior at the boundary values $j=1$ and $j=N$ in the fixed-size spectrum \eqref{eq:skeletonfixedsizespectrum} of Proposition \ref{prop:skeletonspectrum} and fixed-size result \eqref{eq:skeletonspectrumfixedsizecontinuous} for the continuous model.
             }
             \label{fig:mutaccumlationdifference}
         \end{figure*}

Under continuous mutation accumulation, the expected fixed-time spectrum of the skeleton is given by
    \[
            \E[\tilde{S}_j(\tilde t_N)] = \textstyle (\nu/\lambda_0) N \cdot \int_0^{1-1/N} (1-y)y^{j-1} dy,
\]
which is the same as \eqref{eq:skeletonfixedtimespectrum} of Proposition \ref{prop:skeletonspectrum} with $w/q_0$ replaced by $\nu/\lambda_0$, the effective mutation rate in the continuous-time model.
We can use the same proof as in part (1) of Appendix \ref{app:proofprop1}, simply replacing the mutation rate $wr_0$ by $\nu$ and noting that $q_0 = \lambda_0/r_0$.
However, the expected fixed-size spectrum of the skeleton becomes 
 \begin{align} \label{eq:skeletonspectrumfixedsizecontinuous}
                  & \E[\tilde S_j(\tilde\tau_N)] = \begin{cases}  (\nu/\lambda_0)N \cdot \textstyle 1/(j(j+1)) - (\nu/\lambda_0) \delta_{1,j}, &
                  1 \leq j \leq N-1, \\ \nu/\lambda_0, &  j = N. \end{cases}
\end{align}
In the continuous model, mutations occur at rate $\nu$ per unit time, and the effective type-2 cell divisions occur at rate $\lambda_0$ per unit time.
Thus, for $1 \leq k \leq N-1$, the number of mutations that accumulate on skeleton population size level $k$, prior to the type-2 division that changes levels to $k+1$, has the geometric distribution with support $\{0,1,2,\ldots\}$ and success probability
 \[
     \textstyle {k\lambda_0}/({k\lambda_0+k\nu}) = {\lambda_0}/({\lambda_0+\nu}).
 \]
 The expected number of mutations per level is therefore
 \[
     (\lambda_0+\nu)/\lambda_0 -1= \nu/\lambda_0,
 \]
 which applies to all levels $k$ with $1 \leq k \leq N-1$.
In particular, there are $\nu/\lambda_0$ $(=w/q_0)$ clonal mutations in the continuous model, as opposed to $wp_0/q_0 = w/q_0-w$ clonal mutations in the discrete model.
Recall that in the latter model, mutations coincide with cell divisions, and clonal mutations come from the type-1 divisions that occur before the first type-2 division in the process.
The first type-2 division adds $w$ mutations, but it also changes levels, so mutations occurring on this division are not clonal.
In the continuous model, all mutations occur in between cell divisions, which is why there is no such boundary effect.
Similarly, in the discrete model, the very last type-2 division that changes the skeleton size to $N$ adds $w$ mutations on average that each ends up in one cell.
Since this does not occur in the continuous model, the extra term $-\nu/\lambda_0$ ($=-w/q_0$) that appears for $j=1$ in \eqref{eq:skeletonspectrumfixedsizecontinuous} differs from the extra term $-wp_0/q_0=-w/q_0+w$ in the discrete model by $w$.
These key differences between mutation accumulation in the two models are diagramed in Figure \ref{fig:mutaccumlationdifference}.

\subsection{Total population spectrum}

Under continuous mutation accumulation, the expected fixed-time spectrum of the total population is given by
\[
        \E[S_j(t_N)|Z_0(t_N)>0] =\textstyle (\nu/r_0)N \cdot \int_0^{1-1/N} (1-p_0y)^{-1} (1-y) y^{j-1} dy,
\]
which is the same as \eqref{eq:fixedtimeresultgeneral} of Proposition \ref{prop:spectrum} with $w$ replaced by $\nu/r_0$.
We can use the same proof as in part (1) of Appendix \ref{app:proofprop2}, replacing the mutation rate $wr_0$ by $\nu$.
However, the expected fixed-size spectrum of the total population becomes, for $1 \leq j \leq N$,
    \[
     \E[S_j(\tau_N)|\tau_N<\infty]  = (\nu/\lambda_0) \cdot \textstyle \sum_{k=1}^{N-1} (1-p_0^{N-k}) \cdot h^{(j,N-j)}_{(1,k-1)}.
\]
In the continuous model, mutations no longer coincide with changes in population level, so instead of counting level changes $\Lambda_{k,k+1}$ as we did in the proof of part (2) of Proposition \ref{prop:spectrum}, we need to compute the expected time spent at level $k$.
We already know from \eqref{eq:numtimesstatek} that the number of visits to level $k$ in the embedded discrete-time chain $(X_n)_{n \geq 0}$ has expected value
\[
    \textstyle \E_1[\Lambda_k] = \frac{(1+p_0)(1-p_0^{N-k})}{1-p_0^N}.
\]
During each visit to state $k$ in the discrete-time chain, the time spent at population level $k$ in the continuous-time process $(Z_0(t))_{t \geq 0}$ is exponentially distributed with rate $k(r_0+d_0) = kr_0(1+p_0)$. It follows that the mean time spent on level $k$ for $1 \leq k \leq N-1$ is
\[
    \textstyle \frac1{kr_0(1+p_0)} \cdot \frac{(1+p_0)(1-p_0^{N-k})}{1-p_0^N} = \frac{1-p_0^{N-k}}{kr_0(1-p_0^N)}.
\]
Since mutations occur at rate $k\nu$ per unit time on level $k$, we obtain
         \begin{align} \label{eq:fixedsizecont}
                          \E[S_j(\tau_N)|\tau_N<\infty] &= \textstyle \frac{1-p_0^N}{q_0} \cdot \sum_{k=1}^{N-1}  k\nu \cdot \frac{1-p_0^{N-k}}{kr_0(1-p_0^N)} \cdot  h_{(1,k-1)}^{(j,N-j)} \nonumber \\
             &= \textstyle (\nu/\lambda_0) \cdot \sum_{k=1}^{N-1} (1-p_0^{N-k}) \cdot h_{(1,k-1)}^{(j,N-j)},
         \end{align}
         where we use that $q_0 = \lambda_0/r_0$. We now have $h_{(1,k-1)}^{(j,N-j)}$ in the sum instead of $h_{(1,k)}^{(j,N-j)}$ in \eqref{eq:fixedsizeresultgeneral} of Proposition \ref{prop:spectrum} since mutations no longer coincide with population level changes.
         The first equality in \eqref{eq:fixedsizecont} can be obtained rigorously following the same line of reasoning as in the proof of part (2) of Proposition \ref{prop:spectrum}.

        \section{Proof of Proposition \ref{prop:nummut}} \label{app:proofprop3}

In this section, we prove Proposition \ref{prop:nummut} on the total mutational burden of the tumor, both under the fixed-time and fixed-size spectrum.
The result follows from Proposition \ref{prop:spectrum} using simple calculations.

\begin{proof}[Proof of Proposition \ref{prop:nummut}]
     \begin{enumerate}[(1)]
         \item  Define $M_j(t) := \sum_{k \geq j} S_k(t)$ as the cumulative number of mutations found in $\geq j$ cells at time $t$.
         For fixed $j \geq 1$, by \eqref{eq:fixedtimeresultgeneral} of Proposition \ref{prop:spectrum} and Fubini's theorem, the expected cumulative fixed-time spectrum can be written as
\begin{align} \label{eq:cumspectrumgeneralsum}
        \E[M_j(t_N)|Z_0(t_N)>0] 
        &= \textstyle wN \cdot \sum_{k=j}^\infty \big(\int_0^{1-1/N}(1-p_0y)^{-1} (1-y) y^{k-1}dy\big) \nonumber \\
        &= \textstyle wN \cdot \int_0^{1-1/N}(1-p_0y)^{-1} (1-y) \big(\sum_{k=j}^\infty  y^{k-1}\big) dy \nonumber \\
    &= wN \textstyle \cdot \int_0^{1-1/N} (1-p_0y)^{-1} y^{j-1}dy \nonumber \\
    &= wN \textstyle \cdot \sum_{k=0}^\infty p_0^k \big(\int_0^{1-1/N} y^{j+k-1}dy\big) \nonumber \\
    &= \textstyle wN \cdot \sum_{k=0}^\infty \frac{p_0^k}{j+k} (1-\frac1N)^{j+k}.
    \end{align}
    To obtain the desired result, set $j=1$ in \eqref{eq:cumspectrumgeneralsum} and use that $\sum_{k=1}^\infty x^k/k = -\log(1-x)$.
    \item By \eqref{eq:fixedsizeresultgeneral} of Proposition \ref{prop:spectrum},
    \begin{linenomath*}
     \begin{align*} 
            \E[M_1(\tau_N)|\tau_N<\infty] 
            &= \textstyle \sum_{j=1}^N \big((w/q_0) \cdot \sum_{k=1}^{N-1} (1-p_0^{N-k}) \cdot h^{(j,N-j)}_{(1,k)}\big) \\
            &= (w/q_0) \cdot \textstyle \sum_{k=1}^{N-1} (1-p_0^{N-k}) \cdot \big(\sum_{j=1}^N h^{(j,N-j)}_{(1,k)}\big),
    \end{align*}
    \end{linenomath*}
    and the result follows from the fact that $\sum_{(r,s) \in A} h_{(1,k)}^{(r,s)} = 1$. \qedhere
     \end{enumerate}
\end{proof}
         
\section{Fixed-time vs.~fixed-size total population spectrum} \label{app:comparetwospectra}

Here, we present a simple heuristic argument for why the fixed-size spectrum of the total population can be approximated by the fixed-time spectrum on $j \ll N$ when $N$ is sufficiently large.
As we discussed in Section \ref{sec:lawoflargenumbers} of the main text, conditional on the nonextinction event $\Omega_\infty$, the tumor eventually grows at exponential rate $\lambda_0$.
If $s$ is the time it takes to go from population level $k$ to population level $N$, and we assume that $s$ can be treated as deterministic, we can write $ke^{\lambda_0s} = N$ i.e.~$e^{\lambda_0s} = N/k$, following e.g.~\citet{iwasa2006evolution}.
We can then make the approximation $h_{(1,k)}^{(j,N-j)} \approx p_j(s)$, where $h_{(1,k)}^{(j,N-j)}$ is defined as in the proof of part (2) of Proposition \ref{prop:spectrum}, and $(p_j(s))_{j \geq 0}$ is the size-distribution at time $s$ for a single-cell derived clone.
Applying \eqref{eq:sizedisgeneral} with $e^{\lambda_0s} = N/k$, we obtain
\[
    \textstyle h_{(1,k)}^{(j,N-j)} \approx p_j(s) = \frac{q_0^2 N k}{(N-p_0k)^2} \cdot \big(\frac{N-k}{N-p_0k}\big)^{j-1}.
\]
Then, observing that $1-p_0^{N-k} \approx 1$ when $k \ll N$, we can write
\begin{align*}
    & (w/q_0) \cdot \textstyle \sum_{k=1}^{N-1} (1-p_0^{N-k}) \cdot h^{(j,N-j)}_{(1,k)} \approx (w/q_0) \cdot \textstyle \int_1^N \frac{q_0^2 N k}{(N-p_0k)^2} \cdot \big(\frac{N-k}{N-p_0k}\big)^{j-1}dk.
\end{align*}
Using the substitution $y := (N-k)/(N-p_0k)$ and writing $N-p_0 \approx N$, this becomes the fixed-time spectrum \eqref{eq:fixedtimeresultgeneral} of Proposition \ref{prop:spectrum}.

\section{Derivation of expressions \eqref{eq:transitionlargefreq} and \eqref{eq:largep0limit}} \label{app:asymptotics}

Here, we establish the asymptotic expressions \eqref{eq:transitionlargefreq} and \eqref{eq:largep0limit} in the main text.
To establish \eqref{eq:transitionlargefreq}, fix $0<p_0<1$ and set
\[
\textstyle f_j(k) := p_0^k \cdot \frac{j(j+1)}{(j+k)(j+k+1)}, \quad j \geq 1, \; k \geq 0.
\]
Clearly, $f_j(k) \leq p_0^k$ for all $j \geq 1$ and $k \geq 0$. Since $\sum_{k=0}^\infty p_0^k = 1/q_0 < \infty$, it follows from the dominated convergence theorem that
\[
    \textstyle \lim_{j \to \infty} \sum_{k=0}^\infty f_j(k) = 1/q_0,
\]
from which it follows that
\[
    \textstyle wN \cdot \sum_{k=0}^\infty \frac{p_0^k}{(j+k)(j+k+1)} \sim (w/q_0)N \cdot 1/(j(j+1)), \quad j \to \infty.
\]
To establish \eqref{eq:largep0limit}, fix $j\geq1$ and set
\[
    \textstyle f_{p_0}(k) := \frac{p_0^k}{(j+k)(j+k+1)}, \quad 0<p_0<1, \; k \geq 0.
\]
Clearly, $f_{p_0}(k) \leq 1/((j+k)(j+k+1))$ for all $0<p_0<1$ and $k \geq 0$. Since
\[
    \textstyle \sum_{k=0}^\infty \frac1{(j+k)(j+k+1)} = \sum_{k=0}^\infty \big(\frac1{j+k}-\frac1{j+k+1}\big) = 1/j < \infty, 
\]
it follows from the dominated convergence theorem that
\[
    \textstyle \lim_{p_0 \to 1} \sum_{k=0}^\infty f_{p_0}(k) = 1/j,
\]
from which it follows that
\[
    \textstyle  wN \cdot \sum_{k=0}^\infty \frac{p_0^k}{(j+k)(j+k+1)} \textstyle  \sim wN \cdot 1/j, \quad p_0 \to 1.
\]

\section{Derivation of expression \eqref{eq:nummutonecell}} \label{app:s1derivation}

Here, we establish expression \eqref{eq:nummutonecell} in the main text.
By \eqref{eq:fixedtimeresultgeneral} of Proposition \ref{prop:spectrum} and Fubini's theorem, we can write
\begin{align} \label{eq:s1general}
        & \E[S_1(t_N)|Z_0(t_N)>0] \nonumber \\
        &= \textstyle wN \cdot \int_0^{1-1/N}(1-p_0y)^{-1} (1-y) dy \nonumber \\
        &= \textstyle wN \cdot \sum_{k=0}^\infty p_0^k \big(\int_0^{1-1/N} y^k (1-y) dy\big) \nonumber \\
        &= \textstyle wN \cdot \sum_{k=0}^\infty p_0^k \big(\frac1{k+1} (1-\frac1N)^{k+1} - \frac1{k+2} (1-\frac1N)^{k+2}\big) \nonumber \\
                &= \textstyle wN \cdot \sum_{k=0}^\infty \frac{p_0^k}{k+1} (1-\frac1N)^{k+1} - wN \cdot \sum_{k=0}^\infty \frac{p_0^k}{k+2} (1-\frac1N)^{k+2}.
    \end{align}
    Using that $\sum_{k=1}^\infty x^k/k = -\log(1-x)$, the former term can be computed as
    \[
        \textstyle wN \cdot \sum_{k=0}^\infty \frac{p_0^k}{k+1} (1-\frac1N)^{k+1} = -wN \cdot (1/p_0) \log(q_0+p_0/N),
    \]
    and the latter term can be computed as
    \begin{linenomath*}
    \begin{align*}
        \textstyle wN \cdot \sum_{k=0}^\infty \frac{p_0^k}{k+2} (1-\frac1N)^{k+2} 
        &= \textstyle wN \cdot (1/p_0^2) \sum_{k=0}^\infty \frac{p_0^{k+2}}{k+2} (1-\frac1N)^{k+2} \\
        &= \textstyle wN \cdot (1/p_0^2) \big(-\log(q_0+p_0/N)-p_0(1-1/N)\big).
    \end{align*}
    \end{linenomath*}
    Combining with \eqref{eq:s1general}, we obtain
    \begin{align*}
                & \E[S_1(t_N)|Z_0(t_N)>0] 
                = wN \cdot (1/p_0)(1-1/N+(q_0/p_0)\log(q_0+p_0/N)),
    \end{align*}
    the desired result.
    
     \section{Laws of large numbers} \label{app:lawoflargenumbers}     
        
        Here, we present simple calculations in support of the conjectured laws of large numbers \eqref{eq:lawoflargenumbersgeneralfixedtime} and \eqref{eq:lawoflargenumbersgeneralfixedsize} of the main text.
        As stated in the main text, conditional on the nonextinction event $\Omega_\infty$, we have $Z_0(t) \sim Ye^{\lambda_0t}$ as $t \to \infty$ almost surely, where $Y$ follows the exponential distribution with mean $1/q_0$ (Theorem 1 of \citet{durrett2015branching}).
For the fixed-time spectrum, the number of mutations that accumulate in $[0,t_N]$ and are found in $j \geq 1$ cells at time $t_N$ is then approximately
\[
    \textstyle S_j(t_N) \approx 
    \textstyle \int_0^{t_N} w r_0 \cdot Ye^{\lambda_0t} \cdot p_j(t_N-t)dt.
\]
From \eqref{eq:generalresultgeneralform} in the proof of Proposition \ref{prop:spectrum}, we know that
\begin{linenomath*}
\begin{align*}
    &\textstyle \int_0^{t_N} wr_0e^{\lambda_0t}p_j(t_N-t)dt \\
    &= \textstyle  w e^{\lambda_0t_N} \cdot \int_0^{1-q_0/(e^{\lambda_0t_N}-p_0)} (1-p_0y)^{-1} (1-y) y^{j-1} dy \\
    &\sim \textstyle w q_0 N \cdot \int_0^{1} (1-p_0y)^{-1} (1-y) y^{j-1} dy, \quad N \to \infty,
\end{align*}
\end{linenomath*}
where we use that $e^{\lambda_0 t_N} = q_0N+p_0$ by the definition of $t_N$ in \eqref{eq:fixedtime}.
This implies that
\[
    \textstyle S_j(t_N) \approx \textstyle q_0 Y \cdot w N \cdot \int_0^{1} (1-p_0y)^{-1} (1-y) y^{j-1} dy
\]
for large $N$.
Since $Y$ has the exponential distribution with mean $1/q_0$, $q_0Y$ has the exponential distribution with mean $1$.
This suggests \eqref{eq:lawoflargenumbersgeneralfixedtime} in the main text.

For the fixed-size spectrum, note that if $N$ is large, then at time $\tau_N-t$,  we can write $Z_0(\tau_N-t) \approx N e^{-\lambda_0t}$.
The number of mutations that accumulate in $[0, \tau_N]$ and are found in $j \geq 1$ cells at time $\tau_N$ is then approximately
\begin{align*}
    \textstyle S_j(\tau_N) &\approx \textstyle \int_0^{\tau_N} w r_0 \cdot Ne^{-\lambda_0t} \cdot p_j(t)dt \\
    &= \textstyle Ne^{-\lambda_0\tau_N} \cdot \int_0^{\tau_N} w r_0  e^{\lambda_0t}  p_j(\tau_N-t)dt.    
\end{align*}
Again using \eqref{eq:generalresultgeneralform} from the proof of Proposition \ref{prop:spectrum}, we can write
\begin{align*}
        \textstyle \int_0^{\tau_N} w r_0  e^{\lambda_0t}  p_j(\tau_N-t)dt 
    &= \textstyle w e^{\lambda_0 \tau_N} \cdot \int_0^{1-q_0/(e^{\lambda_0 \tau_N}-p_0)} (1-p_0y)^{-1} (1-y) y^{j-1} dy,
\end{align*}
from which it follows that
\begin{linenomath*}
\begin{align*}
    \textstyle S_j(\tau_N) &\approx  \textstyle wN \cdot \int_0^{1-q_0/(e^{\lambda_0 \tau_N}-p_0)} (1-p_0y)^{-1} (1-y) y^{j-1} dy \\
    &\approx \textstyle wN \cdot \int_0^1 (1-p_0y)^{-1} (1-y) y^{j-1}
\end{align*}
\end{linenomath*}
for $N$ large.
This suggests \eqref{eq:lawoflargenumbersgeneralfixedsize} in the main text.

We finally mention that it is straightforward to prove a law of large numbers for a simplified version of our model, where the tumor bulk grows deterministically ($Z_0(t) = e^{\lambda_0t}$), mutant clones arise at stochastic rate $wr_0$, 
and mutant clones grow stochastically.
To state the result, let $\hat{S}_j(\tilde t_N)$ denote the number of mutations found in $j \geq 1$ cells at time $\tilde t_N$ under the simplified model, where $\tilde t_N$ is given by \eqref{eq:fixedtimeskeleton}, i.e.~$e^{\lambda_0 \tilde t_N}=N$.
We want to show that
    \begin{align} \label{eq:llnsimplified}
            \textstyle \hat{S}_j(\tilde t_N) \sim  wN \cdot \int_0^1 (1-p_0y)^{-1} (1-y) y^{j-1} dy
    \end{align}
    as $N \to \infty$ almost surely.
Note that the limit is a constant since we assume deterministic growth of the tumor bulk.
For $0 \leq t \leq \tilde t_N$, let $\hat{N}_{j,\tilde t_N}(t)$ denote the number of mutant clones created in $[0,t]$ that have size $j \geq 1$ at time $\tilde t_N$.
    Then $(\hat{N}_{j,\tilde t_N}(t))_{0 \leq t \leq \tilde t_N}$ is an inhomogeneous Poisson process with rate function $\hat\lambda(t) = wr_0 e^{\lambda_0t} p_j(\tilde t_N-t)$ and mean function
\[
    \hat{m}(t) = \textstyle \int_0^{t} \hat\lambda(s) ds, \quad 0 \leq t \leq \tilde t_N.
\]
Set $\hat{N}_{j}(\tilde t_N) := \hat{N}_{j,\tilde t_N}(\tilde t_N)$.
By \eqref{eq:generalresultgeneralform} in the proof of Proposition \ref{prop:spectrum}, and the fact that $e^{\lambda_0 \tilde t_N} = N$,
\begin{linenomath*}
\begin{align*}
    \hat{m}(\tilde t_N) &= \textstyle wN \cdot \int_0^{1-q_0/(N-p_0)} (1-p_0y)^{-1} (1-y) y^{j-1} dy \\
    &\sim \textstyle wN \cdot \int_0^1 (1-p_0y)^{-1} (1-y) y^{j-1} dy, \quad N \to \infty.
\end{align*}
\end{linenomath*}
Then, by a simple Poisson concentration inequality, see Theorem 1 of \citet{Canonne2017},
\begin{linenomath*}
\begin{align*}
    \P\big(|\hat{N}_j(\tilde t_N)/\hat{m}(\tilde t_N)-1|>(\hat{m}(\tilde t_N))^{-1/3}\big) 
    &= \P\big(|\hat{N}_j(\tilde t_N)-\hat{m}(\tilde t_N)|>(\hat{m}(\tilde t_N))^{2/3}\big) \\
    &\leq 2\exp\big(\!-\!(\hat{m}(\tilde t_N))^{1/3}\big/\big(2(1+(\hat{m}(\tilde t_N))^{-1/3})\big)\big).
\end{align*}
\end{linenomath*}
Since $\hat{m}(\tilde t_N)$ is of order $N$ as $N \to \infty$, it follows from the Borel-Cantelli lemma that $\hat{N}_j(\tilde t_N)/\hat{m}(\tilde t_N) \to 1$ as $N \to \infty$ almost surely.
Since $\hat S_j(\tilde t_N) = \hat N_j(\tilde t_N)$, we have the result.

\bibliographystyle{plainnat}
\bibliography{epi}

\begin{thebibliography}{63}
\providecommand{\natexlab}[1]{#1}
\providecommand{\url}[1]{\texttt{#1}}
\expandafter\ifx\csname urlstyle\endcsname\relax
  \providecommand{\doi}[1]{doi: #1}\else
  \providecommand{\doi}{doi: \begingroup \urlstyle{rm}\Url}\fi

\bibitem[Achaz(2009)]{achaz2009frequency}
Guillaume Achaz.
\newblock Frequency spectrum neutrality tests: one for all and all for one.
\newblock \emph{Genetics}, 183\penalty0 (1):\penalty0 249--258, 2009.

\bibitem[Andrews et~al.(1999)Andrews, Askey, and Roy]{andrews1999special}
George~E Andrews, Richard Askey, and Ranjan Roy.
\newblock \emph{Special functions}.
\newblock Number~71. Cambridge university press, 1999.

\bibitem[Antal and Krapivsky(2011)]{antal2011exact}
Tibor Antal and PL~Krapivsky.
\newblock Exact solution of a two-type branching process: models of tumor
  progression.
\newblock \emph{J.~Stat.~Mech.~Theory Exp.}, 2011\penalty0 (08):\penalty0
  P08018, 2011.

\bibitem[Armitage and Doll(1954)]{armitage1954age}
Peter Armitage and Richard Doll.
\newblock The age distribution of cancer and a multi-stage theory of
  carcinogenesis.
\newblock \emph{Br.~J.~Cancer}, 8\penalty0 (1):\penalty0 1, 1954.

\bibitem[Armitage and Doll(1957)]{armitage1957two}
Peter Armitage and Richard Doll.
\newblock A two-stage theory of carcinogenesis in relation to the age
  distribution of human cancer.
\newblock \emph{Br.~J.~Cancer}, 11\penalty0 (2):\penalty0 161, 1957.

\bibitem[Avanzini and Antal(2019)]{avanzini2019cancer}
Stefano Avanzini and Tibor Antal.
\newblock Cancer recurrence times from a branching process model.
\newblock \emph{PLoS Comput.~Biol.}, 15\penalty0 (11):\penalty0 e1007423, 2019.

\bibitem[Bozic et~al.(2010)Bozic, Antal, Ohtsuki, Carter, Kim, Chen, Karchin,
  Kinzler, Vogelstein, and Nowak]{bozic2010accumulation}
Ivana Bozic, Tibor Antal, Hisashi Ohtsuki, Hannah Carter, Dewey Kim, Sining
  Chen, Rachel Karchin, Kenneth~W Kinzler, Bert Vogelstein, and Martin~A Nowak.
\newblock Accumulation of driver and passenger mutations during tumor
  progression.
\newblock \emph{Proc.~Natl.~Acad.~Sci.~USA}, 107\penalty0 (43):\penalty0
  18545--18550, 2010.

\bibitem[Bozic et~al.(2013)Bozic, Reiter, Allen, Antal, Chatterjee, Shah, Moon,
  Yaqubie, Kelly, Le, et~al.]{bozic2013evolutionary}
Ivana Bozic, Johannes~G Reiter, Benjamin Allen, Tibor Antal, Krishnendu
  Chatterjee, Preya Shah, Yo~Sup Moon, Amin Yaqubie, Nicole Kelly, Dung~T Le,
  et~al.
\newblock Evolutionary dynamics of cancer in response to targeted combination
  therapy.
\newblock \emph{eLife}, 2:\penalty0 e00747, 2013.

\bibitem[Bozic et~al.(2016)Bozic, Gerold, and Nowak]{bozic2016quantifying}
Ivana Bozic, Jeffrey~M Gerold, and Martin~A Nowak.
\newblock Quantifying clonal and subclonal passenger mutations in cancer
  evolution.
\newblock \emph{PLoS Comput.~Biol.}, 12\penalty0 (2):\penalty0 e1004731, 2016.

\bibitem[Bozic et~al.(2019)Bozic, Paterson, and Waclaw]{bozic2019measuring}
Ivana Bozic, Chay Paterson, and Bartlomiej Waclaw.
\newblock On measuring selection in cancer from subclonal mutation frequencies.
\newblock \emph{PLoS Comput.~Biol.}, 15\penalty0 (9):\penalty0 e1007368, 2019.

\bibitem[Burrell et~al.(2013)Burrell, McGranahan, Bartek, and
  Swanton]{burrell2013causes}
Rebecca~A Burrell, Nicholas McGranahan, Jiri Bartek, and Charles Swanton.
\newblock The causes and consequences of genetic heterogeneity in cancer
  evolution.
\newblock \emph{Nature}, 501\penalty0 (7467):\penalty0 338--345, 2013.

\bibitem[Cannone(2017)]{Canonne2017}
C.~Cannone.
\newblock A short note on {P}oisson tail bounds.
\newblock 2017.
\newblock
  \url{http://www.cs.columbia.edu/~ccanonne/files/misc/2017-poissonconcentration.pdf}.

\bibitem[Caravagna et~al.(2020)Caravagna, Heide, Williams, Zapata, Nichol,
  Chkhaidze, Cross, Cresswell, Werner, Acar, et~al.]{caravagna2020subclonal}
Giulio Caravagna, Timon Heide, Marc~J Williams, Luis Zapata, Daniel Nichol,
  Ketevan Chkhaidze, William Cross, George~D Cresswell, Benjamin Werner, Ahmet
  Acar, et~al.
\newblock Subclonal reconstruction of tumors by using machine learning and
  population genetics.
\newblock \emph{Nat.~Genet.}, 52\penalty0 (9):\penalty0 898--907, 2020.

\bibitem[Champagnat et~al.(2012)Champagnat, Lambert, and
  Richard]{champagnat2012birth}
Nicolas Champagnat, Amaury Lambert, and Mathieu Richard.
\newblock Birth and death processes with neutral mutations.
\newblock \emph{Int.~J.~Stoch.~Anal.}, 2012, 2012.

\bibitem[Cheek and Antal(2018)]{cheek2018mutation}
David Cheek and Tibor Antal.
\newblock Mutation frequencies in a birth--death branching process.
\newblock \emph{Ann.~Appl.~Probab.}, 28\penalty0 (6):\penalty0 3922--3947,
  2018.

\bibitem[Davis et~al.(2017)Davis, Gao, and Navin]{davis2017tumor}
Alexander Davis, Ruli Gao, and Nicholas Navin.
\newblock Tumor evolution: Linear, branching, neutral or punctuated?
\newblock \emph{Biochim.~Biophys.~Acta Rev.~Cancer}, 1867\penalty0
  (2):\penalty0 151--161, 2017.

\bibitem[Del~Monte(2009)]{del2009does}
Ugo Del~Monte.
\newblock Does the cell number 109 still really fit one gram of tumor tissue?
\newblock \emph{Cell Cycle}, 8\penalty0 (3):\penalty0 505--506, 2009.

\bibitem[Dinh et~al.(2020)Dinh, Jaksik, Kimmel, Lambert, Tavar{\'e},
  et~al.]{dinh2020statistical}
Khanh~N Dinh, Roman Jaksik, Marek Kimmel, Amaury Lambert, Simon Tavar{\'e},
  et~al.
\newblock Statistical inference for the evolutionary history of cancer genomes.
\newblock \emph{Stat.~Sci.}, 35\penalty0 (1):\penalty0 129--144, 2020.

\bibitem[Durrett(2008)]{durrett2008probability}
Richard Durrett.
\newblock \emph{Probability models for DNA sequence evolution}.
\newblock Springer Science \& Business Media, 2008.

\bibitem[Durrett(2015)]{durrett2015branching}
Richard Durrett.
\newblock Branching process models of cancer.
\newblock In \emph{Branching Process Models of Cancer}, pages 1--63. Springer,
  2015.

\bibitem[Durrett(2013)]{durrett2013population}
Rick Durrett.
\newblock Population genetics of neutral mutations in exponentially growing
  cancer cell populations.
\newblock \emph{Ann.~Appl.~Propab.}, 23\penalty0 (1):\penalty0 230, 2013.

\bibitem[Fay and Wu(2000)]{fay2000hitchhiking}
Justin~C Fay and Chung-I Wu.
\newblock Hitchhiking under positive darwinian selection.
\newblock \emph{Genetics}, 155\penalty0 (3):\penalty0 1405--1413, 2000.

\bibitem[Fu and Li(1993)]{fu1993statistical}
Yun-Xin Fu and Wen-Hsiung Li.
\newblock Statistical tests of neutrality of mutations.
\newblock \emph{Genetics}, 133\penalty0 (3):\penalty0 693--709, 1993.

\bibitem[George(1973)]{george1973nested}
Alan George.
\newblock Nested dissection of a regular finite element mesh.
\newblock \emph{SIAM J.~Numer.~Anal.}, 10\penalty0 (2):\penalty0 345--363,
  1973.

\bibitem[Griffiths and Pakes(1988)]{griffiths1988infinite}
Robert~C Griffiths and Anthony~G Pakes.
\newblock An infinite-alleles version of the simple branching process.
\newblock \emph{Adv.~Appl.~Propab.}, pages 489--524, 1988.

\bibitem[Heide et~al.(2018)Heide, Zapata, Williams, Werner, Caravagna, Barnes,
  Graham, and Sottoriva]{heide2018reply}
Timon Heide, Luis Zapata, Marc~J Williams, Benjamin Werner, Giulio Caravagna,
  Chris~P Barnes, Trevor~A Graham, and Andrea Sottoriva.
\newblock Reply to ‘neutral tumor evolution?’.
\newblock \emph{Nat.~Genet.}, 50\penalty0 (12):\penalty0 1633--1637, 2018.

\bibitem[Iwasa et~al.(2006)Iwasa, Nowak, and Michor]{iwasa2006evolution}
Yoh Iwasa, Martin~A Nowak, and Franziska Michor.
\newblock Evolution of resistance during clonal expansion.
\newblock \emph{Genetics}, 172\penalty0 (4):\penalty0 2557--2566, 2006.

\bibitem[Jones et~al.(2008)Jones, Chen, Parmigiani, Diehl, Beerenwinkel, Antal,
  Traulsen, Nowak, Siegel, Velculescu, et~al.]{jones2008comparative}
Si{\^a}n Jones, Wei-dong Chen, Giovanni Parmigiani, Frank Diehl, Niko
  Beerenwinkel, Tibor Antal, Arne Traulsen, Martin~A Nowak, Christopher Siegel,
  Victor~E Velculescu, et~al.
\newblock Comparative lesion sequencing provides insights into tumor evolution.
\newblock \emph{Proc.~Natl.~Acad.~Sci.~USA}, 105\penalty0 (11):\penalty0
  4283--4288, 2008.

\bibitem[Keller and Antal(2015)]{keller2015mutant}
Peter Keller and Tibor Antal.
\newblock Mutant number distribution in an exponentially growing population.
\newblock \emph{J.~Stat.~Mech.~Theory Exp.}, 2015\penalty0 (1):\penalty0
  P01011, 2015.

\bibitem[Kessler and Levine(2013)]{kessler2013large}
David~A Kessler and Herbert Levine.
\newblock Large population solution of the stochastic {L}uria--{D}elbr{\"u}ck
  evolution model.
\newblock \emph{Proc.~Natl.~Acad.~Sci.~USA}, 110\penalty0 (29):\penalty0
  11682--11687, 2013.

\bibitem[Kessler and Levine(2015)]{kessler2015scaling}
David~A Kessler and Herbert Levine.
\newblock Scaling solution in the large population limit of the general
  asymmetric stochastic {L}uria--{D}elbr{\"u}ck evolution process.
\newblock \emph{J.~Stat.~Phys.}, 158\penalty0 (4):\penalty0 783--805, 2015.

\bibitem[Kimura(1968)]{kimura1968genetic}
Motoo Kimura.
\newblock Genetic variability maintained in a finite population due to
  mutational production of neutral and nearly neutral isoalleles.
\newblock \emph{Genet.~Res.}, 11\penalty0 (3):\penalty0 247--270, 1968.

\bibitem[Kimura(1969)]{kimura1969number}
Motoo Kimura.
\newblock The number of heterozygous nucleotide sites maintained in a finite
  population due to steady flux of mutations.
\newblock \emph{Genetics}, 61\penalty0 (4):\penalty0 893, 1969.

\bibitem[Kingman(1982{\natexlab{a}})]{kingman1982genealogy}
John~FC Kingman.
\newblock On the genealogy of large populations.
\newblock \emph{J.~Appl.~Propab.}, 19:\penalty0 27--43, 1982{\natexlab{a}}.

\bibitem[Kingman(1982{\natexlab{b}})]{kingman1982coalescent}
John Frank~Charles Kingman.
\newblock The coalescent.
\newblock \emph{Stoch.~Process.~Their Appl.}, 13\penalty0 (3):\penalty0
  235--248, 1982{\natexlab{b}}.

\bibitem[Knudson(1971)]{knudson1971mutation}
Alfred~G Knudson.
\newblock Mutation and cancer: statistical study of retinoblastoma.
\newblock \emph{Proc.~Natl.~Acad.~Sci.~USA}, 68\penalty0 (4):\penalty0
  820--823, 1971.

\bibitem[Komarova et~al.(2007)Komarova, Wu, and Baldi]{komarova2007fixed}
Natalia~L Komarova, Lin Wu, and Pierre Baldi.
\newblock The fixed-size luria--delbruck model with a nonzero death rate.
\newblock \emph{Math.~Biosci.}, 210\penalty0 (1):\penalty0 253--290, 2007.

\bibitem[Kuipers et~al.(2017)Kuipers, Jahn, Raphael, and
  Beerenwinkel]{kuipers2017single}
Jack Kuipers, Katharina Jahn, Benjamin~J Raphael, and Niko Beerenwinkel.
\newblock Single-cell sequencing data reveal widespread recurrence and loss of
  mutational hits in the life histories of tumors.
\newblock \emph{Genome Res.}, 27\penalty0 (11):\penalty0 1885--1894, 2017.

\bibitem[Lambert(2009)]{lambert2009allelic}
Amaury Lambert.
\newblock The allelic partition for coalescent point processes.
\newblock \emph{Markov Process.~Relat.~Fields}, 15\penalty0 (3):\penalty0
  359--386, 2009.

\bibitem[Ling et~al.(2015)Ling, Hu, Yang, Yang, Li, Lin, Chen, Dong, Cao, Tao,
  et~al.]{ling2015extremely}
Shaoping Ling, Zheng Hu, Zuyu Yang, Fang Yang, Yawei Li, Pei Lin, Ke~Chen, Lili
  Dong, Lihua Cao, Yong Tao, et~al.
\newblock Extremely high genetic diversity in a single tumor points to
  prevalence of non-darwinian cell evolution.
\newblock \emph{Proc.~Natl.~Acad.~Sci.~USA}, 112\penalty0 (47):\penalty0
  E6496--E6505, 2015.

\bibitem[McDonald et~al.(2018)McDonald, Chakrabarti, and
  Michor]{mcdonald2018currently}
Thomas~O McDonald, Shaon Chakrabarti, and Franziska Michor.
\newblock Currently available bulk sequencing data do not necessarily support a
  model of neutral tumor evolution.
\newblock \emph{Nat.~Genet.}, 50\penalty0 (12):\penalty0 1620--1623, 2018.

\bibitem[McGranahan and Swanton(2017)]{mcgranahan2017clonal}
Nicholas McGranahan and Charles Swanton.
\newblock Clonal heterogeneity and tumor evolution: past, present, and the
  future.
\newblock \emph{Cell}, 168\penalty0 (4):\penalty0 613--628, 2017.

\bibitem[Nowell(1976)]{nowell1976clonal}
Peter~C Nowell.
\newblock The clonal evolution of tumor cell populations.
\newblock \emph{Science}, 194\penalty0 (4260):\penalty0 23--28, 1976.

\bibitem[O'Connell(1993)]{o1993yule}
Neil O'Connell.
\newblock Yule process approximation for the skeleton of a branching process.
\newblock \emph{J.~Appl.~Propab.}, 30\penalty0 (3):\penalty0 725--729, 1993.

\bibitem[Ohtsuki and Innan(2017)]{ohtsuki2017forward}
Hisashi Ohtsuki and Hideki Innan.
\newblock Forward and backward evolutionary processes and allele frequency
  spectrum in a cancer cell population.
\newblock \emph{Theor.~Popul.~Biol.}, 117:\penalty0 43--50, 2017.

\bibitem[Pakes(1989)]{pakes1989infinite}
Anthony~G Pakes.
\newblock An infinite alleles version of the markov branching process.
\newblock \emph{J.~Aust.~Math.~Soc.}, 46\penalty0 (1):\penalty0 146--169, 1989.

\bibitem[Rew and Wilson(2000)]{rew2000cell}
DA~Rew and GD~Wilson.
\newblock Cell production rates in human tissues and tumours and their
  significance. part ii: clinical data.
\newblock \emph{Eur.~J.~Surg.~Oncol.}, 26\penalty0 (4):\penalty0 405--417,
  2000.

\bibitem[Simkin and Roychowdhury(2011)]{simkin2011re}
Mikhail~V Simkin and Vwani~P Roychowdhury.
\newblock Re-inventing willis.
\newblock \emph{Phys.~Rep.}, 502\penalty0 (1):\penalty0 1--35, 2011.

\bibitem[Sottoriva et~al.(2015)Sottoriva, Kang, Ma, Graham, Salomon, Zhao,
  Marjoram, Siegmund, Press, Shibata, et~al.]{sottoriva2015big}
Andrea Sottoriva, Haeyoun Kang, Zhicheng Ma, Trevor~A Graham, Matthew~P
  Salomon, Junsong Zhao, Paul Marjoram, Kimberly Siegmund, Michael~F Press,
  Darryl Shibata, et~al.
\newblock A big bang model of human colorectal tumor growth.
\newblock \emph{Nat.~Genet.}, 47\penalty0 (3):\penalty0 209--216, 2015.

\bibitem[Tajima(1989)]{tajima1989statistical}
Fumio Tajima.
\newblock Statistical method for testing the neutral mutation hypothesis by dna
  polymorphism.
\newblock \emph{Genetics}, 123\penalty0 (3):\penalty0 585--595, 1989.

\bibitem[Tarabichi et~al.(2018)Tarabichi, Martincorena, Gerstung, Leroi,
  Markowetz, Spellman, Morris, Lingj{\ae}rde, Wedge, and
  Van~Loo]{tarabichi2018neutral}
Maxime Tarabichi, I{\~n}igo Martincorena, Moritz Gerstung, Armand~M Leroi,
  Florian Markowetz, Paul~T Spellman, Quaid~D Morris, Ole~Christian
  Lingj{\ae}rde, David~C Wedge, and Peter Van~Loo.
\newblock Neutral tumor evolution?
\newblock \emph{Nat.~Genet.}, 50\penalty0 (12):\penalty0 1630--1633, 2018.

\bibitem[Tomasetti et~al.(2013)Tomasetti, Vogelstein, and
  Parmigiani]{tomasetti2013half}
Cristian Tomasetti, Bert Vogelstein, and Giovanni Parmigiani.
\newblock Half or more of the somatic mutations in cancers of self-renewing
  tissues originate prior to tumor initiation.
\newblock \emph{Proc.~Natl.~Acad.~Sci.~USA}, 110\penalty0 (6):\penalty0
  1999--2004, 2013.

\bibitem[Turajlic et~al.(2019)Turajlic, Sottoriva, Graham, and
  Swanton]{turajlic2019resolving}
Samra Turajlic, Andrea Sottoriva, Trevor Graham, and Charles Swanton.
\newblock Resolving genetic heterogeneity in cancer.
\newblock \emph{Nat.~Rev.~Genet.}, 20\penalty0 (7):\penalty0 404--416, 2019.

\bibitem[Venkatesan and Swanton(2016)]{venkatesan2016tumor}
Subramanian Venkatesan and Charles Swanton.
\newblock Tumor evolutionary principles: how intratumor heterogeneity
  influences cancer treatment and outcome.
\newblock \emph{Am.~Soc.~Clin.~Oncol.~Educ.~Book}, 36:\penalty0 e141--e149,
  2016.

\bibitem[Vogelstein et~al.(2013)Vogelstein, Papadopoulos, Velculescu, Zhou,
  Diaz, and Kinzler]{vogelstein2013cancer}
Bert Vogelstein, Nickolas Papadopoulos, Victor~E Velculescu, Shibin Zhou,
  Luis~A Diaz, and Kenneth~W Kinzler.
\newblock Cancer genome landscapes.
\newblock \emph{Science}, 339\penalty0 (6127):\penalty0 1546--1558, 2013.

\bibitem[Watterson(1975)]{watterson1975number}
GA~Watterson.
\newblock On the number of segregating sites in genetical models without
  recombination.
\newblock \emph{Theor.~Popul.~Biol.}, 7\penalty0 (2):\penalty0 256--276, 1975.

\bibitem[Werner et~al.(2018)Werner, Williams, Barnes, Graham, and
  Sottoriva]{werner2018reply}
Benjamin Werner, Marc~J Williams, Chris~P Barnes, Trevor~A Graham, and Andrea
  Sottoriva.
\newblock Reply to ‘currently available bulk sequencing data do not
  necessarily support a model of neutral tumor evolution’.
\newblock \emph{Nat.~Genet.}, 50\penalty0 (12):\penalty0 1624--1626, 2018.

\bibitem[Werner et~al.(2020)Werner, Case, Williams, Chkhaidze, Temko,
  Fern{\'a}ndez-Mateos, Cresswell, Nichol, Cross, Spiteri,
  et~al.]{werner2020measuring}
Benjamin Werner, Jack Case, Marc~J Williams, Ketevan Chkhaidze, Daniel Temko,
  Javier Fern{\'a}ndez-Mateos, George~D Cresswell, Daniel Nichol, William
  Cross, Inmaculada Spiteri, et~al.
\newblock Measuring single cell divisions in human tissues from multi-region
  sequencing data.
\newblock \emph{Nat.~Commun.}, 11\penalty0 (1):\penalty0 1--9, 2020.

\bibitem[Williams et~al.(2016)Williams, Werner, Barnes, Graham, and
  Sottoriva]{williams2016identification}
Marc~J Williams, Benjamin Werner, Chris~P Barnes, Trevor~A Graham, and Andrea
  Sottoriva.
\newblock Identification of neutral tumor evolution across cancer types.
\newblock \emph{Nat.~Genet.}, 48\penalty0 (3):\penalty0 238, 2016.

\bibitem[Williams et~al.(2018)Williams, Werner, Heide, Curtis, Barnes,
  Sottoriva, and Graham]{williams2018quantification}
Marc~J Williams, Benjamin Werner, Timon Heide, Christina Curtis, Chris~P
  Barnes, Andrea Sottoriva, and Trevor~A Graham.
\newblock Quantification of subclonal selection in cancer from bulk sequencing
  data.
\newblock \emph{Nat.~Genet.}, 50\penalty0 (6):\penalty0 895--903, 2018.

\bibitem[Wu and Kimmel(2013)]{wu2013modeling}
Xiaowei Wu and Marek Kimmel.
\newblock Modeling neutral evolution using an infinite-allele markov branching
  process.
\newblock \emph{Int.~J.~Stoch.~Anal.}, 2013:\penalty0 1--10, 2013.

\bibitem[Yule(1925)]{yule1925ii}
George~Udny Yule.
\newblock I{I}.~—-{A} mathematical theory of evolution, based on the
  conclusions of {Dr. J.C.~Willis, F.R.~S}.
\newblock \emph{Philos.~Trans.~R.~Soc.~Lond., B, Biol.~Sci.}, 213\penalty0
  (402-410):\penalty0 21--87, 1925.

\bibitem[Zeng et~al.(2006)Zeng, Fu, Shi, and Wu]{zeng2006statistical}
Kai Zeng, Yun-Xin Fu, Suhua Shi, and Chung-I Wu.
\newblock Statistical tests for detecting positive selection by utilizing
  high-frequency variants.
\newblock \emph{Genetics}, 174\penalty0 (3):\penalty0 1431--1439, 2006.

\end{thebibliography}

\end{document}